# A Spline-based Volumetric Data Modeling Framework and Its Applications

Bo Li

Master Thesis

Department of Computer Science
Stony Brook University

Advisor: Professor Hong Qin

January 30, 2012

## Abstract

The strategic technical vision of this thesis proposal is to seek to systematically trailblaze a novel volumetric modeling framework/methdology to represent 3D solids. The rapid advances in 3D scanning and acquisition techniques have given rise to the explosive increase of volumetric digital models in recent years. The strong need to explore more efficient and robust 3D modeling techniques has become prominent. Although the traditional surface representation (e.g., triangle meshes) has many attractive properties, it is incapable of expressing the solid interior space and materials. Such a serious drawback overshadows many potential modeling and analysis applications. Consequently, volumetric modeling techniques become the well-known solution to this problem. Nevertheless, many unsolved research issues still remain outstanding when developing an efficient modeling paradigm for existing 3D models, including complex geometry (fine details and extreme concaveness), arbitrary topology, heterogenous materials, large-scale data storage and processing, etc.

In this thesis proposal, we concentrate on the challenging research issue of developing a spline-based modeling framework, which aims to convert the conventional data (e.g., surface meshes) to tensor-product trivariate splines. This methodology can represent both boundary/volumetric geometry and real volumetric physical attributes in a compact and continuous matter. The regularly-defined tensor-product structure enables our newly-developed methods to be embedded into the CAD-design industry standards such as NURBS and B-splines seamlessly. These properties make our techniques highly preferable in many physically-based applications including mechanical analysis, shape deformation and editing, reverse engineering, hexahedral meshing, virtual surgery training, etc.

Using tensor-product trivariate splines to reconstruct existing 3D objects is highly challenging, which always involves component-aware decomposition, volumetric parameterization, and trivariate spline approximation. This thesis proposal seeks accurate and efficient technical solutions to these fundamental and important problems, and demonstrates their efficiencies in modeling 3D objects of arbitrary topology.

First, in order to achieve a "from surface model to trivariate splines" transformation, we define our new splines upon a novel parametric domain called generalized poly-cubes (GPCs), which comprise a set of regular cube-like domains topologically glued together.

We then further improve our trivariate splines to support arbitrary topology by allowing the divide-and-conquer scheme, i.e., the user can decompose the model into components and represent them using trivariate spline volumetric patches. We design algorithms and prove valuable properties for our powerful merging strategy that can glue tensor-product spline solids together, while preserving many attractive modeling advantages.

We also develop an effective method to reconstruct discrete volumetric datasets (e.g., volumetric images) into continuous trivariate splines. To capture fine features in the data, we construct an as-smooth-as-possible frame field based on locally-computed, 3D principal curvatures to align with a sparse set of directional features. The frame field naturally yields a volumetric parameterization that aids a trivariate spline definition, construction, and approximation.

Finally, we focus on broadening visual computing applications of our powerful modeling techniques from geometric modeling and scientific visualization. We present a novel methodology based on geometric deformation metrics to simulate magnification lens that can be utilized for the Focus+Context (F+C) visualization.

Through our extensive experiments, we hope to demonstrate that our framework is an effective and powerful tool that augments existing modeling techniques. In this thesis proposal, we also outline our on-going research work such as feature-sensitive, edge-aware volumetric splines and volumetric lens sim-ulation, which will naturally lead toward our near-future plans and comprise a part of our PhD dissertation topics.

# Contents

# Chapter 1

# Introduction

## 1.1 Problem Statements

Since the starting point of computer graphics research and application, surface shape modeling and designing have always been the central issue, mainly because shape design keeps acting as the core applications in industry as while as lacking real 3D data.

In the recent years, we have witnessed a great potential of paradigm shift from surface-only to volume data. Behind this are the rapidly developing 3D data acquisition techniques and the urgent 3D analysis requirement: proliferation of modern 3D scanning devices and shape modeling technologies give rise to the huge number of available high quality 3D dataset. As a result, the need to the ability of making good use of existing models has gained the prominence; Many computer graphics research and industry applications benefit tremendously from this trend: As a direct downstream application, we can now, in the first time, efficiently and robustly adapt real heterogenous material data onto 3D objects, which will significantly improve the physical analysis. Consequently, these newly emerged 3D datasets, as a novel data platform, may lead to a revolutionary transformation and update from existing graphics (e.g., deformation, simulation), visualization (e.g., rendering) and modeling (e.g., multivariate splines) techniques.

Consequently, we now desire to explore more efficient and robust 3D volume data modeling framework to suffice the exciting age of discovery in the above topics. This direction is always accompanied by many challenges. In detail, the difficulties arise from the fact that the quality criteria are diverse and their optimization often requires the consideration of tradeoff on specific applications. The most common quality aspects involve: The representation format must be flexible and powerful to describe complex shape and arbitrary topology; From perspective of analysis brings out the request that it should be simple and analytic; In physical analysis and texturing, we also need to represent both geometry and materials in our data structure; Meanwhile, a volumetric data normally has very large data scales with an explosive growth in the time and memory cost.

In our proposal, we specifically advocate a spline-based framework which can trade-off above requirements well. A key concern of engineering design industry is: These data have to be converted to continuous, compact representations to enable geometric design and downstream product development processes (e.g., finite element analysis, physical simulation, virtual surgery, etc) in CAD environments. As the natural correspondence, spline schemes and relative techniques have been extensively investigated during the recent past to fulfill the aforementioned goal. The material data can also be easily adapted by using multivariate splines. We will achieve a more compact representation of curves, surfaces or volumes at different scales in terms of data size, the number of control points, the user-specified threshold error, and other relevant criteria. We can compute all the differential quantities such as geodesics, curvatures, tensor fields without resorting to any numerical approximations via linear interpolation and/or local algebraic fitting. The rapid and precise evaluations of local and global differential properties will facilitate many applications such as finite element analysis, image registration/segmentation, shape modificaton/integration, surface quality analysis and control, and scientific visualization etc.

We observe that current spline prototypes are frequently based on 2-manifolds geometry and topology (i.e., "surface splines"). Typically, this representation describes the boundary of a solid model. However, a volumetric spline scheme has gathered growing interest from both analysis and CAD research communities, due to its computational advantage over traditional surface-based analysis method and its promise to alleviate the burden of creating effective 3D interior analysis-ready domains in many solid modeling and volume graphics applications. Iso-geometric analysis is an example to illustrate this necessity. NURBS based isogeometric analysis leverages the possible advantages of closer integration of CAD and FEA in isogeometric analysis. However, the critical challenge is how to convert a volumetric NURBS from its original boundary shape NURBS. This is because that any accurate physical analysis approach is based on a volumetric formulation where trivariate NURBS solids are needed for the analysis of three-dimensional (3D) problems while CAD systems use a boundary representation where only a surface representation is available. Although several numerical techniques such as boundary integration are amenable to several specific problems, it is more useful to provide the real trivariate NURBS geometry and material for generalized applications.

Existing volumetric spline techniques generally follow two different trends: (1) Many recent methods divide the volume space into a tetrahedral mesh domain then construct a trivariate spline (like super spline or box spline) on each tetrahedra domain. These unregular-domain spline theories have just emerged recently, and have not been recognized by the communities outside computer graphics. At present, the regular tensor-product B-splines (NURBS) are still the prevailing industrial standard for freeform surface representation. (2) In contrast, many recent techniques [86], [87], [166] attempt to convert each part into splines defined on a cylinder/tube domain, because they can intuitively use the shape skeleton to produce a tube domain and reveal the global structure and topology. A severe limitation of such approaches is that points on the tube centerline are all singular. Also, the shape of tube is very simple such that it can not support complex shape and preserve any sharp edge and point feature when it serves as the domain.

An ideal volumetric spline modeling framework should have the following properties:

(1) Singularity free. A *singular point* in volumetric domain is a node with valence larger than four on an iso-parametric plane (Fig. 3.1(a-b)). Handling singularity with tensor-product splines is extremely challenging. It is desirable to have a global one-piece spline defined on a globally-connected singularity-free domain.

(2) The proposed domain construction method must be sufficient for surface with boundaries/complex shapes/arbitrary topology/long branches. The only feasible way is to introduce additional cuts and decompose the model into reasonable elements. Each element should abstract a component-aware part in a geometrically meaningful way thus make the following spline fitting process accurate and numerically stable. Also, the separate elements must be glued in a simple and singularity-free fashion.

(3) A practical volumetric parameterization technique must preserve shape feature. Specifically, in areas with well-pronounced consistent curvature directions, patch parametric lines should follow the curvature and patch boundaries should be aligned with sharp features and smooth surface boundaries. Moreover, an improved parameterization method should develop an efficient and systematical framework to better address the heterogenous model with various interior materials.

(4) In our new designed trivariate spline scheme, we desire to inherit the attractive properties of prevailing industrial standard NURBS. For example, NURBS have local support, i.e., moving one control point will only affect its immediate neighborhood. This makes intuitive design with NURBS possible; The basis functions of NURBS are non-negative, have the property of partition-of-unity, thus are qualified as basis functions required by finite element method; Non-uniform knot can confine the basis function inside the domain completely.

(5) We urgently need to design a more efficiently fitting pipeline to handle large scale computation during trivariate spline approximation. For example, a genus-0 solid bounded by 6 simple four-sided B-spline surfaces has originally $6 \times 1024^2$ control points (DOFs). The size of DOFs increases drastically to $1024^3$ or even larger when we naively convert it to a volumetric spline representation. This exponential increase during volumetric spline conversion poses a great challenge in terms of both storage and fitting costs.

In conclusion, our modeling framework involves 3 main challenges and all above requirements can be categorized into them:

(1) Mesh decomposition.

- How to decompose them into component-aware parts?
- How to design a practical or automatic scheme to generate consistent partitioning, with a small number of parts and spline-friendly domain shapes and gluing types ?

(2) Volumetric parameterization.

- How to reduce the computation complexity of volumetric mapping and make it more robustly?
- How to analyze and restrict the mapping distortion?
- How to integrate the shape feature (like sharp edges, corners), or even various materials (like density value) into our parameterization result?

(3) Trivariate splines.

- How to preserve the critical properties of NURBS surface like partition-of-unity, local refinement and boundary confinement?
- How to decrease the control point number to adapt huge number of degree-of-freedom in trivariate splines?
- How to accelerate fitting efficiency and save fitting cost (time and storage).
- How to handle multivariate splines for many applications like vector volume imaging.

To overcome the above modeling and design difficulties and address the topological issue, we seek novel modeling techniques based on tensor-product spline schemes that would allow designers to directly define continuous spline models over any manifold (serving as parametric domain). Such a global approach would have many modeling benefits, including no need of the transition from local patch definition to global surface construction via gluing and abutting, the elimination of non-intuitive segmentation and patching process, and ensuring the high-order continuity requirements. More importantly, we can expect a true one-piece representation for shapes of complicated topology, with a hope to automate the entire reverse engineering process (by converting points and/or polygonal meshes to spline surfaces with high accuracy) without human intervention.

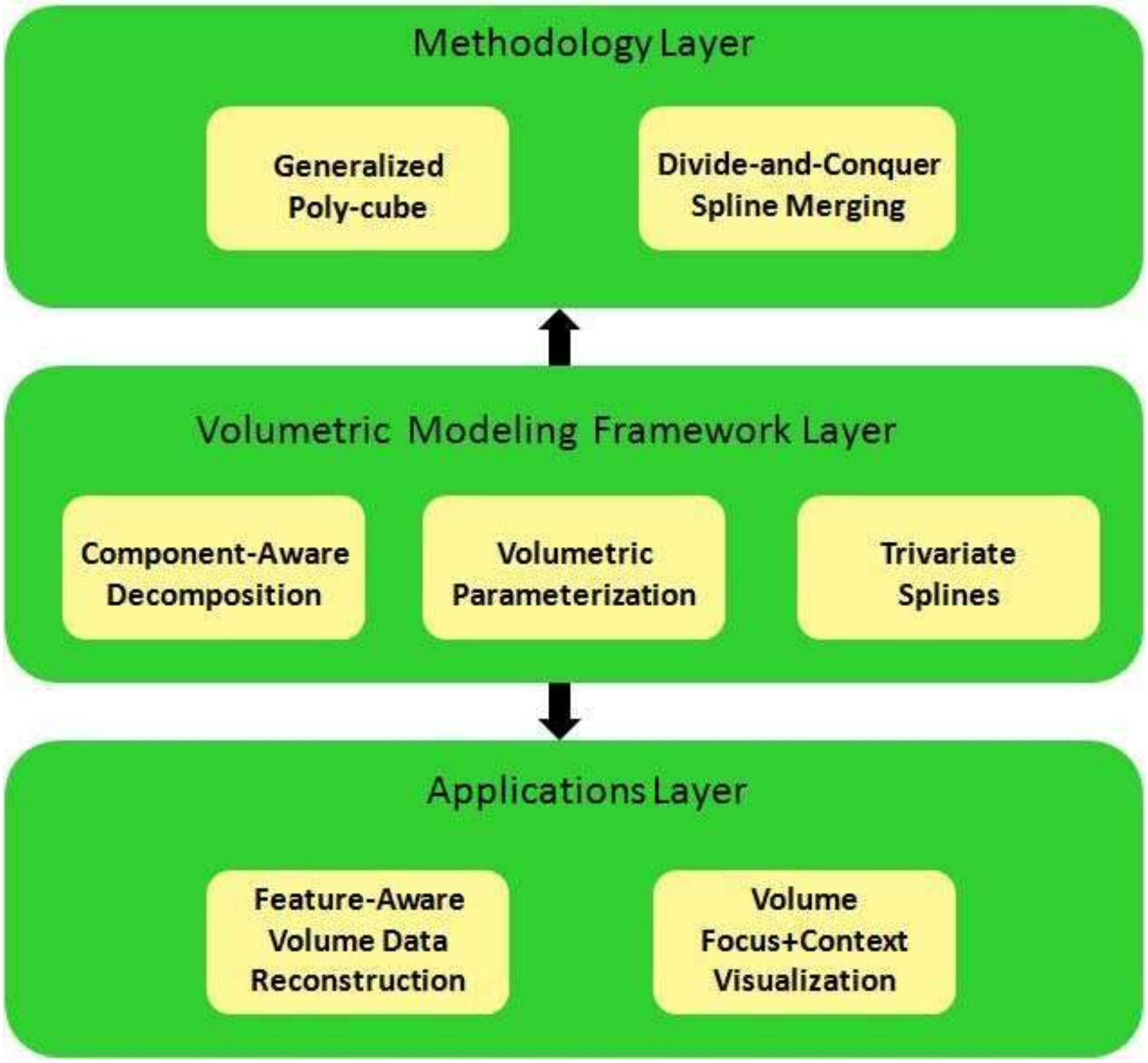

Figure 1.1: Hierarchy of our research contents. Key streamline of our framework (middle row); Main techniques for trivariate spline modeling (upper row); Utilized applications (bottom row).

Towards this goal, we present a novel spline-based solid modeling framework. Figure 1.1 illustrates the conceptual hierarchy of above discussions and the whole proposal. This framework integrates a few projects (first row) and targets on key challenging problems (third row). By solving these key difficulties we have improved the effectiveness and efficiency of shape mapping computation, and are able to utilize this framework into various applications (bottom row).

Through our experiments, we hope to demonstrate that the proposed data modeling framework is very flexible and can potentially serve as a geometric standard for product data representation and model conversion in shape design and geometric processing.

## 1.2 Contributions

In this thesis proposal, we present a spline-based framework to solve 3D objects modeling problems. Particularly, we emphasize our research interest on regular domain ("cuboid") tensor-product splines, because of their favorite advantages. Combining volumetric decomposition, parameterization with trivariate splines, we successfully and effectively solve a variety problems in the areas of geometric shape design and modeling.

Our specific contributions include:

1. We propose a new concept of *"Generalized poly-cube"* (GPC). A GPC comprises a set of regular cube domains topologically glued together. Compared with conventional poly-cubes (CPCs), GPC is much more powerful and flexible and has improved numerical accuracy and computational efficiency. We propose an automatic method to construct a GPC domain and we develop a novel volumetric parameterization and spline construction framework based on the resulting domain, which is an effective modeling tool for converting surface meshes to volumetric splines (Chapter 3).

2. We design a novel component-aware shape modeling methodology based on tensor-product trivariate splines for solids with arbitrary topology. Instead of using conventional top-down method, our framework advocates a divide-and-conquer strategy: The model is first decomposed into a set of components and then each component is natually modeled as tensor-product trivariate splines. The key novelty lies at our powerful merging strategy that can glue tensor-product spline solids together subject to high-order global continuities, meanwhile preserving boundary restriction and semi-standardness (Chapter 4).

3. We propose a systematic framework that transforms discrete volumetric raw data from scanning devices directly into continuous spline representation with regular tensor-product structure. To achieve this goal, we propose a novel volumetric parameterization technique that constructs an as-smooth-as-possible frame field, satisfying a sparse set of directional constraints, and we compute a globally smooth parameterization with iso-parameter curves following the frame field directions. The proposed method can efficiently reconstruct model with multi-layers and heterogenous materials, which are usually extremely difficult to be handled by the traditional techniques 5.

4. Aiming to promote new applications of our powerful modeling techniques in visual computing, we present a novel methodology based on geometric deformation metrics to simulate magnification lens that can be utilized for Focus+Context (F+C) visualization. Compared with conventional optical lens design (such as fish-eyes, bi-focal lens), our geometric modeling based method are much more capable of preserving shape features (such as angles, rigidities) and minimizing distortion Chapter 6.

## 1.3 Research Plan

To date, we have completed the following work:

### 1.3.1 Summary of Current Work

1. We design an interface to incorporate multivariate spline schemes with a variational framework for solving general numerical problems by using finite-element method.

2. We have developed a set of toolkits to facilitate image segmentation, feature extraction and pairing, and other procedure in image registration.

3. We incorporate thin-shell elastic models in our framework and successfully solve Kirchhoff-Love equations on manifold domain, with triangular B-spline finite element method.

4. We extend the construction algorithm of T-spline from 2D to 3D, and write a library to build volumetric T-spline in octree domain. The evaluation of spline value and relevant derivatives are also implemented.

### 1.3.2 Summary of Remaining Work

To complete the spline-based data modeling framework, we will continue to conquer the following problems:

1. Our primary goal is to construct a volumetric parameterization/remeshing that maps to a conventional/gneralized poly-cube domain, while mapping important features (sharp lines/feature points) to the cuboid domain edges/corners. Currently some existing research methods only focus on feature-aware quadrangulation on surface mesh. We will strengthen current framework to facilitate volumetric poly-cube generation, and more importantly, enforce points-to-corners mapping and lines-to-edges mappings.

2. We extend our core framework as a “reverse-parameterization” process to volume data. Instead of mapping a high-dimension object into a low-dimension space, we attempt to reversely map a low-dimension object into a high-dimension space, such that the visual information is enlarged. Since we have demonstrated efficiency and robustness of our metric-based framework, now we generalize the same idea onto 3D volume images.

## 1.4 Thesis Proposal Organization

The remainder of the thesis proposal is organized in the following fashion. In Chapter 2, we begin with the detailed review prior research work related to component-aware mesh decomposition, volumetric parameterization and trivariate splines with regular structures. In Chapter 3, we present a novel modeling concept “Generalized poly-cube”, and develop an automatic modeling framework using GPC to convert a surface mesh into volumetric splines. In Chapter 4, we propose a new bottom-up paradigm that decomposes a surface model into separate spline patches and then integrates them into a global continuous formulation. We design a new spline merging algorithm to guarantee high-order continuities while keeping all other spline properties. In Chapter 5, we propose a trivariate spline-based approach that is able to reconstruct discrete volumetric data directly acquired from scanning devices into regular tensor-product spline representation. We study a new volumetric frame field and parameterization generation method to achieve reconstruction. In Chapter 6, we apply our geometric modeling method into a visualization application: lens design problem. We integrate a flexible geometric metric to simulate the optical lens and our method is much more capable of preserving shape features (angles and rigidities) and minimizing distortion. Finally, we conclude in Chapter 7 with the summery and the discussion on future research directions. We articulate all useful theoretical propositions and proofs about trivariate splines we develop in this proposal.

# Chapter 2

# Background Review

As we have introduced in Chapter 1, the hierarchy of this thesis includes 3 main steps: decomposition, parameterization and spline construction. Spline and parameterization consist of our primary research topics thus we review them first. We notice that many researchers have explored and studied deeply topics in $R^2$, and since our focus is on volumetric modeling, here we only introduce basic techniques and theories about surface study and main review the work on $R^3$.

## 2.1 Splines

Splines normally refer to smooth, piecewise polynomials. They are ideal tools for applications where continuous representations are critical. Their most common quality aspects involve: The fitting can be piece-wised; The data is highly compressed; The analytic computation is very easy; The format is widely accepted by most design softwares.

The first study on splines goes back to 1946 by Schoenberg. Since then, splines become a very active research because of the fast development of industry application and computer science. Between the 1960's and early the 1970's, Birkhoff, Garabedian and deBoor have studied and established a series of theories on Cartesian regular tensor product splines to represent surface. It is well known that now these types of spline functions become the industry standard and play very important roles in many engineering design applications. Although there are huge number of literatures on many extension types of splines to combat the shortcoming of regular splines (like triangular B-splines, Powell-Sabin splines, etc), their applications only exit in theoretical study and the whole industry still insists on regular splines. Therefore, we shall briefly explain the relative concepts of regular tensor-product splines in the following section. Then, we will pay attention on existing trivariate spline techniques.

### 2.1.1 Polynomials and Polar Forms

The most fundamental class of splines is the class of parametric polynomials. In the context of CAGD and computer graphics, splines are best studied with the help of a classical theoretical foundation like "Polar Form" [108],[121]. All spline theories are covered and generated from the polar form theory. Therefore, we here simply brief the basic idea of the polar form.

**Polar Forms.** The parametric polynomials are the fundamental basis for splines. The polar form is a very important tool for polynomials and thus spline study. The definition of polar form are as follows [122]:

**Definition 2.1.1** (**Affine Map**). *A map* $f : \mathbb{R}^k \to \mathbb{R}^t (k \geq 1)$ *is affine, if and only if it preserves affine combinations, i.e., if and only if f satisfies* $f(\sum_{i=0}^{m} \alpha_i u_i) = \sum_{i=0}^{m} \alpha_i f(u_i)$ *for all scalars* $\alpha_0, \ldots, \alpha_m \in \mathbb{R}$ *with* $\sum_{i=0}^{m} \alpha_i = 1$.

**Definition 2.1.2** (**Symmetric, Multi-Affine**). *Let* $F$ *be an n-variable map.* $F$ *is symmetric if and only*

$$F(u_1, u_2, \cdots, u_n) = F(u_{\pi(1)}, u_{\pi(2)}, \cdots, u_{\pi(n)}).$$

*For all permutations* $\pi \in \sum_n$, *The map* $F$ *is multi-affine if and only if* $F$ *is affine in each argument and the others are held fixed.*

Blossoming principle is a very important express that indicates that any polynomial is equivalent to its polar form [108]:

**Theorem 2.1.3** (**Blossoming Principle**). *Polynomials* $F : \mathbb{R}^k \to \mathbb{R}^t (k \geq 1)$ *of degree* $n$, *and a symmetric multi-affine map* $f : (\mathbb{R}^k)^n \to \mathbb{R}^t$ *are equivalent. Given a map of either type, unique map of the other type exists that satisfies the identity* $F(u) = f(\underbrace{u, \cdots, u}_{n})$. *The map* $f$ *is called the multi-affine polar form or blossom of* $F$.

The property of blossoming principle is used to define deCasteljau algorithm and de Boor algorithm in the following sections.

### 2.1.2 Regular Tensor Product Splines

**Bézier Splines.** Among all regular splines, a Bézier representation in its most common form is the most widely accepted equation that can be used in any number of useful ways. Bézier curves have obtained dominance in the typesetting industry since 1970's. A Bézier spline can be defined as:

**Theorem 2.1.4** (**Bézier Curve**). *Given a set of* $n + 1$ *control points* $P_0, P_1, \ldots, P_n$, *the corresponding*

**Bézier Curve** *is given by*

$$C(t) = \sum_{i=0} P_i B_{i,n}(t),$$

*where* $B_{i,n}(t)$ *is a* Bernstein polynomial $B_{i,n}(t) = C_i^n t^i (1 - t)^{n-i}$ *and* $t \in [0, 1]$.

As we mentioned in the last section, we can also represent Bézier splines of a polynomial $F$ from its polar form like [21]:

**Theorem 2.1.5.** ***(Bézier Points and de Casteljau algorithm)*** *Let* $\Delta = [r, s]$ *be an arbitrary interval. Every polynomial* $F : \mathbb{R} \to \mathbb{R}^t$ *can be represented as a Bézier polynomial w.r.t.* $\Delta$. *The Bézier points are given as*

$$b_j = f(\underbrace{r, \ldots, r}_{n-j}, \underbrace{s, \ldots, s}_{j}),$$

*where* $f$ *is the polar form of* $F$.

Equation above immediately leads to an evaluation algorithm that recursively computes the values

$$\begin{aligned} b_j^l(u) &= f(\underbrace{r,\ldots,r}_{n-l-j},\underbrace{u,\ldots,u}_{l},\underbrace{s,\ldots,s}_{j}) \\ &= \frac{s-u}{s-r} f(\underbrace{r,\ldots,r}_{n-l-j+1},\underbrace{u,\ldots,u}_{l-1},\underbrace{s,\ldots,s}_{j}) + \frac{u-r}{s-r} f(\underbrace{r,\ldots,r}_{n-l-j},\underbrace{u,\ldots,u}_{l-1},\underbrace{s,\ldots,s}_{j+1}) \\ &= \frac{s-u}{s-r} b_j^{l-1}(u) + \frac{u-r}{s-r} b_{j+1}^{l-1}(u) \end{aligned}$$

from the given control points. For $l = n$ we finally compute $b_0^n = f(u,\ldots,u) = F(u)$, which is the desired point on the curve. This algorithm is called *de Casteljau Algorithm* [21].

Formula above also shows that the de Casteljau Algorithm offers a way to subdivide a Bézier curve: suppose that we wish to subdivide a Bézier curve F over a given interval $\Delta = [s,t]$ at an arbitrary parameter $u \in \Delta$. The new Bézier points of the left and right segments $F_l$ and $F_r$ with respect to the subintervals $\Delta_l = [r,u]$ and $\Delta_r = [u,s]$ are given as

$$b_0^l = f(r,\ldots,r), b_1^l = f(r,\ldots,r,u),\ldots,b_n^l = f(u,\ldots,u),$$

and

$$b_0^r = f(u,\ldots,u), b_1^r = f(u,\ldots,u,s),\ldots,b_n^r = f(s,\ldots,s).$$

**B-Splines.** B-splines (short for Basis Splines) go back to Schoenberg who introduced them in 1946 [116, 117] for the case of uniform knots. B-splines over nonuniform knots go back to a review article by Curry in 1947. De Boor derived the recursive evaluation of B-spline curves [10]. It was this recursion that made B-splines a truly viable tool in CAGD. Before its discovery, B-splines were defined using a tedious divided difference approach which was numerically unstable. Later on, Gordon and Riesenfeld realized that de Boor's recursive B-spline evaluation is the natural generalization of the de Casteljau algorithm and Bézier curves are just subset of B-spline curves. Versprille [148] generalization of B-spline curves to NURBS (non-uniform rational B-spline) which has become the standard curve and surface form in the CAD/CAM industry [99].

**Definition 2.1.6** (**B-Spline**). *Let a vector known as the knot vector defined as*

$$T = \{t_0, t_1, \ldots, t_m\}$$

*where* $T$ *is a nondecreasing sequence with* $t_i \in [0,1]$ , *and define control points* $P_0,\ldots,P_n$ . *Define the degree as*

$$p \equiv m - n - 1$$

*The knots* $t_{p+1}, ..., t_{m-p-1}$ *are called internal knots.*
*Define the basis functions as*

$$N_{i,0}(t) = \begin{cases} 1 & \text{if } t_i \le t < t_{i+1} \text{ and } t_i < t_{i+1}; \\ 0 & \text{otherwise}. \end{cases}$$

$$N_{i,p}(t) = \frac{t - t_i}{t_{i+p} - t_i} N_{i,p-1}(t) + \frac{t_{i+p+1} - t}{t_{i+p+1} - t_{i+1}} N_{i+1,p-1}.$$

*Then the curve defined by*

$$C(t) = \sum_{i=0}^{n} P_i N_{i,p}(t)$$

*is a **B-Spline**.*

The B-spline basis functions are positive and form a partition of unity. In addition, they have local support given by $N_i^n(u) = 0$ for $u \notin [t_i, t_{i+n+1}]$ .The knot values determine the extent of the control of the control points.

The B-spline can be divided into different types with respect to knot values:

**Uniform B-spline.** When the knots are equidistant the B-spline is called uniform. The uniform B-spline has a succinct definition:

$$b_{j,n} = b_n(t - t_j),$$

with

$$b_n(t) = \frac{n+1}{n} \sum_{i=0}^{n+1} \mu_{i,n}(t - t_i)_+^n,$$

and

$$\mu_{i,n} = \prod_{j=0, j\neq i}^{n+1} \frac{1}{t_j - t_i}.$$

where $(t - t_i)_+^n$ is the truncated power function:

$$F_+^n = \begin{cases} F^n & \text{if } F \geq 0 \\ 0 & \text{otherwise} \end{cases}$$

**Open-uniform B-spline.** The difference between uniform spline and open-uniform spline is that there exists k degree at the start and end points of the vector knots. This open-uniform B-spline defines the open-uniform basis function. The motivation of open-uniform B-spline comes from the difference of B-spline and Bézier spline. The B-spline can not preserve one property of Bézier spline that the start and end points of the curve are the same points of the first control point and the last control point. Open-uniform B-spline can solve this problem. For instance, if we set the knot vector as (0,0,0,1,1,1), it can be directly proved that the basis function generated from this vector is equal to the degree-2, with 3 control point Bézier curve's basis function. (0,0,0,0,1,1,1,1) is another example that is the same as cubic, with 4 control point Bézier curve.

**Non-uniform B-spline.** B-spline basis function with arbitrary knot vector that follows the definition requirements. Uniform B-spline is special cases of no-uniform.

**Degree of B-spline.** B-spline allows arbitrary degree of B-spline. In practical use the degree is rarely more than 3. So the basis function computing can be specialized for each degree. Figure 2.1illustrates the basis functions in degree 0,1,2.

- Constant B-spline: The constant B-spline is the simplest B-spline. It is defined on only one knot span.

$$N_{j,0}(t) = 1_{[t_j, t_{j+1})} = \begin{cases} 1 & \text{if } t_j < t < t_{j+1}; \\ 0 & \text{otherwise.} \end{cases}$$

- Linear B-spline: The linear B-spline is defined on two knot spans.

$$N_{j,1}(t) = \begin{cases} \frac{t - t_j}{t_{j+1} - t_j} & \text{if } t_j < t < t_{j+1}; \\ \frac{t_{j+2} - t}{t_{j+2} - t_{j+1}} & \text{if } t_{j+1} < t < t_{j+2} \\ 0 & \text{otherwise.} \end{cases}$$

- Uniform quadratic B-spline: the un-uniform quadratic B-spline does not have the uniform expression. Here we write out the blending function for uniform type.

$$N_{j,2}(t) == \begin{cases} \frac{1}{2}t^2 \\ -t^2 + t + \frac{1}{2} \\ \frac{1}{2}(1 - t)^2 \end{cases}$$

.

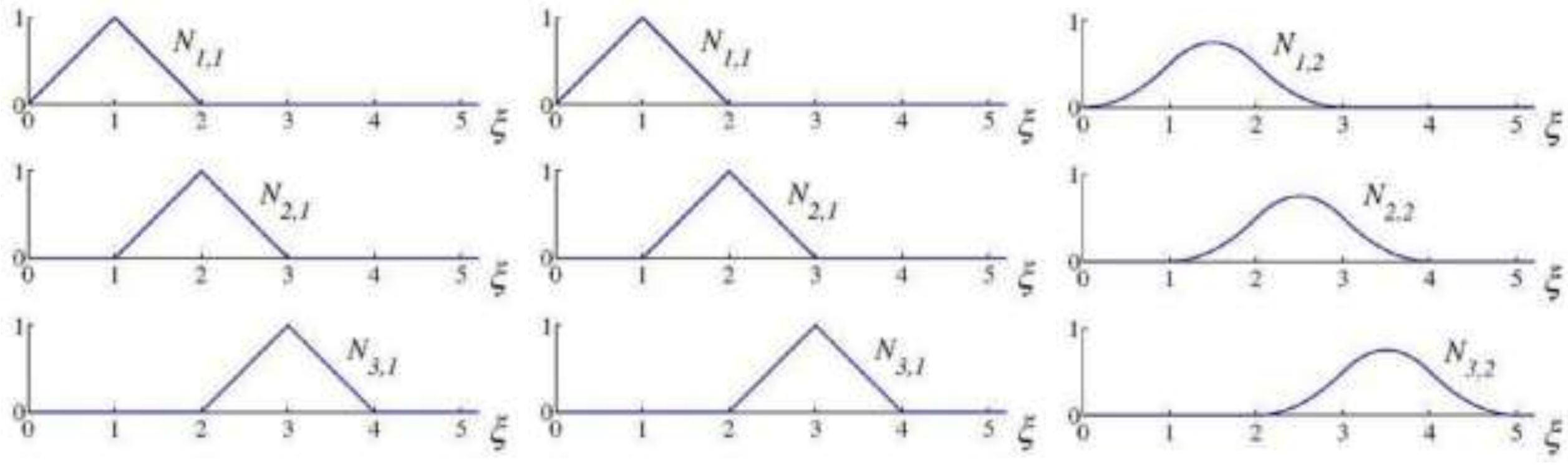

Figure 2.1: Basis functions for B-spline with degree 0,1,2 from left to right.

As we mentioned in the last section, we can also represent B-Splines of a polynomial F from its polar form [11, 107].

**Theorem 2.1.7.** ***(De Boor Points and De Boor Algorithm)*** *Every polynomial*

$$F : \mathbb{R} \to \mathbb{R}^t$$

*can be represented as a B-spline segment over a non-decreasing knot sequence*

$$r_n \leq \ldots \leq r_1 < s_1 \leq \ldots \leq s_n.$$

*The de Boor points are given as*

$$d_j = f(r_1, \ldots, r_{n-j}, s_1, \ldots, s_j),$$

*where* $f$ *is the polar form of* $F$.

**Tensor Product B-spline.** We can extend the B-spline from curve to surface. Tensor product surfaces are the most popular surface design method in theory and industry: Given a curve scheme

$$F(u) = \sum_{i=0} B_i(u) b_i, \quad b_i \in \mathbb{R}^t,$$

the corresponding tensor product scheme is defined as

$$F(u, v) = \sum_{i=0} \sum_{j=0} B_i(u) B_j(v) b_{ij}, \quad b_{ij} \in \mathbb{R}^t,$$

which can also be written as

$$F(u, v) = \sum_{i=0} B_i(u) b_{i_v},$$

with

$$b_{i_v} = b_i(v) = \sum_{j=0} B_j(v) b_{ij}.$$

The last equation demonstrates that tensor product surfaces may be considered as curves of curves.

**NURBS.** B-spline shows that it is a powerful tool for free form curve and surface shape design. However, it has the drawback that can not express exactly the regular shape. The invention of non-uniform rational B-spline (NURBS) is to solve this problem.

**Definition 2.1.8 (NURBS).** *Let a vector known as the knot vector be defined*

$$T = \{t_0 \le t_1 \le \ldots \le t_{k+n} \le t_{k+n+1}\},$$

*with the restriction that the interior knots have at most multiplicity* $n$*, that is* $t_i < t_{i+n}$ *for* $i = 1, 2, \ldots, k$*, define control points* $P_0, \ldots, P_k \in \mathbb{E}^d$*, and define positive weights* $w_0, w_1, \ldots, w_k$*, associated to the control points* $P_i$.
*The analytic representation of the corresponding NURBS curve* $R$ *of degree n in* $\mathbb{E}^d$ *is given by*

$$R(u) = \frac{\sum_{i=0}^{k} w_i P_i N_i^n(u)}{\sum_{i=0}^{k} w_i N_i^n(u)}, \quad u \in [t_0, t_{k+n+1}],$$

*where* $N_i^n, i = 0, 1, \ldots, k$ *are the normalized B-spline basis functions of degree n corresponding to the knot vector* $T$.

Another advantage is that it is invariant under projective transformation (only affine invariance holds for its integral counterpart). Additionally, there are weights which can be used to control shapes in a manner similar to shape parameters. Geometrically, a rational curve can be viewed as the projection of an integral curve from a vector space of one higher dimension. The NURBS curve can be obtained by projecting the B-spline curve $\hat{R}$ in $\mathbb{E}^{d+1}$ having the same knot vector and control points $\hat{P}_i = (w_i P_i, w_i)$. As a consequence, the NURBS inherit all the nice properties from B-splines, and can represent conic sections.

**NURBS Surfaces.** If we extend equation in two parametric directions we obtain a surface with the same properties as the NURBS curve:

$$F(u, v) = \frac{\sum_{i=0}^{n} \sum_{j=0}^{m} w_i P_i B_i(u) B_j(v)}{\sum_{i=0}^{n} \sum_{j=0}^{m} w_i B_i(u) B_j(v)}.$$

The surface does not have to be of equal degree in both directions. Observe the surface in its rendered form in where we clearly see the local control property.

NURBS generalize the nonrational parametric form. Like nonrational B-splines, the rational basis functions of NURBS sum to unity, they are infinitely smooth in the interior of a knot span, and at a knot they are at least $C^{k-1-r}$ continuous with knot multiplicity $r$, which enables them to satisfy different smoothness requirements. They inherit many of the properties of uniform B-splines, such as the strong convex hull property, variation diminishing property, local support, and invariance under standard geometric transformations. More material of NURBS and further detailed discussion of its properties can be found in [8, 29, 97, 145].

### 2.1.3 Hierarchical Schemes

Forsey and Bartels have presented the hierarchial B-spline [34], in which a single control point can be added without covering an entire row or column of control points. In their work two concepts are introduced: local refinement using an efficient representation, and multi-resolution editing. These notions can be generalized to any surface such as subdivision surface. Meanwhile, the localized hierarchical splines have been proposed by Gonzalez-Ochoa and Peters [40], which extend the hierarchial spline paradigm to surfaces of arbitrary topology. Kraft [66] has constructed a hierarchical B-splines with a multilevel spline space which is a linear span of tensor product B-splines on different, hierarchically ordered grid levels. Charms [42] have extended this scheme in a more general setting and adapted it to more applications. Weller et al. [156] have studied spaces of piecewise polynomials with an irregular, locally refinable knot structure (thus it is called “semi-regular bases”). Deng et al. [23] have introduced a new type of splines-polynomial splines over hierarchical T-meshes (called PHT-splines) to model geometric objects. PHT-splines are a generalization of B-splines over hierarchical T-meshes. Song et al. [132] have presented the method to approximate the signed distance function of a surface by using polynomial splines over hierarchical T-meshes. In particular, they compute on closed parametric curves in the plane and implicitly defined surfaces in space.

T-splines, developed by [120], are the most important scheme in our proposal. T-splines are generalizations of NURBS surfaces that are capable of significantly reducing the number of superfluous control points by using the T-junction mechanism. The main difference between a T-spline control mesh and a NURBS control mesh is that T-splines allow a row or column of control points to terminate at anywhere without strictly enforcing the rectangular grid structure throughout the parametric domain. Consequently, T-splines enable much better local refinement capabilities than NURBS. Furthermore, using the techniques presented in [120], we are able to merge adjoining T-spline surfaces into a single T-spline without adding new control points. Sederberg et al. have also developed a simplified algorithm to convert NURBS surfaces into T-spline surfaces, in which a large percentage of superfluous control points are eliminated [118].

T -spline is a P B-spline for which some order has been imposed on the control points by means of a control grid called a textit T-mesh. A T -mesh is basically a rectangular grid that allows T -junctions. Each edge in T -mesh is a line segment of constant $s$ (which is called s-edge) or constant $t$ (which is called $t$-edge). A T -junction is a vertex shared by one s-edge and two $t$-edges, or by one $t$-edge and two s-edges. For example, $P_1$ (see Fig.2.2(b)) is a T -junction. Each edge in a T -mesh is labeled with a knot interval, constrained by the following rules:

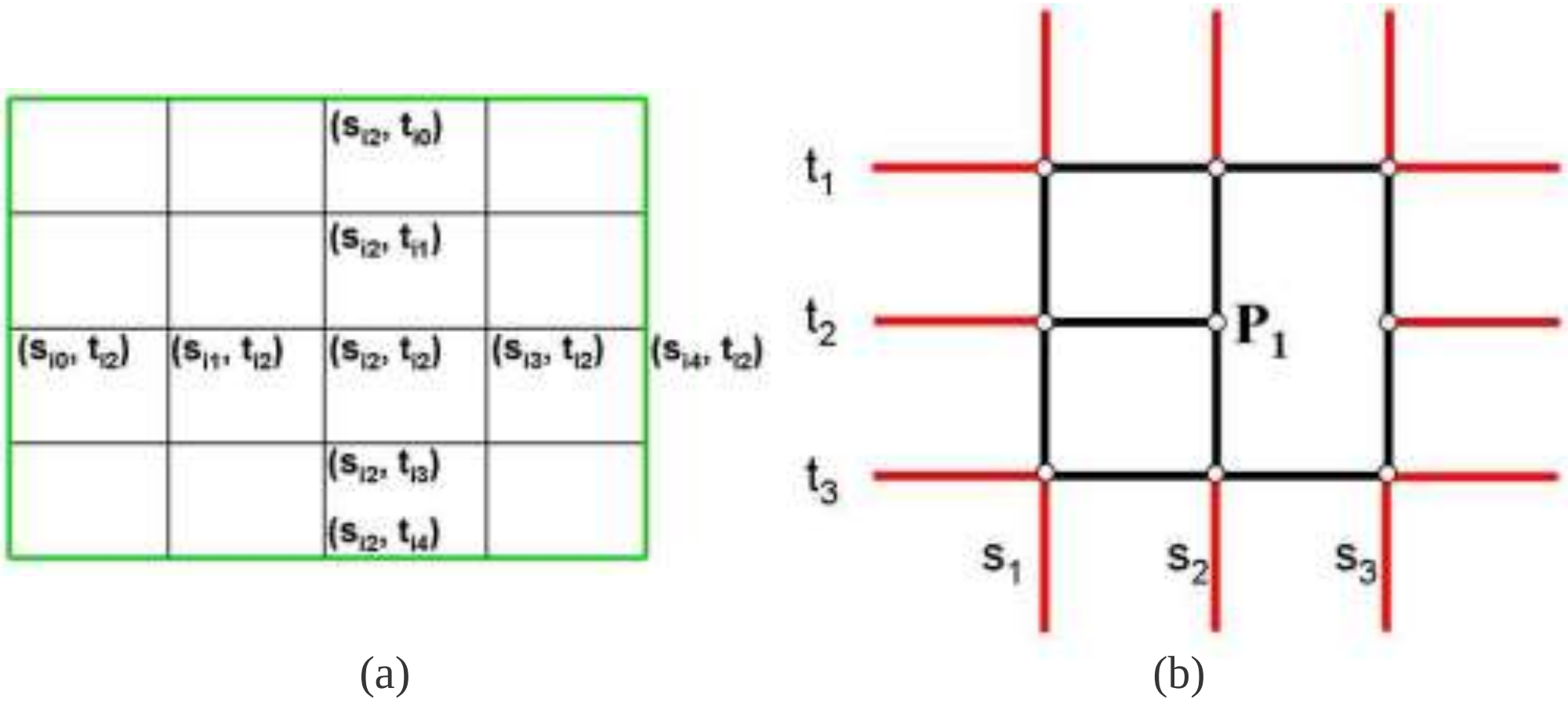

Figure 2.2: (a) Local knot lines for basis function $B_i(s,t)$; (b) $\mathbf{P}_i$ is a T-junction.

1. The sum of knot intervals on opposing edges of any face must be equal.
2. If a T-junction on one edge of a face can be connected to a T-junction on an opposing edge of the face (thereby splitting the face into two faces) without violating Rule 1, the edge must be included in the T-mesh.

In contrast to tensor-product B-spline that uses a rectangular grid of control points, P B-spline is point-based and requires no topological relationship among control points. The equation for a P B-spline is given by:

$$P(s,t) = \frac{\sum_{i=1}^{n} \mathbf{P}_i B_i(s,t)}{\sum_{i=1}^{n} B_i(s,t)} \quad (s,t) \in D,$$

where the $\mathbf{P}_i$ are control points. The $B_i(s,t)$ are basis functions written as

$$B_i(s,t) = N_{i0}^3(s) N_{i0}^3(t),$$

where $N_{i0}^3(s)$ is the cubic B-spline basis function associated with the knot vector $s_i = [s_{i0}, s_{i1}, s_{i2}, s_{i3}, s_{i4}]$ and $N_{i0}^3$ is associated with the knot vector $\mathbf{t_i} = [t_{i0}, t_{i1}, t_{i2}, t_{i3}, t_{i4}]$ as illustrated in Fig. 2.2(a). Every control point has its influence domain $D_i = (s_{i0}, s_{i4}) \times (t_{i0}, t_{i4})$. The T-spline equation is very similar to the equation for a tensor-product rational B-spline surface, except that knot vectors $s_i$ and $\mathbf{t_i}$ are deduced from the T-mesh neighborhood of $\mathbf{P}_i$.

Knot vector $s_i$ and $\mathbf{t_i}$ for the basis function $B_i(s,t)$ are determined as follows. Let $(s_{i2}, t_{i2})$ are the knot coordinate of $\mathbf{P}_i$. Consider a ray in parameter space $R(\alpha) = (s_{i2} + \alpha, t_{i2})$. Then $s_{i3}$ and $s_{i4}$ are the s coordinates of the first two s-edges intersected by the ray. The other knots can be found in like manner.

In computer graphics T-splines have been applied to many applications. For example, Song et al. [131] have generalized a T-spline scheme to weighted T-spline and demonstrated its applicability in 3D free-form deformation. Lévy et al. [76] have utilized T-splines for surface reconstruction.

### 2.1.4 Global Splines V.S. Spline Merging

Spline merging techniques always involve the following steps. In order to model an arbitrary manifold in 3D using conventional spline schemes, current approaches will segment the manifold to many smaller open patches, then cover each patch by a single coordinate system, so that each patch can be modeled by a spline surface. Finally, any generic approach must glue all the spline patches together by adjusting the control points and the knots along their common boundaries in order to ensure continuity of certain degree. It requires the merging of splines defined over different local domains. Surface patch merging has been thoroughly discussed first in [120, 57] and later is used in [119], in order to glue the trimmed region to form a single spline. However, it is far more complicated to design semi-standard trivariate splines which demand much more in-depth studies. During spline merging, handling singularity with still high-order continuity is extremely difficult in spline research. For surface modeling, Loop and Scheafer in [84] have given an example of a $G^2$ polynomial construction with general connectivity to accommodate singularities. On the other hand, Peters and Fan [98] have introduced rational linear maps to replace affine linear atlas and handle singularities between charts.

Spline merging also has many shortcomings. The entire segmenting and patching process is primarily performed manually, and it requires users' knowledge and skills, and for non-trivial topology and complicated geometry this task is laborious and error-prone. To overcome the above modeling and design difficulties and address the topological issue, many researchers seek novel modeling techniques that would allow designers to directly define continuous spline models over any manifolds (serving as parametric domains). Such a global approach would have many modeling benefits, including no need of the transition from local patch definition to global surface construction via gluing and abutting, the elimination of a non-intuitive segmentation and patching process, and ensuring the high-order continuity requirements. More importantly, we can expect a true "one-piece" representation for shapes of complicated topology, with a hope to automate the entire reverse engineering process.

Li et al. have presented an automatic technique to convert polygonal meshes to T-splines using periodic global parameterization [76, 109]. Li et al.'s method can be also viewed as manifold splines since the transition functions of the periodic global parameterization are compositions of translations and rotations. Grimm et al. [41] have pioneered a generic method to extend B -splines to surfaces of arbitrary topology, based on the concept of overlapping charts. Cotrina et al. have proposed a $C^k$ construction on a manifold [17, 18]. Ying and Zorin [162] have presented a manifold-based smooth surface construction method which has high-order continuities with explicit nonsingular parameterizations only in the vicinity of regions of interest. Gu et al. [43] have developed a general theoretical framework of manifold splines in which spline surfaces, defined over planar domains, can be systematically generalized to any manifold domain of arbitrary topology (with or without boundaries). He et al. have further developed modeling techniques for applications of manifold splines using triangular B -splines [47].

### 2.1.5 Trivariate Splines

Spline-based volumetric modeling and analysis have gained much attention recently with many applications. For geometric processing, Song et al. [131] have employed trivariate splines with non-uniform weights to model free-form deformation. For physical analysis, Hughes et al. [56] have proposed isogeometric analysis on surface, using bivariate NURBS for modeling smooth geometry and physical attributes together, and conducting physical analysis simultaneously. For virtual surgery, Tan et al. [140] have utilized spherical volumetric simplex splines to model and simulate the human brain. In visualization, Rössl et al. [112] have utilized trivariate super splines to model and render multi-dimensional material attributes for solid objects. A modeling technique introduced in [169] has been developed to model skeletal muscle with anisotropic attributes and conduct FEM analysis directly on NURBS solid. Martin et al. [87] have presented a method to fit a solid model using a cylindrical trivariate NURBS and support continuum force analysis. However, these existing spline schemes tend to handle only simple inputs like genus-0 surfaces. For more complicated shapes, Zhang et al. [166] have proposed the method to convert the long-branch/bifurcations dominant shapes. Martin et al. [86] have studied shapes with a symmetry (called "mid-face") structure. These methods always attempt to transform the model through a top-down scheme, which inspires us to research a new method in a divide-and-conquer fashion.

Compared with surface splines designed to extract features (e.g., [76, 91]), trivariate splines mainly focus on finding part-aware component structures. Besides poly-cube domains, another commonly-used part-aware domain is cylinder (tube) like [87]. Martin et al. in [86] have extended this domain to mimic more complex shapes. However, in terms of spline construction, the cylinder (tube) domain inevitably produces singular points along the tube axis.

## 2.2 Parameterization

Model parameterization is the fundamental basis and powerful geometry processing tool with versatile application, such as detail mapping, such as spline fitting and CAD, meshing processing, FEM analysis, visualization etc. In this thesis research proposal, parameterization is the first and un-avoided step during enabling data-to-spline conversion. In this section, we first briefly outline its mathematical foundations and describe recent methods for parameterization. Second, since our research mainly focuses on trivariate spline construction, it is necessary to discuss some recent emerging study interest on volumetric parameterization. Finally, we demonstrate feature-aware parameterization specifically because an efficient feature-aware technique leads to better spline fitting result.

### 2.2.1 Theory and Techniques

In this section, we outline the mesh parameterization including its mathematical foundations, versatile local parameterization techniques on different domains. In [126, 50], authors also have discussed this topic. Our review starts with an introduction to the general idea of parameterization and the state-of-art is reviewed by summarizing the motivation and major idea of several important approaches. Since we mainly consider the representation of volumetric information, we also discuss the emerging tools for regular global parameterization and volumetric parameterization.

**Metric and Distortion Minimization.** Parameterization can be viewed as a procedure of energy/distortion metric minimization procedure. Energy (distortion metric) gives rise to the solution from the degree of global energy field, that the spring model will converge at a balance state when the global spring energy is minimized. The advantage of these ideas involves that once we set the energy field function , we can solve the parameterization by numerical energy minimization tools directly.

Now we need to specify the energy, or define distortion metrics. The distortion derives from the stretching during the mapping $F$ between the surface $(x, y, z)$ and the domain $(u, v)$. Suppose $(x, y, z) = F(u, v)$ is a center point $P$ of an infinitesimal planar circle. Then, one point on this circle $F(u + \delta u, v + \delta v)$ is approximated given by first order Taylor expansion:

$$\underline{F}(u + \delta u, v + \delta v) = F + F_u(u, v)\delta u + F_v(u, v)\delta v,$$

or

$$\underline{F}(u + \delta u, v + \delta v) = P + F_u(u, v)\delta u + F_v(u, v)\delta v = P + J_f(\delta u, \delta v),$$

where $J = [F_u, F_v]$ is a $3 \times 2$ mapping matrix (normally it is also called Jacobian matrix). Using singular value decomposition, we have:

$$J_f = U\Sigma V^T = U \begin{pmatrix} \sigma_1 & 0 \\ 0 & \sigma_2 \\ 0 & 0 \end{pmatrix} V^T.$$

Then we can define the conception of isometric, conformal and equiareal (See details in [20]. The computer language friendly explanation can be found in [42]).

**Theorem 2.2.1.** *For a planar mapping* $f : R \to R$, *the following equivalence gives:*

1. *f is isometric* $\Leftrightarrow \sigma_1 = \sigma_2 = 1$
2. *f is conformal* $\Leftrightarrow \sigma_1/\sigma_2 = 1$
3. *f is equiareal* $\Leftrightarrow \sigma_1\sigma_2 = 1$

So it is $\sigma_1$ and $\sigma_2$ that directly influence the stretch (and the distortion metric energy) of the mapping. So we have

$$E(f) = \int_\sigma E(\sigma_1(u, v), \sigma_2(u, v))dudv.$$

This equation should be defined here in different methods. Malliot et al. [85] have proposed the method which minimizes "Green-Lagrange deformation tensor". This tensor is given by:

$$E = (\sigma - 1)^2 + (\sigma - 1)^2.$$

Hormann et al. [52] have presented another method call "Mostly Isometric Parameterization of Surfaces" (MIPS) for parameterization. This method is based on the minimization of the ratio between two direction stretching: $\frac{\sigma_1}{\sigma_2}$. Since minimizing this energy is a difficult numerical problems, they replace it with another simple metric $\frac{\sigma_1^2+\sigma_2^2}{\sigma_1\sigma_2}$. Sander et al. [113, 114] have studied a reversed parameterization method that their formalism uses the inverse function to map the parametric space onto the surface. For this reason, their energy can be expressed as $\sqrt{(\frac{1}{\sigma_1})^2 + (\frac{1}{\sigma_2})^2}$. Sokine et al. [134] have proposed a method based on the remark that shrinking and stretching should be treated the same. their method uses the following energy to minimize $Max(\frac{1}{\sigma_1}, \sigma_2)$.

To introduce more flexibility in these methods, some researchers focus on blending these method together in a spectrum. Degener et al. [22] have proposed to use a combined energy, with a term that penalizes area deformations, and another term that penalizes angular deformations. Wang et al. [82] have invented a family of metrics that can flexibly blend the LSCM method [73] and ARSP method [133].

**Barycentric Coordinates.** Barycentric coordinates solve the parameterization procedure from another degree. Retrospect to the simple spring model, barycentric coordinates consider the converge from local region: every vertex and its local neighbors are averaged by the special designed spring force of the connected edge. The motivation of barycentric coordinates derives from the affine combination parameterizing. A succinct idea of this method is based on simple physical model: We constrain the boundary of the mesh onto the boundary of the parameter domain which we target to map to (for simplicity, the domain here is planar rectangular). Suppose two vertices $V_i$ and $V_j$ are connected by Edge $E_{ij}$ and we imagine this edge as a spring. Then, the mesh is transformed to a spring system and the parameterization solving transform to spring energy converge equation: we give each vertex a parameter that where the vertex stop in the domain.

The most important issue here is to specify the spring energy. Barycentric coordinates is one of the spring force representation. Each vertex is represented as the weighted average of the neighbor vertex as:

$$x_i = \sum_{j \in N_i} \lambda_{ij} x_j,$$

and

$$\sum_{j \in N_i} \lambda_{ij} = 1,$$

here the $\lambda_{ij}$ is defined as barycentric coordinates. In some cases the coordinates $w_{ij}$ are determined independently and $\sum_{j \in N_i} w_{ij} = 1$. Then for normalization we set

$$\lambda_{ij} = \frac{w_{ij}}{\sum_{j \in N_i} w_{ij}},$$

where we call $w_{ij}$ homogeneous coordinates. One advantage of inventing $w_{ij}$ includes that we can focus on computing coordinates from geometry information without considering the normalization property.

The earliest generalization of barycentric coordinates goes back to Wachspress [149]. It focuses on finite element analysis and suggests to set the homogeneous coordinates as follows:

$$w_{ij} = \frac{\cot \alpha_{ji} + \cot \beta_{ij}}{r_{ij}^2},$$

where $r_{ij}$ is the edge length. Desbrun et al. [24] have utilized them for parameterization. Meyer et al. [89] for interpolating density values inside convex polygons.

Another set of barycentric coordinates also stems from finite element solving. It actually arises from linear approximation of Laplace equation and is utilized to parameterization, which is given by:

$$w_{ij} = \cot \gamma_{ij} + \cot \gamma_{ji}.$$

Pinkall et al. [102] have also utilized it to compute discrete minimal surfaces. In the area of mesh deformation and interpolation, Sorkine et al. [133] have generalized this coordinates to preserve the surface details.

Another set of coordinates "Mean value coordinates" is proposed by in [32]. The coordinates are given by:

$$w_{ij} = \frac{\tan \frac{\alpha_{ij}}{2} + \tan \frac{\beta_{ji}}{2}}{r_{ij}}.$$

contrary to other coordinates, one advantage of mean value coordinates is that it guarantees that $w_{ij}$ is positive. The negative coordinates may lead to flip-over phenomena and violate injectivity property. Hormann et al. [51] have presented that mean value coordinates have many useful application in computer graphics.

There still exist some other coordinates. [72] have studied and modified the continuity of barycentric coordinates. Lipman et al. [81] have proposed Green Coordinates for closed polyhedral cages. They respect both the vertices position and faces orientation such that it lead to space deformations with shape preserving. Joshi et al. [61] have proposed a character-based barycentric coordinates as practical means to manipulate 3D models by operating to their cages. As indicated in [61], the rigid spatial topological structure of the FFD latices makes the deformation less flexible. Many papers have attempted to analyze the principle of existed coordinates and attempt to give a comprehensive image to all. Ju et al. [62] have analyzed and compared three coordinates (Wachspress, Harmonic, Mean value). They view stokes theory as the root of all three methods. From respect of stock theory, the difference between three coordinates is the chosen of unit element shape: Wachspress use polar dual, mean value use unit circle and Harmonic use original polygon. following the same motivation and pipeline, all 2D polygon barycentric coordinates can extended to arbitrary polyhedron in $R^3$, which is necessary for our volumetric parameterization. [63, 33, 71] have extended the mean value coordinates from 2D polygon to 3D polyhedron. [70] have developed the spherical coordinates specifically used for spherical polygons.

### 2.2.2 Volumetric Parameterization Techniques

We have already reviewed many surface parameterization techniques. As a very closely relevant topic to our proposal, here we briefly review the relevant volumetric parameterization techniques. Volumetric parametrization aims to compute a one-to-one continuous map between a 3-manifold and a target domain (or a given surface with interior space) with low distortions. Volumetric parametrization has been gaining greater interest in recent years, a few related techniques have been conducted towards various applications such as shape registration [152, 79], volume deformation [63, 61, 168], and spline construction [87]. Wang et al. [152] have parameterized solid shapes over solid sphere by a variational algorithm that iteratively reduces the discrete harmonic energy defined over tetrahedral meshes, the harmonic energy is rigorously deducted but the optimization is prone to getting stuck on local minima and it only focuses on spherical like solid shapes such as human brain datasets. Ju et al. [63] have generalized the mean value coordinates [32] from surfaces to volumes for a smooth volumetric interpolation. Joshi et al. [61] have presented harmonic coordinates for volumetric interpolation and deformation purposes. Their method guarantees the non-negative weights and therefore leads to a more pleasing interpolation result in concave regions compared with that in [63]. Martin et al. [87] have computed the precise (u, v, w) coordinates for genus-zero tetrahedral meshes, and the target domain is a cylinder. Li et al. [79] have used the fundamental solution method to map solid shape onto general target domains. The current existing methods always attempt to map the model to a standard or simple domain primitives. Thus, how to handle the complex model volumetric mapping is very intriguing. [86] have used a “mid-surface” in combination with harmonic functions to decompose the object into a small number of volumetric tensor-product patches. However, all these methods can not eliminate singularities. Zhang et al. [166] have proposed a method to handle long branches: The algorithm divides possible bifurcations of a vascular system into different cases to solve. Zeng et al. [164] have studied the volumetric parameterization of cylinder wall. In the paper, the differential operator is extended from 2D to 3D. In a similar idea, Xia et al. [159] have utilized Green’s function for parameterizing star-shaped volumes. Han et al. [45] have proposed the method to construct the shell space using the distance field and then parameterize the shell space to a poly-cube.

### 2.2.3 Spline-Friendly and Feature-Aware Methods

In this section we briefly review the parameterization techniques that are “Spline-Friendly”. “Spline-Friendly” here means “feature-aware”. Preserving feature in the parameterization result is very important to spline approximation because it will allow splines to approximate more accurately around the feature region.

Many quadrangulation methods are actually based on parameterization techniques. One important property in quad-mesh generation research is edge-preserving. [9, 91] have constructed an as smooth as possible symmetric cross field that satisfying a sparse set of directional feature edge constraints. Then Daniels et al. [19] have proposed a template-based approach for generating quad-only meshes, which offers a flexible mechanism to allow external input, through the definition of alignment features that are respected during the mesh generation process. [39] have introduced the concept of an exoskeleton as a new abstraction of shapes that succinctly conveys the structure of a 3D model. Here “exoskeleton” actually is the important feature edges on the model surface. Xia et al. [158] have proposed an editable poly-cube parameterization techniques that optional sketched features can be mapped to the corresponding edges on the domain. Huang et al. [54] have presented a extended spectral-based approach. In contrast to the original scheme, it can provide flexible explicit controls of the shape, size, orientation and feature alignment of the quadrangular faces. Zhang et al. [165] have proposed a new method which constructs a special standing wave on the surface to generate the global quadrilateral structure. The wave-equation based method is capable of controlling the quad size in two directions and precisely aligning the quads with feature lines.

### 2.2.4 Global Parameterization and Poly-cube

The motivation of global parameterization comes from the requirement of B-spline. B-Spline fitting demands that the parameter of each local domain keeps regular (tensor-product). It also requires the consistence between different local domains. Another important issue concerns that we expect to construct volumetric spline so that each parameter domain is a $\mathsf{R}^3$ space. The surface meshes cover the boundaries of all $\mathsf{R}^3$ domains seamlessly and consistently. The way of keeping this property includes choosing a domain (may be composed by a set of sub-domain) that has the same topology but with simplified geometry feature. The most simple way is to map the genus-0 model to a sphere without considering its geometry feature like [125, 105]. However, for more complex topology and geometry feature, more complex domains and parameterization techniques have been developed in the last decades.

The linear discrete harmonic theory is interesting and rich, attractive computationally and enormously useful in applications. The ideas inform contemporary notions of discrete conformality and harmonicity that are based on linear conditions on the vertex coordinates. Examples of applications include [44, 73, 146]. Another set of theories considers the analysis and modification of some key metric (e.g., curvatures). [59, 136, 3] have proposed the similar methods based on this theory: First compute a metric for the image mesh and only then a set of vertex positions and then solve the Laplace-Beltrami operator about the metric to flatten a mesh.

Another group of parameterization techniques utilizes curvature directions to drive the parameterization result. For example, in [1, 67], they have proposed an anisotropic polygonal remeshing method, which is the direct application of parameterization,by extracting and smoothing the curvature tensor field and use lines of minimum and maximum curvatures to determine appropriate edges for the remeshed version in anisotropic regions. Meanwhile in some other techniques like [109, 64], they generate two orthogonal piecewise linear vector fields defined over the input mesh (typically the estimated principal curvature directions) and then compute two piecewise linear periodic functions, aligned with the input vector fields, by minimizing an objective function.

Spectral-based parameterization methods study the eigenfunctions of operators (or eigenvectors of matrices in the discrete setting). Dong et al. [26] have used the Laplacian to decompose a mesh into quadrilaterals in a way that facilitates constructing a globally smooth parameterization. Huang et al. [54] have presented an extended spectral-based approach. In contrast to the original scheme, it can provide flexible explicit controls of the shape, size, orientation and feature alignment of the quadrangular faces. Zhang et al. [165] have proposed a new method which constructs a special standing wave on the surface to generate the global quadrilateral structure. The wave-equation based method is capable of controlling the quad size in two directions and precisely aligning the quads with feature lines.

**Poly-cube.** In [141], they have represented a method to map model with arbitrary shape and geometry to a domain-called poly-cube. Poly-cube is a domain composed by gluing small cubes together. Each segment of input surface mesh maps to one of six surfaces of one cube. The advantage of this mapping method is that the mapping is seamless and each mapping patch is tensor-product regular. The parameter between neighboring patches can transform consistently to each other simply by linear parameter transformation or rotation. So it guarantees consistence between patches by setting the resolution and sampling set of parameter between two patch the same.

Meanwhile, several methods have been developed to improve user control: The user can easily control the mapping by specifying optional features on the model and their desired locations on the poly-cube domain. For instance, Wang et al. [154] have presented a technique where the user can interactively control the desired locations and the number of corners of the poly-cube map; Xia et al. [158] have used user sketches as constraints to control the poly-cube map. Automatic poly-cube construction is always extremely difficult due to the complexity of the input shape. Lin et al. have used Reeb graph to segment the surface and then developed an automatic method to construct poly-cube map [80]. However, their segmentation method may not work for shapes with complicated topology and geometry and does not guarantee a bijection between the poly-cube and the 3D model. He et al. [48] have proposed an automatic algorithm by slicing the model along one horizontal direction and then gluing together. It can only handle the horizontal, planar features from the 3D model. In fact, none of the current techniques constructs the poly-cube simultaneously following all above criteria.

### 2.2.5 Applications on Visualization

In our proposal, one of the important applications about parameterization is on focus+constext (F+C) visualization. Also, a critical part of remaining work involves volume data F+C visualization based on volumetric parameterization. Therefore, it is necessary to introduce and review the related work on this research topic.

Various F+C visualization techniques have already been proposed on many types of informatics inputs, such as trees [130, 90], treemaps [30], graphs [36], tables [143], and city maps [170]. Plaisant et al. [103] have defined the SpaceTree as a novel tree browser to support exploration in the large node link tree. The algorithm applies dynamic re-scaling of branches to best fit the space and includes integrated search and filter functions. For the seamless F+C, Shi et al. [128] have proposed a distortion algorithm that increases the size of a node of interest while shrinking its neighbors. Ying et al. [147] have also presented a seamless multi-focus and context technique, called Balloon Focus, for treemap. Gansner et al. [36] have presented a topological fisheye view for the visualization of large graphs. A method to cope with map and route visualization has been proposed by Ziegler et al. [170]. They depicted navigation and orientation routes as a path between nodes and edges of a topographic network. Recently, Karnick et al. [65] have presented a novel multifocus technique to generate a printable version of a route map that shows the overview and detail views of the route within a single, consistent visual frame. Different from the above methods with specific pre-defined targets, our framework is capable of handling various information or visualization-based applications.

The key component in F+C visualization is to design an efficient lens. Optical effects, such as fisheye [35] for the nonlinear magnification transformation with multi-scale, have been widely used. Fisheye views can enlarge the ROI while showing the remaining portions with successively less detail. Fisheye lens offers an effective navigation and browsing device for various applications [94]. In addition, InterRing proposed by Yang et al. [161] and Sunburst proposed by Stasko et al. [137] have incorporated multi-focus fisheye techniques as an important feature for radial space-filling hierarchy visualization. The major advantage of the fisheye lens is the ability to display the data in a continuous manner, with a smooth transition between the focus and context regions. Although fisheye lens has advantages in preserving the spatial relation, it creates noticeable distortions towards its edges, which fails to formally control the focused region and preserve the shape features in the context region.

Aiming to cope with the shortcomings of the basic fisheye lens, more sophisticated lenses have been proposed. Bier et al. [6] have presented a user interface that enhances the focal interest features and compresses the less interesting regions using a Toolglass and Magic Lenses. Carpendale et al. [14] have proposed several view-dependent distortion patterns to visualize the internal ROI, where more space is assigned for the focal region to highlight the important features. LaMar et al. [68] have presented a fast and intuitive magnification lens with a tessellated border region by estimating linear compression according to the radius of lenses and texture information. Pietriga et al. [100] have provided a novel sigma lens with new dimensions of time and translucence to obtain diverse transitions. Later, they provided in-place magnification without requiring the user to zoom into the representation and consequently lose context [101]. Their representation-independent system can be implemented with minimal effort in different graphics frameworks. Meanwhile, the deformation methods are recently used for the complicated 3D datasets, including volume data [16, 155] and mesh model [153]. Wang et al. [153] have presented a method for magnifying features of interest while deforming the context without perceivable distortion, using an energy optimization model for large surface models. Later, they further extended this framework into 3D volumetric datasets [155]. Inspired by these methods, we utilize geometric deformation that applies to visualization of 2D data sets, targeting to eliminate the local angle distortion and keep the visual continuity. By comparison with Wang et al. [153], our framework supports more flexible metric design (See Section 6.4.1) to satisfy various requirements. Meanwhile, because we focus on processing the informatics data, multi-scale details are revealed after the magnification rather than the simple interpolation.

Many image deformation techniques have been successfully studied and used for various image manipulation applications like image editing and resizing. For example, Schaefer et al. [115] have utilized moving least squares to fit transformations and achieve image editing. Also, many blending polynomial coordinates have been developed for better shape interpolation with boundary deformation constraints (e.g., biharmonic weights [58], green coordinates [81]). Meanwhile, image resizing [154, 60] is introduced in the literature for retargeting images to displays of different resolutions and aspect ratios. Note that, image resizing has a completely different goal from lens design, since the resizing task requires that important image regions are optimized to scale uniformly while regions with other contents are allowed to be distorted. Also, we observe the fact that all of the above techniques confine their operations as energy minimization in the 2D space only. Therefore, it is very attractive to explore a new deformation method that utilizes 3D geometric modeling techniques and broaden the scope of geometric modeling to help the visualization process.

## 2.3 Component-Aware Decomposition

Segmenting 3D surface meshes has been widely studied in graphics and digital geometry processing community. A thorough and detailed discussion on these surface segmentation techniques is beyond the scope of this work, we refer the interested readers to Shamir's great survey [123]. Among these segmentation methods, our volumetric spline conversion task demands to decompose shapes into meaningful volumetric parts, simulating how our vision identifies perceptual parts. "Perceptual" stresses that part-aware decomposition is inspired by research in perception, in particular by the idea that the human visual system understands shapes in terms of parts [49, 5, 129]. Guided by this observation, a lot of part-aware decomposition methods have attempted to encode the appropriate parts-aware metrics to agree with human visual perception and thus get the part-aware parts. For instance, these methods include the slippage [38], shape diameter function [124], interior visual region difference [83], intrinsic symmetry [104, 95, 160], modal analysis [55], etc. Meanwhile, particularly relevant to our requirement, skeletons are commonly used global perceptual-part structure representation tools. A lot of skeleton extraction techniques have been presented and thus can be used for part-aware decomposition (e.g., Mesh contraction [15], Reeb graphs [96], Thinning [2], etc). Finally, a part-aware decomposition can be manually edited by simple user interactions on the original surface [167, 28]. However, these methods mainly focus on designing suitable part-aware metrics, none of them has analyzed the segmentation results from the spline modeling view, with respect to criteria such as regularity, controllable corners, patch numbers, etc.

# Chapter 3

# Generalized Poly-cube Splines

As we have introduced in Chapter 2, the engineering design industry frequently pursues data transformation from discrete 3D data to spline formulations because of their compactness and continuous representation. As the newly merged research topic, we want to study the method to construct the volumetric splines in this chapter. The main challenge here is to handle arbitrary topology and complex geometry, which gives rise to our novel idea of "Generalized Poly-cube".

## 3.1 Motivation

Compared with the commonly-used *"surface model to surface spline"* paradigm, volumetric splines can represent both boundary geometry and real volumetric and physical/material attributes. This property makes volumetric representation highly preferable in many physically-based applications including mechanical analysis [56], shape deformation and editing, virtual surgery training, etc. However, converting arbitrary meshes to volumetric splines is extremely challenging because of many conflicting requirements for volumetric parametric domain construction. Attractive volumetric splines should have the following properties.

1. **Structural Regularity.** Tensor-product splines (e.g., NURBS) are defined over regular "cube-like" domains. Compared with the unstructured domain (e.g., polygonal regions covered by tetrahedral meshes), regular domain supports more efficient evaluation and refinement, and GPU acceleration can also be applied directly to spline representation with regular structure. Also, spline-based physical analysis (e.g., isogeometric analysis [56]) has a preference for "cube-shaped" domain.

2. **Singularity-free.** A *singular point* in volumetric domain is a node with valence larger than four on an iso-parametric plane (Fig. 3.1(a-b)). Handling singularity with tensor-product splines is extremely challenging. It is desirable to have a global one-piece spline defined on a globally-connected singularity-free domain.

3. **Controllable Ill-points.** We define an *ill-point* as the point where the domain of a basis function spans across nearby cubes. Fig. 3.1(e-h) illustrate all possible types of ill-points in red (note that they might not be singularities in volumetric splines, even though they are singular for surface spline construction). Ill-points may have a serious side-effect on the physical analysis of volumetric splines. As an example in Fig. 3.1(c-d), a spline's control point is attached to the ill-point and its basis function covers the space outside the cube domain [151]. It is much more desirable to control the number and

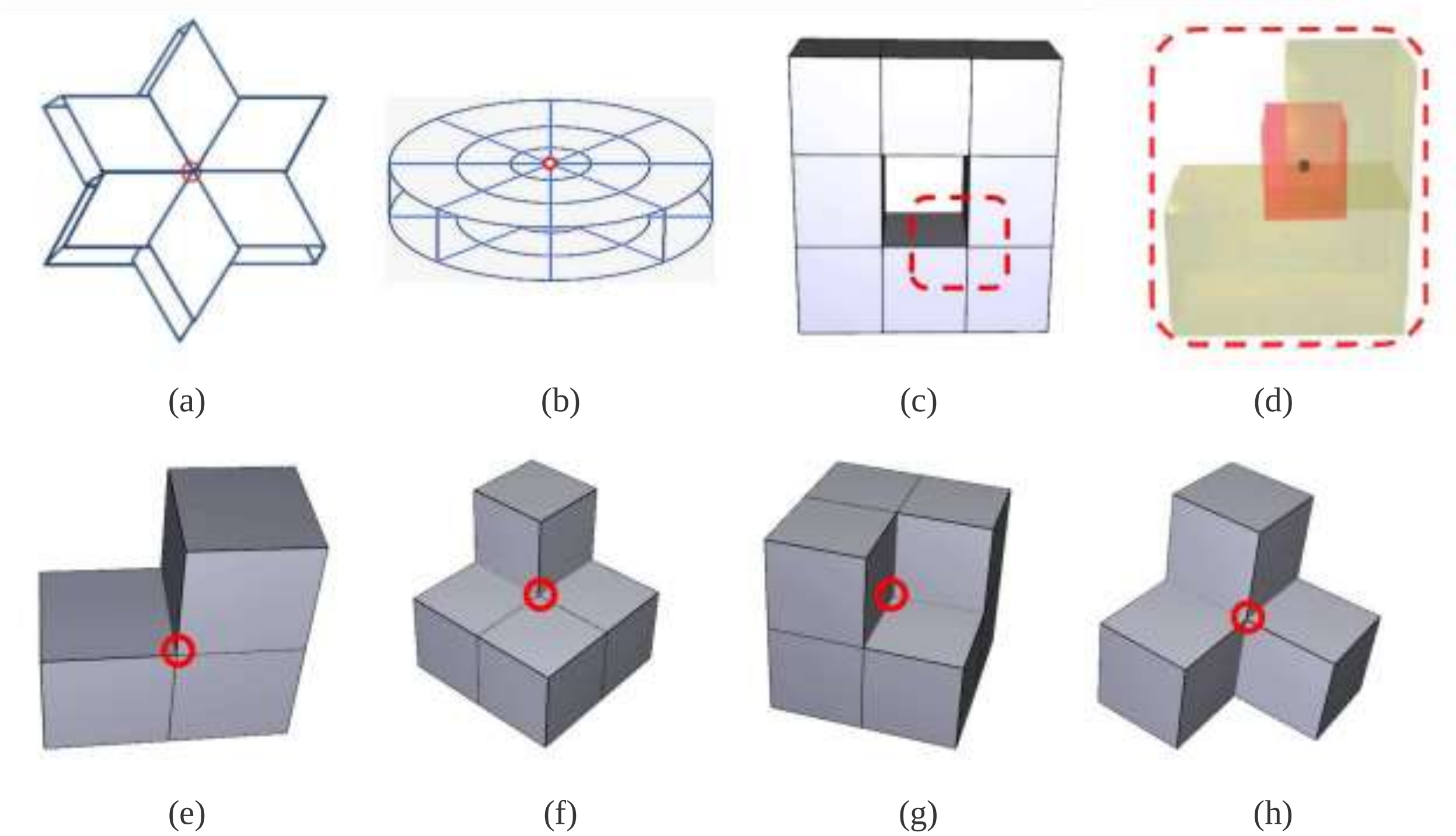

Figure 3.1: Singularity and ill-points. (a-b) show two cases of singular points. (c) highlights one ill-point. (d) shows that the basis function around the ill-point has influence outside the cube boundary. (e-h) show different types of ill-points: “Type-1” to “Type-4”. types of ill-points in the parametric domain. In practice, we hope to restrict ill-points to “Type-1” only shown in Fig. 3.1(e), since it is the easiest type and we can simply modify and restrict its “boundary” basis function [75].

4. **Shape Awareness.** Each spline patch should abstract the shape in a geometrically meaningful way, and it should reveal the shape's key perceptual parts and topological structures (e.g., skeleton-like representation).

Existing volumetric spline techniques generally follow two different trends: (1) many recent methods [86], [87], [166] convert each part into splines defined on a cylinder/tube domain (e.g., Fig. 3.1(b)), because they can intuitively use the shape skeleton to produce a tube domain and reveal the global structure and topology. A severe limitation of such approaches is that points on the tube centerline are all singular; (2) In contrast, poly-cube splines [150], [48] are defined on domains assembled by multiple cubes, which avoid the central line singularity problem. Such splines are flexible to resemble the shape of the given mesh and are capable of capturing the large scale features with low-distortion mapping. However, the gluing of many cubes may produce many uncontrollable ill-points. Limitations from both categories of splines have inspired us to develop a new method that is superior to both types of splines.

The main contributions of this chapter are as follows.

1. We propose a novel concept of *Generalized poly-cube (GPC)* to serve as the parametric domain for spline construction. GPC combines advantages of existing primitives: (a) GPC is powerful and flex-ible for representing complex models; (b) GPC provides a simple and regular domain with no singularity and controllable ill-point numbers and types.

2. We develop an effective, automatic GPC construction and parameterization framework. User interactions are also permitted while respecting both the global structure and the geometric features of 3D shapes.

3. We present a global "one-piece" volumetric spline scheme without stitching/trimming for general volumetric models. Unlike conventional spline schemes, our conversion does not require global coordinates everywhere, and piecewise local coordinates suffice. GPC therefore becomes an ideal parametric domain. We also design an efficient volumetric hierarchical spline fitting algorithm to support recursive refinement with improved accuracy and reduced number of control points.

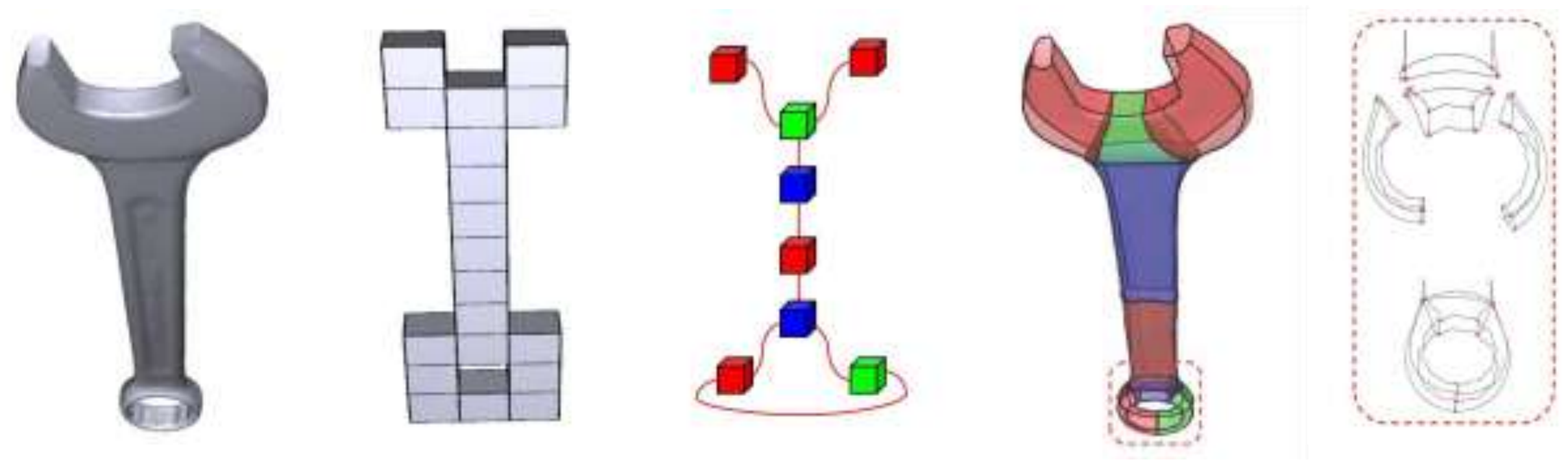

(a) (b) (c) (d) (e)

Figure 3.2: Generalized poly-cubes: (a) The wrench model; (b) The conventional poly-cube (CPC); (c) The generalized poly-cube (GPC) as a topological graph; (d-e) The cuboid edges are overlaid onto the model to visualize the GPC global structure.

## 3.2 Generalized Poly-cubes (GPCs)

**Conventional poly-cube (CPC)** is a shape composed of axis-aligned unit cubes that abut with each other. This way, cubes are glued and realized in a global 3D world coordinate system. CPC usually uses unit cubes as the building block. All cubes are glued together and embedded in the 3D space; Any point in a cube is associated with a unique global coordinate. Fig. 3.2(b) shows an example of CPC constructed for a wrench model in (a). Constructing effective (good approximation, coherent topology) CPC for volumetric models with relatively complicated geometry and topology usually requires extensive user involvement. Such a parametric domain is inadequate. Our hope is that, its construction can be less tedious, and the number of ill-points can be reduced.

**Generalized poly-cube (GPC)** is a set of topologically-connected cuboid (no need to be unit cube anymore) domains without the need of any 3D realization in a global 3D world coordinate system. We allow that any pair of two distinct cuboid faces to be glued together if these faces have the same size. Fig. 3.2(c-e) show the GPC constructed for the wrench model (Fig. 3.2(a)). GPC is different from CPC in the following aspects. First, every GPC cuboid is not required to be a unit box; It can be a general cuboid with rectangular faces. Each cuboid has its local coordinate system; A cuboid can deform (bend or twist) in order to glue with each other to form a global topological structure. To represent each cuboid, we project the 12 cuboid edges onto the model to visualize different faces (see Fig. 3.2(d-e)). Second, cuboids in GPC can be glued together through arbitrary two faces, and it is even possible that they are from the same cuboid. The topology of GPC can be represented using a topological graph, which we denote as a *GPC-graph*. Each node in the GPC-graph represents a local cuboid while each edge between two adjacent nodes indicates the gluing of corresponding sub-parts. Fig. 3.2(c) illustrates a GPC graph of Fig. 3.2(d). GPC has several advantages over CPC when serving as parametric domain.

**Generality.** GPC can serve as the parametric domain for a more general category of solid shapes that CPC can not easily mimic. Several types of examples are illustrated in Fig. 3.3. One category of models that CPC can not well approximate is the twisted or highly curved model, such as the twirl (a) and möbius band with thickness (d). Unlike CPC whose cubes are all axis-aligned, GPC can glue adjacent cuboids after twisting them so that sharp features on the models can still be accurately modeled as the cuboid edges (b,e). The GPC can also use much fewer cuboids than CPC in modeling such shapes. Taking (d) as an example, for such a möbius band (note that the band now has thickness), we can hardly construct a CPC as a useful domain. With GPC, only one cuboid is enough (f). Another category of models that the CPC has difficulty in handling includes models with complex topology especially when handles/voids are relatively small, such as the solid bucky model in Row 3. Beside the above non-axis-aligned problem, small handles/voids also make the construction of CPC highly challenging. In GPC, we can always effectively model them after a correct topological decomposition, and model a handle of the solid bucky model by 5 to 6 T-shapes (h). The pattern of bucky's GPC-graph around one hole is shown in (i).

**Ill-point Controllability.** GPC also has the advantage in ill-point controlling. First, the topological gluing can significantly reduce the number of ill-points (due to the usage of fewer cuboids and simple gluing rules). In Fig. 3.4(a-b), a torus' CPC generates 4 ill-points (in red circles) while a torus' GPC (See the kitten model, Fig. 3.14) has none. Second, our GPC construction algorithm will only generate *Type-1* (Fig. 3.1(e)) ill-points and we can place them in regions with far less geometric details.

**Easier and Better Domain Construction.** Because of its topological simplicity and elegance, the construction of GPC is usually easier than that of CPC. Automatic GPC construction can be developed naturally following the part-aware decomposition of the model. Furthermore, unlike CPC that requires many redundant cubes (to assemble topological handles), cuboids in GPC can be glued in a more structural way similar to the *generalized cylinder* defined by the shape skeleton, nicely encoding the shape structure. Such a GPC domain, which better mimics shape geometry and structure, could desirably lead to *far less parameterization distortion*. For example, a CPC (Fig. 3.4(d)) can merely mimic the shape of the genus-3 model (with a narrow top and wide bottom region) in an axis-aligned domain. Consequently, the narrow top and wide bottom are parameterized onto the equally-sized parametric domain, introducing large volume distortion. A GPC (Fig. 3.5(c)) can fit the shape better and significantly improve the parameterization quality, benefitting the final spline construction.

The following three sections discuss the algorithmic pipeline to construct the GPC (also illustrated in Fig. 3.5). The input model is first decomposed into a few T-shapes. Then, each T-shape is decomposed into 4 volumetric sub-patches and each volumetric sub-patch is parameterized onto cuboid. Each cuboid is mapped to an appropriately-sized volumetric cuboid domain, with aligned coordinates between two cube interfaces. The final step is to convert it to a global one-piece spline representation.

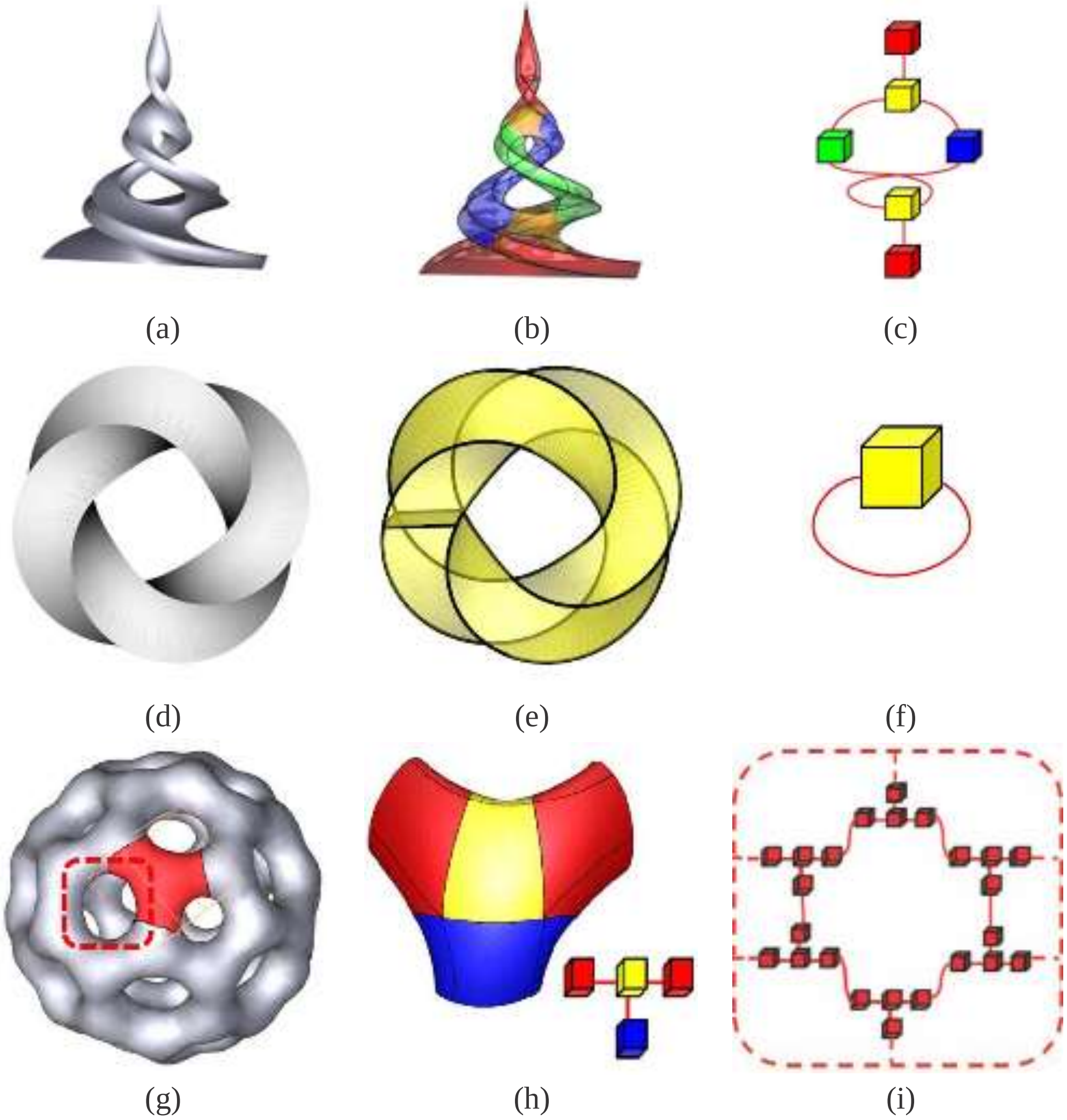

Figure 3.3: GPC can handle more generalized models. Row 1: (a) The highly-twisted swirl model, (b) Its GPC, and (c) Its topological graph. Row 2: (d) The non-axis-aligned möbius model, and (e,f) Its GPC and topological graph. Row 3: (g) The bucky model with complex topology, (h) It is decomposed into small "T-shapes" with 4 cuboids. (i) A subset of the GPC graph around the hole.

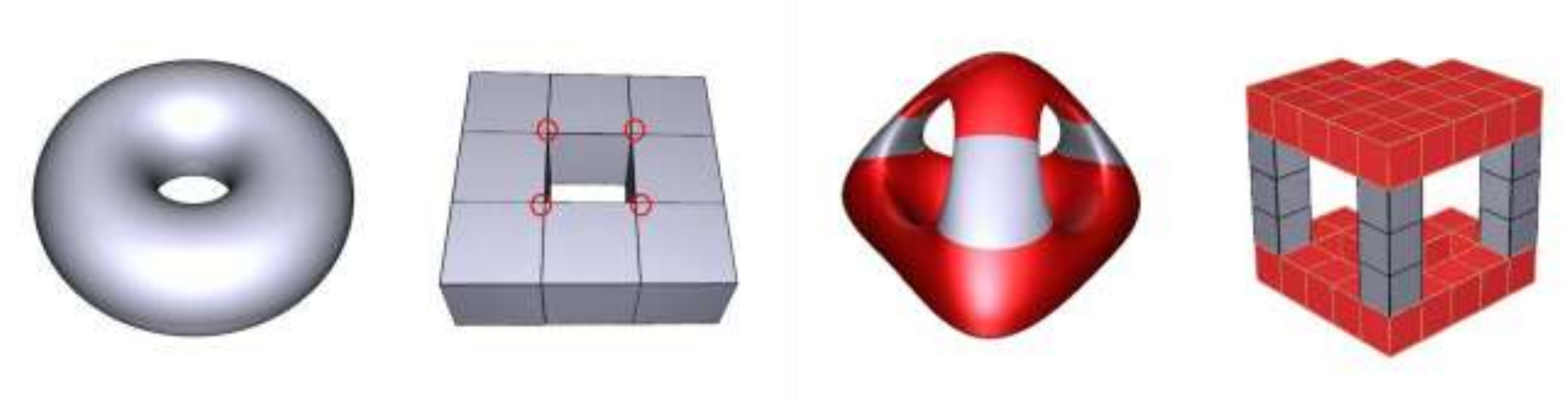

(a) (b) (c) (d)

Figure 3.4: (a) The torus model. (b) Its CPC uses at least 8 cubes and generates 4 ill-points. (c) The genus-3 model with narrow top and wide bottom regions. (d) Its CPC maps two regions onto the equal-sized parameterization domain, leading to large distortion.

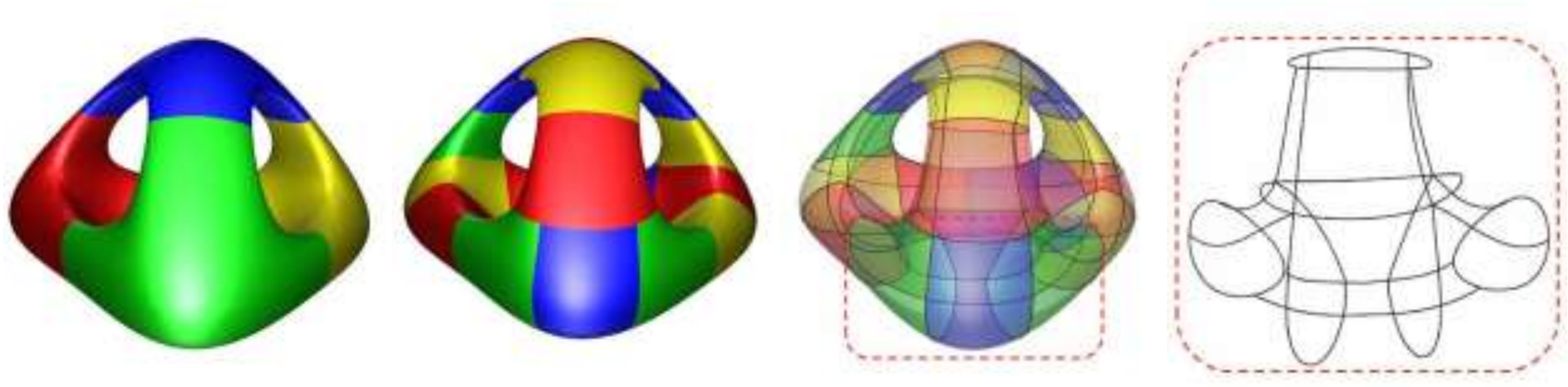

(a) (b) (c) (d)

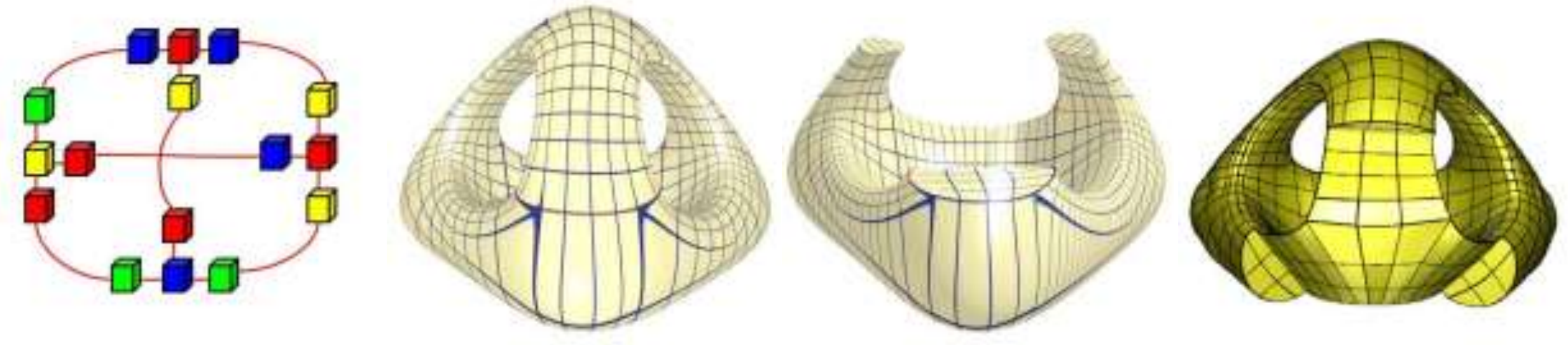

(e) (f) (g) (h)

Figure 3.5: GPC and spline construction pipeline. (a) The input genus-3 model is first decomposed into some "T-shape" patches. (b) Each "T-shape" is further decomposed into 4 cuboids. (c-d) Overlay all cuboid edges onto the model to visualize the global structure. (e) All cuboids comprise a topological GPC. (f-g) Construct the parametric mapping between the input model and its GPC. (h) Transform the model into a volumetric spline representation.

## 3.3 Model Partitioning

The first step to construct the GPC parametric domain is to segment the input model. Suppose a solid region is bounded by a triangle-meshed surface $\partial M$ (note that $\partial M$ can be of high-genus, but as the boundary of a solid object $M$, $\partial M$ is a closed surface), this section focuses the computation of a group of curves $\{c\}$ on $\partial M$. These curves segment $\partial M$ into sub-patches $\partial M_i$, bounding sub-solid regions $M_i^s$ to be parameterized upon GPC cuboids. We denote these traced curves on $\partial M$ as *poly-edges*, as they will be mapped to edges of GPC cuboids serving as parametric domain. Our segmentation includes two main steps:

- Partitioning into T-shapes: we decompose the entire model into a group of T-shaped patches.
- T-to-cube decomposition: we generate poly-edges on each T-shape and decompose it into 4 connected cube-like sub-patches.

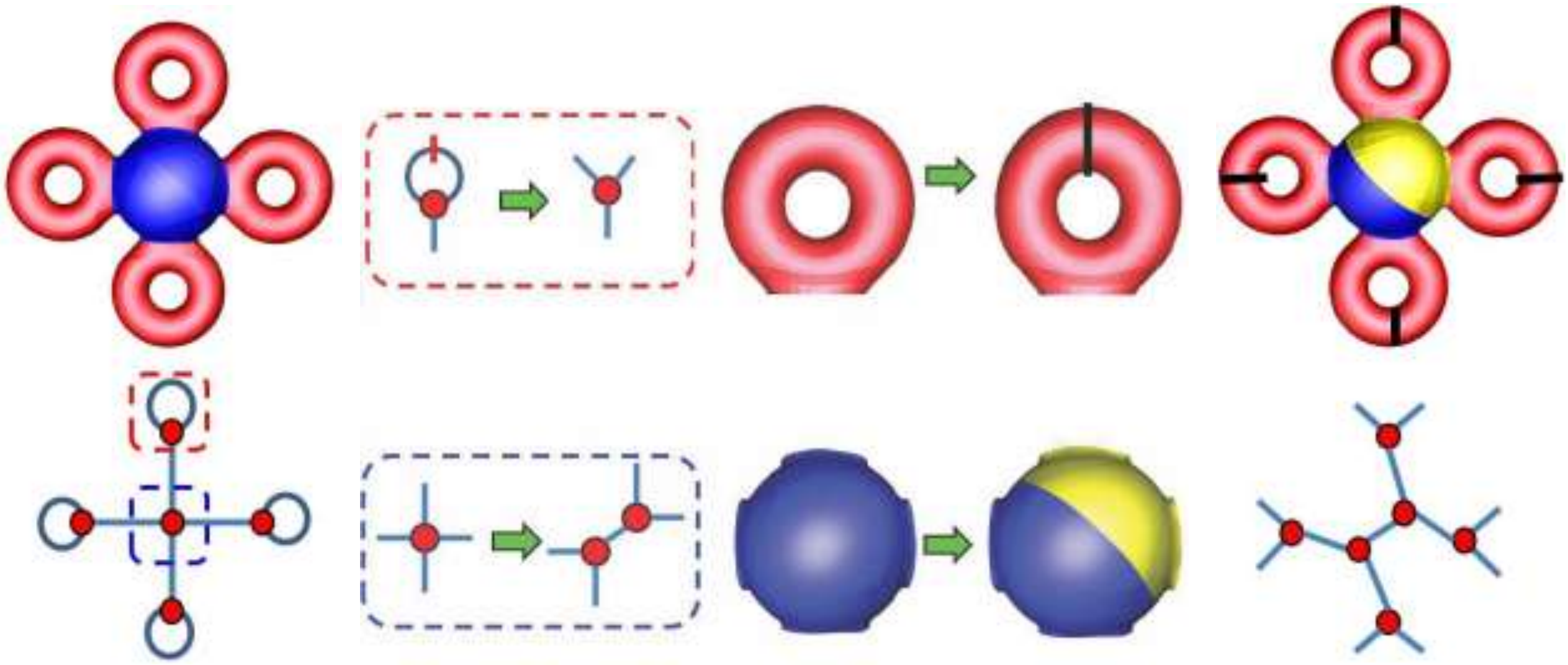

(a) (b) (c)

Figure 3.6: Model segmentation into "T-shape" patches. (a) The part-aware segmentation and its abstraction graph. The nodes in the graph have different cases for edge connection (red and blue regions). (b) For each case, we have corresponding operations on the graph and input model. (c) Our operation guarantees that the resulting nodes in the graph are all degree $d = 3$, and the model is segmented into T-shapes.

### 3.3.1 T-shape Segmentation

We use $\partial T_i$ to represent a T-shape surface and $T_i$ for its bounded volume. Our idea is to partition a given model $M$ into several T-shaped sub-regions $\{T_i\}$. This section discusses how to trace curves on the boundary surface $\partial M$ and partition it to sub-patches $\partial T_i$ or many simpler patches. This pipeline is illustrated in Fig. 3.6. T-shapes are used as the basic primitive in our framework to decompose more complicated solid models. A T-shape has trivial topology, only contains Type-1 ill-points, and represents the simplest 3-branched volume shape.

The challenge is how to segment $\partial M$ with nontrivial topology into $\{\partial T_i\}$, especially on ensuring the partitioning is topologically correct (only T-shapes or simpler) and each patch is geometry-aware (i.e., geometrically similar to a "T" in 3D space, not just topologically). The algorithm has the following steps:

**Step 1.** We first partition the input $\partial M$ into several part-aware patches. Many approaches have been investigated. Since part-awareness is usually subjective and highly dependent on applications, different segmentation methods typically select different criteria such as convexity, rigidity, or shape diameter functions [124]. Many effective manual partitioning schemes are also possible. In order not to restrict our segmentation framework, we remain agnostic as to which method or criteria should be used. We take shape diameter functions [124] and use it to perform partitioning in the next stage.

**Step 2.** Upon a complete decomposition, we construct an abstraction graph. We use a node to represent a patch; then add an edge between two nodes if their corresponding patches are adjacent (if a patch has a handle, an edge is added into the graph with both ends on the same node to form a loop). Fig 3.6(a) shows a 4-torus with colored part-aware segmentation and the resulting abstraction graph.

**Step 3.** We refine the partitioning by further segmenting patches which have many boundaries. Looking at its dual graph, we split nodes with high valance until all nodes have $\leq 3$ incident edges (A graph node with $d = 3$ represents a 3-branch patch, i.e., T-shape, and $d = 1$ or $d = 2$ indicates the patch that bounds a tube). We need to consider the following different cases. For each case we first formulate the operation on the abstraction graph then explain its corresponding computation on the model.

(3.1) *Handle loop* (see Fig 3.6(b), Row 1). In the graph, we cut the loop into two edges. In the model, we generate the shortest handle loop [25] and cut the model along it.

(3.2) *High Valence (*$d > 3$*) branch* (see Fig 3.6(b), Row 2). In the graph, such a node $n$ can split into two connected nodes $n_1$ and $n_2$. The newly-appeared node $n_2$ is incident to 2 existing edges and linked with $n_1$, and is therefore valence-3. The other corresponding node $n_1$ is associated with the remaining $d - 2$ edges, so its valence is $d - 1$. Repeat the split until all newly-generated nodes are valence-3. In the model, this means that we iteratively choose 2 boundaries (a pair with the closest distance) and then generate a loop bounding them [78]. This segments the patch into two patches, one with 3 boundaries (i.e., a T-shape) and another one with $d - 1$ boundaries. Row 2 shows an example of the cutting loop.

**Algorithm 1** Decompose a model into T-shapes.

```
Construct abstract graph G
for all node n ∈ G do
  if HasHandleLoop(n) then
    CutLoop(n)
  end if
  if degree(n) > 3 then
    repeat
      (n1, n2) = Partition(n),
      with degree(n2) = 3
      n = n1
    until degree(n) = 3
  end if
end for
```

After repeating the above operations on every node, we can get a partitioning where every node has its valence equivalent to 3 or less. This computation pipeline is formulated in Algorithm 1. Fig 3.6(c) shows the final T-shape segmentation and the abstraction graph.

**Discussion.** Our T-shaped primitives have desirable advantages over existing approaches based on similar spirit. Tierny et al. [144] have generated similar T-shape primitive (called "2-strip") for decomposition. The resulting primitive types are versatile and not unified, with singularities generated in the later step. In contrast, our algorithm guarantees unified T-shape and Type-1 ill-points only. Li et al. [78] have computed a homology basis of a closed genus-g surface and canonically decomposed the surface into a set of $2g - 2$ "pants-patches" (like a T-shape). However, this method is purely based on the surface topology. The resultant sub-patch may not be part-aware as they may not nicely bound a volumetric part. Therefore, Li's method in [78] is not directly applicable to volumetric data. In contrast, our graph-based method for shape abstraction guarantees that each patch is part-ware.

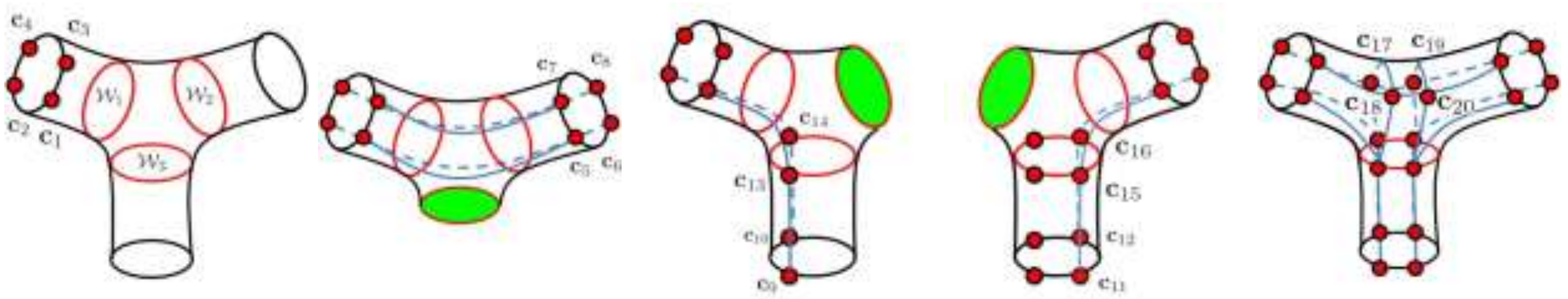

(a) (b) (c) (d) (e)

Figure 3.7: Illustration of T-to-cube segmentation. (a) Generate 3 cutting lines and 4 starting corners. (b-d) Cut along one cutting lines and trace poly-edges (and generate corners) between 2 boundaries. (e) Trace poly-edges on the central cuboid.

### 3.3.2 T-to-cube Segmentation

After the T-shape decomposition discussed in the previous section, we now have a set of T-shape $\partial T_i$ (or perhaps the simpler 2-boundary "Tube"-shaped patches). We shall further partition $\partial T_i$ into 4 sub-patches $\partial M_{ij}$ and generate corners and poly-edges (Recall that poly-edges are the traced curves that will be mapped to the edges of cuboid domains). As shown in Fig. 3.7(e), the lines segment the T-shape into 4 sub-patches, each of which has 8 corners and 12 poly-edges, and bounds a cuboid (Note that each T-shape boundary is divided by 4 corners as 4 poly-edges). Our decomposition algorithm includes the following steps.

1. We first determine the queue of processing sequence for all T-shapes. This queue is necessary because along any boundary two connected T-shapes must share the same 4 corners. Therefore, we need to know the processing sequence such that the first-processed T-shape on one side of the boundary can propagate the computed 4 corners correctly to the un-processed T-shape on the other side. In order to determine this sequence, we use a propagation strategy on the abstraction graph (Algorithm 2): We select the first node/T-shape in the graph (or a near-center one in the graph). Then we iteratively put its 3 unprocessed neighbors into the queue. Now for each boundary, its 4 corner positions are determined during the process of the precedent T-shape in the queue.

2. In order to process a T-shape, we need to compute 3 cutting lines and 4 corners on one boundary (See Fig. 3.7(a), positions of these corners are determined by the previously-processed adjacent T-patch). To generate 3 cutting lines, we detect 3 branches of $\partial T_i$ by extracting associated skeleton [111], with 3 resulting cutting lines $W_1$, $W_2$, and $W_3$. Our processing queue in Step 1 guarantees that each T-shape has a precedent adjacent T-patch, i.e., at least 4 pre-determined corners (except for the first T-shape in the queue, on which we manually set these 4 corners).

3. We generate all poly-edges and corners on a T-shape $\partial T_i$. We compute them separately in 3 passes. Each time we trace the poly-edges between 2 boundaries (as shown in Fig. 3.7(b-d)). Each time there are 3 sub-steps: (1) Cut the T-shape to a cylinder-like shape; (2) Map it to a cylinder domain; and (3) Trace the poly-edges on the cylinder domain.

   - When tracing poly-edges between 2 boundaries, we remove the third long branches by cutting along its cutting lines (e.g., $W_1$ in Fig. 3.7(b)). After filling the cutting hole [142], the resulting surface is a 2-boundary tube-shaped patch $\partial P$.
   - We map the tube shape to a cylinder domain $[\mathbf{u}, \mathbf{v}]$ following the approach of [87]. We shall briefly discuss this algorithm as follows. First, set vertex on one boundary $u = 0$ and $u = 1$ on the other, solve $\Delta \mathbf{u} = 0$ by mean value coordinates [32]. Second, trace an iso-v curve along $\nabla u$ from an arbitrary seed vertex on the boundary $u = 0$ to the other boundary $u = 1$, slice this iso-curve and get two duplicated boundary paths, then set $v = 0$ and $v = 1$ on them respectively and solve $\Delta v = 0$. The $\partial P$ is therefore parameterized onto a cylinder domain.
   - We trace poly-edges on the cylinder-parameterized patch $P$ between a pair of corners (blue lines in Fig. 3.7(b-d)). For example in Fig. 3.7(c), we trace between corners $< c_1, c_9 >$ and $< c_2, c_{10} >$. According to the processing queue (Step 1), the corner positions of $c_9$ and $c_{10}$ are either already determined by other precedent patches or are not yet known. If positions are unknown, we trace the poly-edge from the starting corner (e.g., $c_1$) along the gradient direction $\nabla \mathbf{u}$ to another boundary at a point $c_9$. Otherwise (corner position $c_9$ are predetermined), we map both corners $c_1$ and $c_9$ to the cylinder domain $[\mathbf{u}, \mathbf{v}]$ and trace the straight line on the domain between two corners, then project this parametric straight line back to the patch and get the resulting poly-edge.

4. After tracing poly-edges between 3 boundaries, we still have to generate the poly-edges and corners for the central cuboid (corner $c_{17}, c_{18}, c_{19}, c_{20}$). Similar to Step 3, we cut along $M_3$ and map it to a cylindrical domain. Then we trace the new poly-edges on this cylinder domain between corners $c_{13}, c_{14}$ and $c_{15}, c_{16}$ (They are the intersections of precedent poly-edges and cutting lines). Two resulting poly-edges have 4 intersections with poly-edges $c_3 c_7$ and $c_4 c_8$. These intersections are the resulting corners $c_{17}, c_{18}, c_{19}, c_{20}$. Algorithm 3 documents all above-mentioned 4 steps to handle a T-shape. For more simple input patches like tube-shape patch, the processing is much easier so that we can directly use Step 3.2-3.3 to generate the poly-edges and corners.

**Feature-preserving Segmentation.** Although the above automatic algorithm can handle most of models very well, sometimes users still expect to use several sharp features as the poly-edges. For example, this choice is specially natural and meaningful on the strong symmetric man-made models with sharp features (e.g., CAD models in Fig. 3.3(b-e)). We can modify Algorithm 2 and Algorithm 3, in order to better handle such kind of patches. The user decides if one T-shape has symmetric and dominant sharp features. In Algorithm 2, we put all strong symmetric T-shapes in front of the list $L$ and process them first, thus we determine the poly-edges and corners on strong symmetric T-shapes first, and transfer the resulting corners to other non-symmetric T-shapes. In Algorithm 3, when we trace the poly-edges by `Trace_Line` on a strong symmetric T-shape, we do not find the corners and trace the line along the cylinder domain. Instead, we

**Algorithm 2** Segmenting T-shapes.

Input: $T_0$ (//A starting T-shapes)

$L = \{empty\}$ (//T-shape array)<br>
L.push($T_0$)<br>
**while** $L = \{empty\}$ **do**<br>
  $T^0$ = pop (L)<br>
  T_to_Cube($T^0$) (//Algorithm 3)<br>
  $(T_1, T_2, T_3) = T^0$.Neighbors()<br>
  **for** $i = 1, 2, 3$ **do**<br>
    **if** $IsProccesed(T_i) = False$ **then**<br>
      L.push($T_i$)<br>
    **end if**<br>
  **end for**<br>
**end while**

**Algorithm 3** T-to-Cube Segmentation.

Input: T-shape: $T$, Cutting lines: $\{W_1, W_2, W_3\}$

//-----The first cut----- (Fig. 3.7(b))<br>
$P$ = Cut_to_Cylinder ($T$, $W_3$)<br>
$S_s = \{c_1, c_2, c_3, c_4\}$ //--Starting corners--<br>
$S_e = \{c_5, c_6, c_7, c_8\}$ //--Ending corners--<br>
Trace_Polyedge($P$, $S_s$, $S_e$)<br>
//-----The second cut----- (Fig. 3.7(c))<br>
$P$ = Cut_to_Cylinder ($T$, $W_2$)<br>
$S_s = \{c_1, c_2\}$<br>
$S_e = \{c_9, c_{10}\}$<br>
Trace_Polyedge($P$, $S_s$, $S_e$)<br>
//-----The third cut----- (Fig. 3.7(d))<br>
$P$ = Cut_to_Cylinder ($T$, $W_1$)<br>
$S_s = \{c_5, c_6\}$<br>
$S_e = \{c_{11}, c_{12}\}$<br>
Trace_Polyedge($P$, $S_s$, $S_e$)<br>
//---- The final line tracing (Fig. 3.7(e))<br>
$P$ = Cut_to_Cylinder ($T$, $W_3$)<br>
$S_s = \{c_{13}, c_{15}\}$ $S_e = \{c_{14}, c_{16}\}$<br>
Trace_Polyedge($P$, $S_s$, $S_e$)

can take the dominant sharp edge directly as the resulting poly-edge, and its ending points on the boundary as the corners. The sharp edges can be extracted and processed automatically [163]. Fig. 3.3(a, d, e) and Fig. 3.2(d) show the results with feature-preserving poly-edges.

## 3.4 Parameterization

After the input model $M$ is decomposed into sub-patches $\{\partial M_{ij}\}$, bounding topological solid cuboids $\{M_{ij}\}$, it sets a stage for us to perform cuboid parameterization of $\{M_{ij}\}$. There are two main steps. We first map the patch boundary to the cuboid domain surface. Then we use this mapping as boundary condition to compute the interior volumetric parameterization.

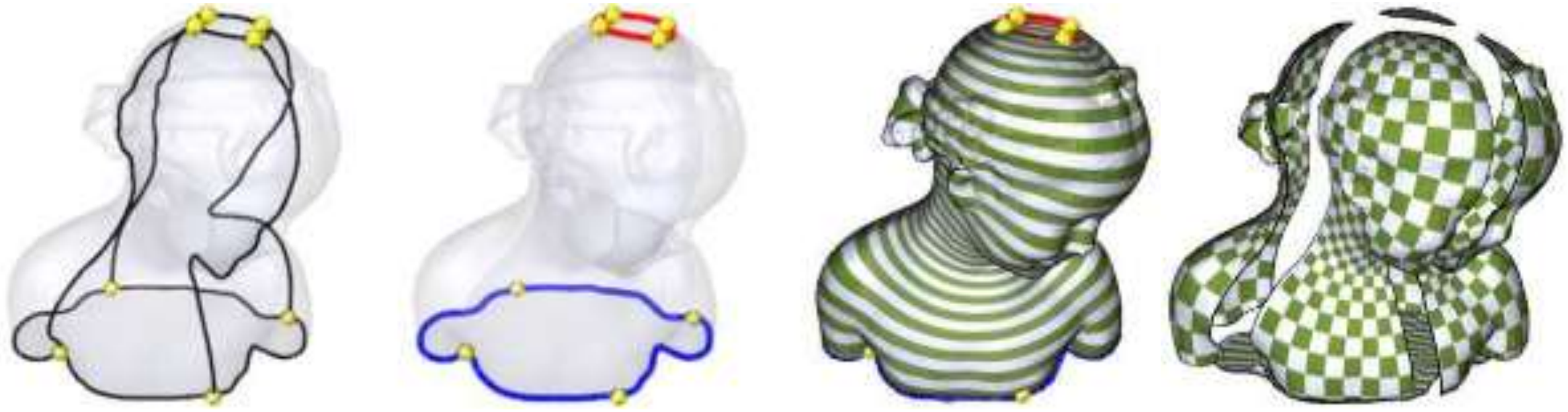

(a) (b) (c) (d)

Figure 3.8: Illustration of surface parameterization.

### 3.4.1 Surface Parameterization

The subpatch $\partial M_{ij}$ computed previously has 8 corners and 12 poly-edges (see Fig. 3.8(a)), we partition $\partial M_{ij}$ into 6 topological rectangles, then solve 3 harmonic mappings $\Delta \mathbf{u} = 0$, $\Delta v = 0$, $\Delta w = 0$ on each rectangle. Each time we map 2 opposite rectangles to 2 iso-plane domains (e.g., iso-u, iso-v, or iso-w) and on each plane compute the parameters of the coordinates.

For example, to solve $\Delta \mathbf{u} = 0$, we select 8 poly-edges on two opposite rectangles (see Fig. 3.8(b)). The 4 red poly-edges bounds an iso-u rectangle ($u = 0$) and the 4 blue poly-edges bounds another iso-u plane ($u = 1$). Then we compute the approximated discrete harmonic map $\Delta \mathbf{u} = 0$ [32] on other regions. Fig. 3.8(c) illustrate the computed u. Similarly, we can compute the harmonic scalar fields of v and w with $\Delta v = 0$ and $\Delta w = 0$, respectively. After solving 3 harmonic mappings, each vertex on the surface patch is mapped to a coordinate $(u_0, v_0, w_0)$ on the cube surface. The surface parameterization is illustrated in Fig 3.8(d).

### 3.4.2 Volumetric Parameterization

We compute the volumetric parameterization of $M_{ij}$ on a set of $n_0 \times n_1 \times n_2$ grid points. These grid points correspond to the uniformly-sampled coordinates in the parametric space (u, v, w). This volumetric parameterization can be considered as finding the locations of these nodes within $M_{ij}$. Similarly, as we discussed in surface parameterization, we need to find the point locations that minimize the equations $\Delta \mathbf{u} = 0$, $\Delta v = 0$ and $\Delta w = 0$ in 3D space.

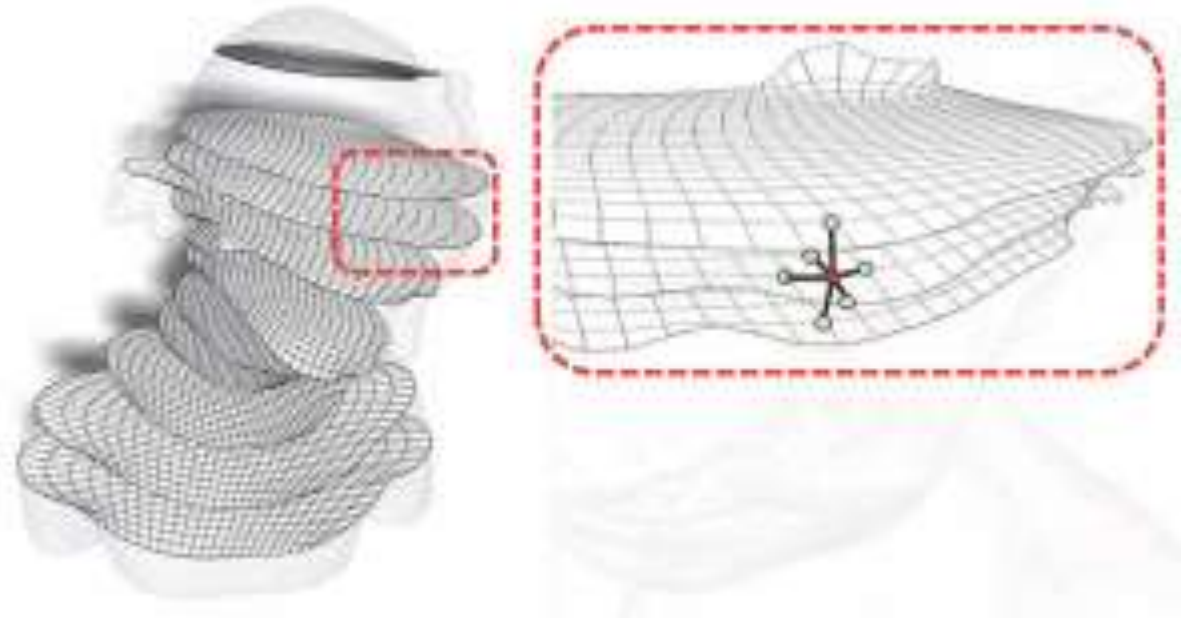

Figure 3.9: Volumetric mapping. We extract sample points as a hexahedral model. Each node has 6 neighbors for solving 3-D Laplacian in Eq. (3.1).

The $n_0 \times n_1 \times n_2$ grid points include two categories: the surface grid points and interior points. We determine their positions as follows.

(1) If the parameter of a grid point $n$ falls on the domain surface, we can always find its location on $\partial M$ by the parameter of $n$, where $n$'s parameter always falls into a triangle $[v_1, v_2, v_3]$ of $\partial M$ on the parametric domain with corresponding barycentric coordinates $\lambda_1, \lambda_2, \lambda_3$, then its spatial location is interpolated as $\sum_{i=1}^{i=3} \lambda_i P(v_i)$, where $P(v)$ denotes the 3D position of vertex $v$.

(2) Keeping the surface points fixed, we compute the interior point position by minimizing 3D Laplacian Eq. (3.1), where $n_{ijk}$ and $N_{ijk}^{\lambda}$ represent the node and its neighbor's spatial positions in $(x, y, z)$ and $w_\lambda$ is the point weight. Here, $w_\lambda$ encodes the string energy between the node and its neighbors. The choice of $w_\lambda$ has been studied in [152], [63]. In our implementation we simply use the uniform weight $w_\lambda = 1/|N_b(i)|$ as suggested in [142] and [135] ($N_b(i)$ is the set of node i's neighbors), and $|N_b(i)| = 6$ in our case as shown in Fig. 3.9.

$$E(n_{ijk}) = \sum_{\lambda} w_\lambda \times ||(n_{ijk} - N_{ijk}^{\lambda})||, \lambda \in N_b(n_{ijk}). \quad (3.1)$$

We move grid points iteratively. The update converges when changes of all node positions are smaller than a threshold during one iteration. Fig. 3.10(a-c) show the computation results of the Femur model after 20, 60, and 80 iterations.

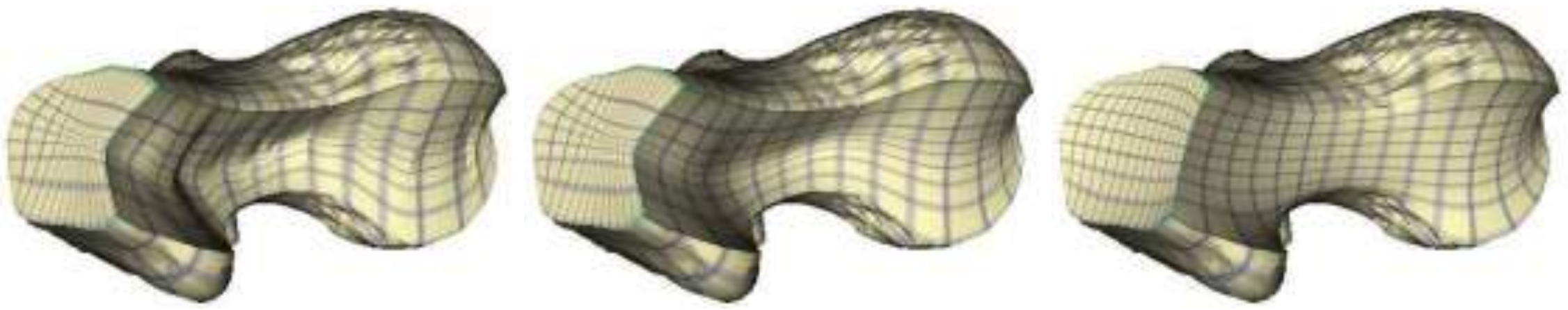

Figure 3.10: Results of the Femur model by solving Eq.(3.1) after 20, 60, and 80 iterations.

**Boundary Modification.** Since the parameterization is performed on piecewise sub-regions individually, across the partitioning boundary the composite global mapping is not smooth, such parameterization is not even aligned. We refine the parameterization on both sides of a partitioning boundary to make it aligned and smooth. The input is 2 connected patches along a partitioning boundary. To reduce the computation, we can cut each patch only a small portion. Then we glue two separate patches together as one patch.

The new patch also has 8 new corners and 12 new poly-edges. Now, we recompute the surface mapping and volumetric mapping on the glued surface, subject to two constraints: (1) parameters on the poly-edges; (2) regions that connect to other cuboids. We keep these parameters unchanged to avoid our modification destroying global parameter consistence.

## 3.5 GPC-Splines

After volumetric parameterization, we shall conduct the mesh-to-spline transformation over GPC. Two challenging issues must be addressed when designing effective spline construction algorithms. First, allowing adaptive refinement without significantly increasing control points is highly desirable since volumetric spline fitting usually requires a large number of control points when we seek high approximation accuracy. Second, unlike conventional B-splines that each control point and its knots are associated with global coordinates, GPC provides only locally-defined parameters in each cuboid domain, and this is because a global realization of GPC parametric domain in 3D Euclidean space is oftentimes impossible for models with complex structure, arbitrary topology, and deformable geometry. Therefore, we plan to design a unique GPC-spline construction algorithm using a point-based scheme.

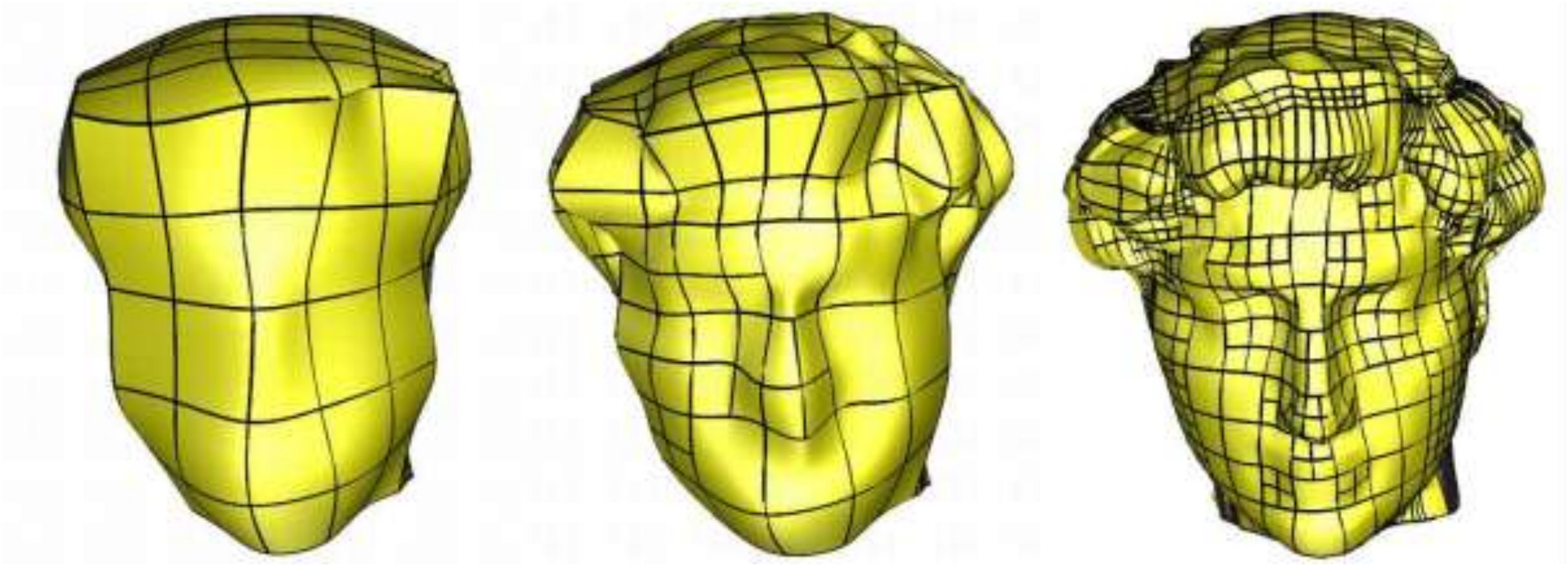

Figure 3.11: Hierarchical spline fitting results at levels 0, 1, and 2, respectively.

In principle, a volumetric cubic spline can be viewed as a point-based spline: Each control point $C_i$ (located in parametric cube $D^j$ with local coordinate $c_i^j$) is associated with three knot vectors along three principal axes: $r = [r_1, r_2, r_3, r_4, r_5]$, $s = [s_1, s_2, s_3, s_4, s_5]$, $t = [t_1, t_2, t_3, t_4, t_5]$, where $c_{ij} = (r_3, s_3, t_3)$. All knots can be determined using a *ray-tracing* strategy [120]. For any sample point with $(u, v, w)$ as its local parameter, the blending function is

$$B_i(u, v, w) = N_r(u) \times N_s(v) \times N_t(w), \tag{3.2}$$

where $N_r$, $N_s$, and $N_t$ are cubic B-spline basis functions associated with the knot vector $r$, $s$, and $t$ respectively. The formulation for point-based splines (PB-splines) is

$$P(u, v, w) = \frac{\sum_0^n C_i B_i(u, v, w)}{\sum_0^n B_i(u, v, w)}. \tag{3.3}$$

We modify the above equation to construct GPC splines. The GPC domain comprises a collection of coordinate charts locally defined in individual cuboid; adjacent local parametric coordinates are transformed coherently by transition functions, which can be encoded in a GPC-graph structure. As a result, the global PB-splines are piecewise rational polynomials defined on GPC, whose transition functions between adjacent cuboids are compositions of simple cuboid translations and rotations of $n\pi/2$, where $n$ is an integer.

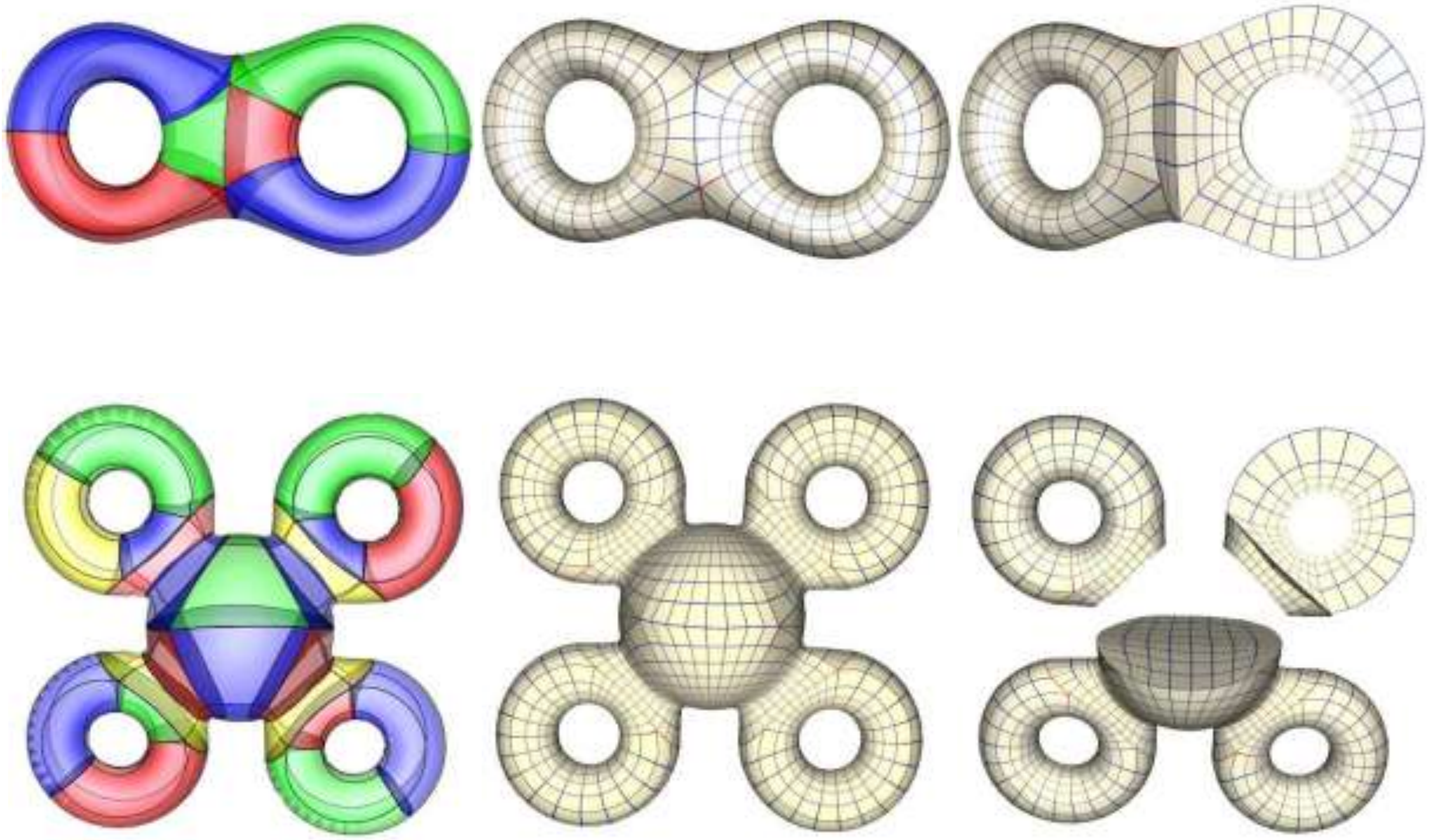

Figure 3.12: The eight/4-sphere model visualized with cuboid organization, poly-edge structure, surface parameterization, and volumetric parameterization.

In a cuboid $D^j$, given an arbitrary parameter $h$, also denoted as $h^j$, the spline approximation can be carried out as follows:

(1) Find all the neighboring cubes $\{D^i\}$ that support $h$ (i.e., it contains control points $C_k$ that may support $h$);

(2) The spline function is:

$$P(h) = \frac{\sum_{k=0}^{n} C_k^i B_k(\varphi^{ij}(h^j))}{\sum_{k=0}^{n} B_k(\varphi^{ij}(h^j))}, \quad (3.4)$$

where $h^j$ is the local parametric coordinate of point $h$ in the cube domain $D^j$, $\varphi^{ij}$ is the transition function from cube domain $D^j$ to $D^i$, and $C_k^i$ denotes the control point $k$ in the cube domain $D^i$.

In theory, a transition function $\varphi^{ij}$ from cube domains $D^j$ to $D^i$ is a composition of translations and rotations following the shortest path from cube $D^j$ to cube $D^i$ in the GPC-graph. Suppose $\widehat{D^iD^j} := D_1 (= D^i) \rightarrow D_2 \ldots \rightarrow D_n(= D^j)$, and the transition function $\Phi_{(i,i+1)}$ (derived by way of cube-gluing) from $D_{i+1}$ to $D_i$ is already known, then $\varphi^{ij}$ is formulated by

$$h^i = \varphi^{ij}(h^j) = \Phi_{1,2}(\Phi_{2,3}(\ldots \Phi_{n-1,n}(h^j))).$$

In practice, however, because most control points only influence a very small local region and do not cut across non-adjacent cubes, we observed that only using a neighboring cube transition function is usually enough.

### 3.5.1 Hierarchical Fitting

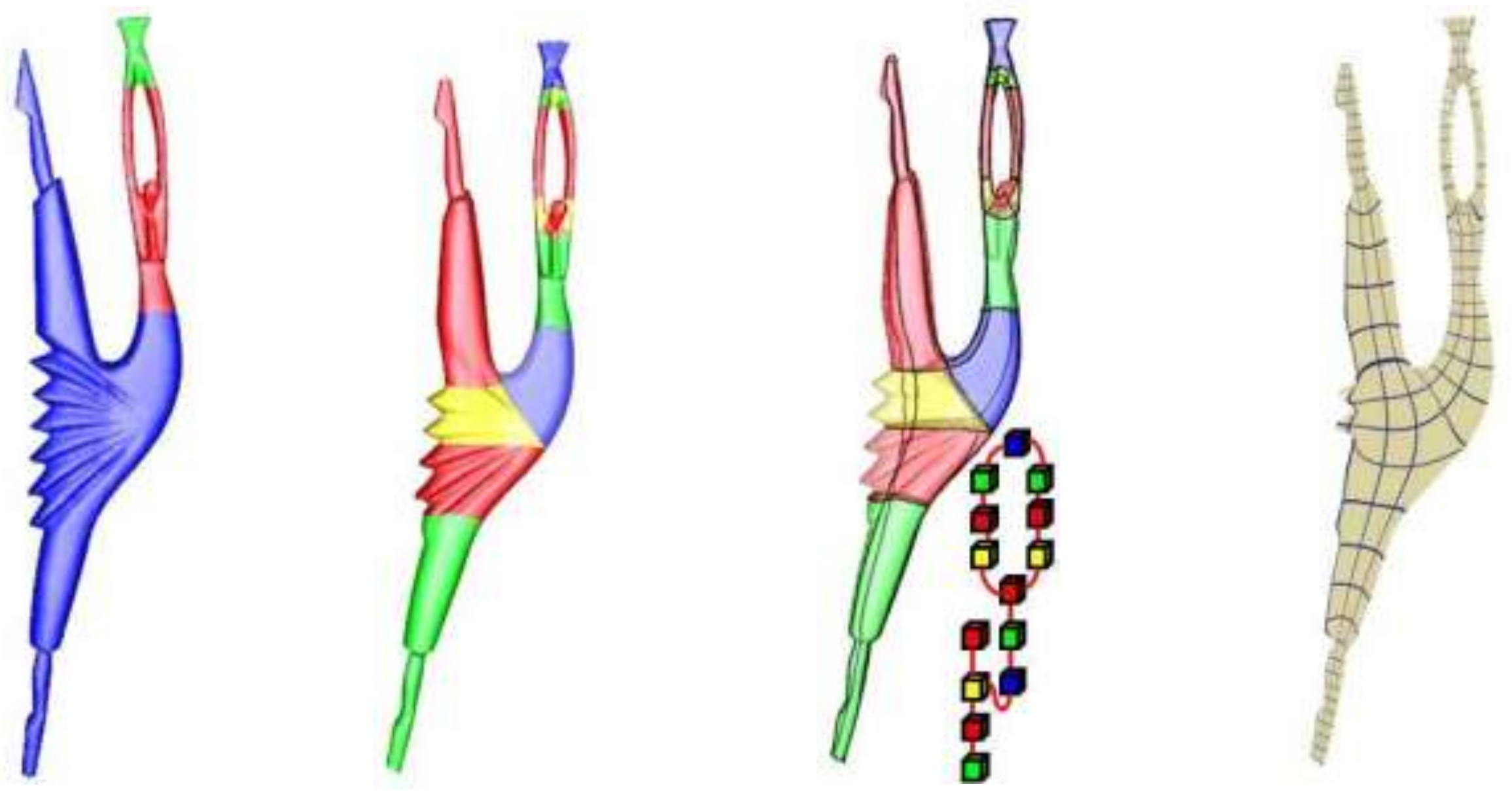

Figure 3.13: The dancer model visualized with T-shape decomposition, cuboid organization, poly-edge structure, GPC-graph, and volumetric parameterization.

Following the above GPC-spline definitions, we develop a hierarchical fitting scheme to approximate volumetric models. For a sample point $\mathbf{f}(h_i)$ in the model whose parametric coordinate is $h_i$ (defined by the volumetric parameterization computed in previous sections), $P(h_i)$ is our GPC-spline representation. We minimize the following equation:

$$E_{dist} = \sum_{i=0} ||P(h_i) - v_i||^2, \tag{3.5}$$

which can be rewritten in matrix format

$$\frac{1}{2} C^T B^T B C^T - V^T BC, \tag{3.6}$$

where $C$ is the vector of control points, $V = v_i$ is the vector of sample points, and $B = B_i(h_i)$ is the matrix of basis functions. This least square problem is not difficult to solve numerically. Given a sample parametric point $\mathbf{h}$ in GPC, in order to decide if we need to refine the approximation, we measure the root-mean-square error (rms) $\sigma(\mathbf{h})$ between its spatial position $\mathbf{f}(\mathbf{h})$ and its spline approximation $P(\mathbf{h})$. Algorithm 4 documents the main steps. The input includes the sample points and an initial control grid with control points. The initial control grid mimics the structure of GPC: Each cube corresponds to a local regular control grid. All local grids are topologically glued coherently following the GPC-graph, generating a one-piece global control grid. The function `KnotVectors` collects the 3 direction knots for each control point. We use the same “ray-tracing” strategy in [120]. `InfluencedSamples` returns all sample points in the influenced region of a control point. `Transition` transports a local parameter from one cube to another cube. `AssembleMatrix` assembles the matrix for Eq. (3.6) and `SolvingEquation` solves it and determines the control point positions. `FittingError` returns the worst fitting result in a small grid. `Subdivision` divides a grid uniformly into 8 smaller sub-grids. Fig. 3.11 illustrates our hierarchical fitting results.

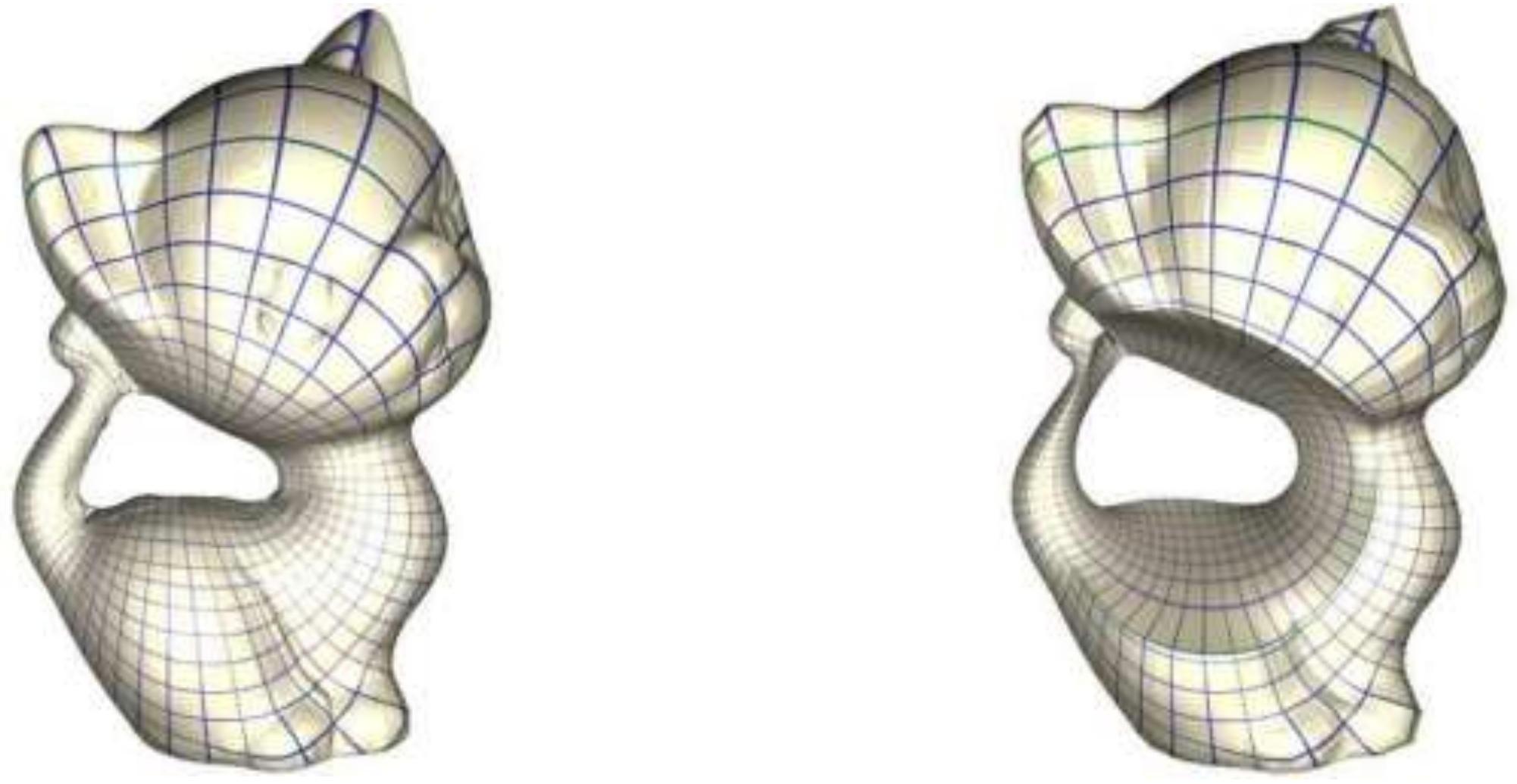

Figure 3.14: The kitten model visualized with surface and volumetric parameterization.

Our fitting scheme has two attractive properties. First, we eliminate a large percentage of superfluous control points by introducing T-junctions. Second, we provide adaptive control to users for proper trade-off between fitting accuracy and computational efficiency.

**Algorithm 4** Hierarchical spline fitting.

```
Input: Initial control grid Lg,
       List of sample points Ls,
       List of control points Lc,
       Fitting error threshold Output:
all control points positions. loop
  //Update control point knot vectors
  for all Lc do
    c = Lc .next()
    c.knots = KnotVectors(c, Lg)
    Ls' = InfluencedSamples(c, Ls)
    for all Ls' do
      s = Ls'.next()
      s.ctrlist.push_back(c)
    end for
  end for
  //Compute basis functions for samples
  for all Ls do
    s = Ls .next() Btotal = 0
    Lc' = s.ctrlist LB = {}
    for all Lc' do
      c = Lc'.next()
      param= Transition(s.cube#,c.cube#,,c.param)
      B= BasisFunction (param,c.knots)
      LB .push_back(B) Btotal = Btotal + B
    end for
    AssembleMatrix (LB, Btotal, s)
  end for
  //Fitting and evaluation
  SolvingEquation()
  for all Lg do
    g = Lg .next()
    if FittingError(g) >  then
      Lg' =Subdivision(g)
      Lg.delete(g)
      Lg.insert(Lg')
    end if
  end for
  Stop if no updated grid
end loop
```

## 3.6 Implementation and Discussion

Our experimental results are implemented on a 3GHz Pentium-IV PC with 4Giga RAM. To demonstrate the versatility of our approach (the flexibility of our computational framework), we construct GPC splines for many models. Our experiments include models with twisted shape: twirl (Fig. 3.3(Row 1)), möbius solids (Fig. 3.3(Row 2)); and with complex topology: bucky (genus 31, Fig. 3.3(Row 3)), genus-3 (Fig. 3.5), eight (genus 2, Fig. 3.12), 4-sphere (genus 4, Fig. 3.12); and with complex conceptual parts: wrench (Fig. 3.2), dancer (Fig. 3.13), and greek and david (Fig. 3.15).

Table 3.1 summarizes the statistics of the GPC construction, including the number of T-shapes, cuboids and ill-points. Compared with CPC, the statistics indicate that our GPC construction significantly reduces the number of cuboids. Moreover, the number of ill-points in GPC is far less than CPC. For example, Kitten (Fig. 3.14) has absolutely no ill-point in our GPC but at least 4 ill-points in any CPC.

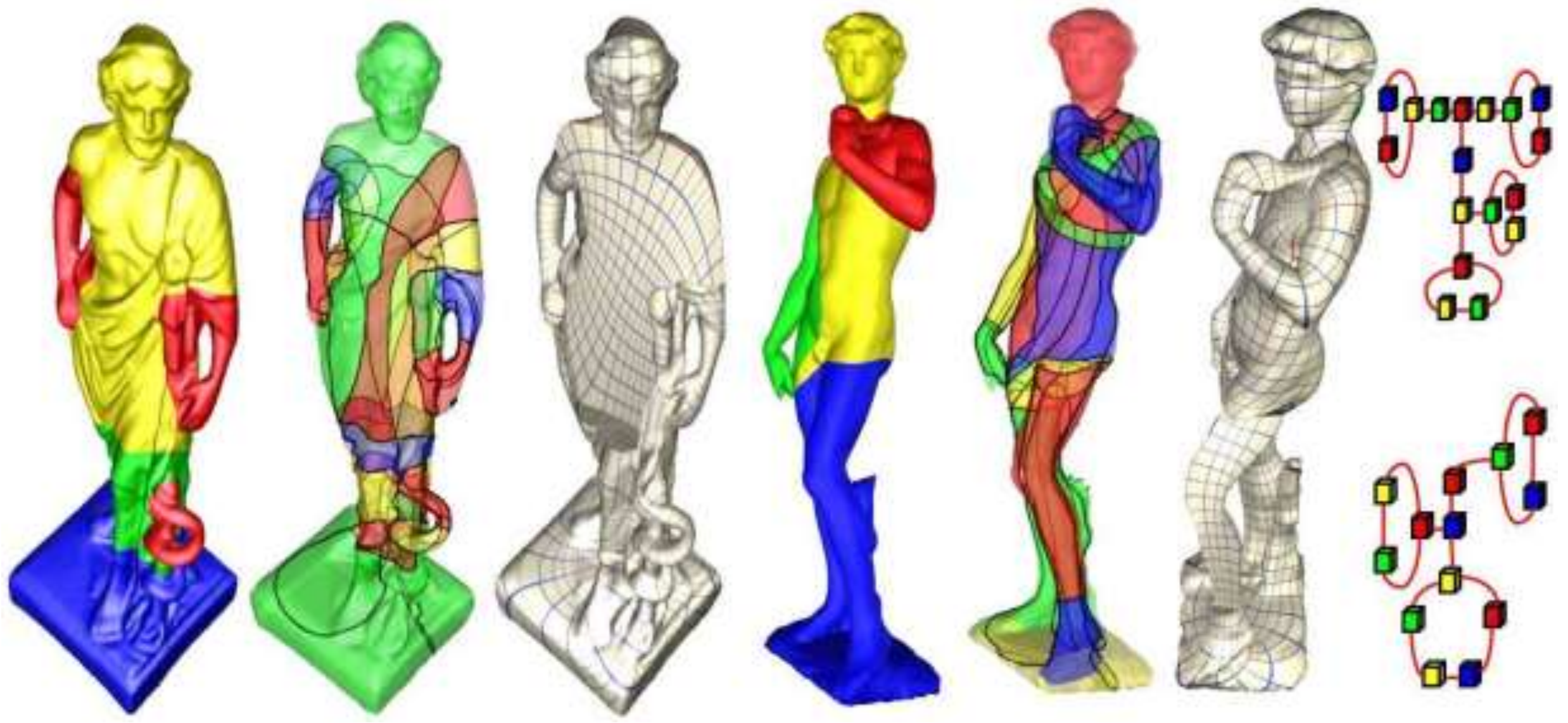

Figure 3.15: The greek and david model visualized with T-shape decomposition, cuboid organization, poly-edge structure, volumetric parameterization and their GPC graphs, respectively.

It may be noted that our parameterization algorithm may not guarantee a globally-minimized angle and volume distortion. However, since our algorithm decomposes the input into part-aware patches, each of which is parameterized on a geometrically similar cuboid, the distortion is satisfactory for our spline construction. The models of eight, dancer, 4-sphere, kitten, greek and david (Fig. 3.12, Fig. 3.13, Fig. 3.14, and Fig. 3.15) demonstrate several surface and volumetric GPC parameterization results. Fig. 3.16 shows several volumetric spline approximation results. We overlay the control grid line (black lines) onto the fitting results, and the T-junctions on the control grid reduce the control point greatly while still preserving the shape details. The statistical results are given in Table 3.2. The table shows that the vertices increases dramatically when we convert a surface model into a volume data. Our spline scheme can significantly reduce the degree of freedom for shape representation. In most of our experiments, approximation with good quality can be achieved within 3 levels of hierarchical refinement. the Fitting qualities are measure by RMS errors normalized to the overall sizes of solid models.

**Comparisons.** We compare our method with other volumetric parametric domain construction and mapping approaches: [141, 48, 86, 166, 80, 152]. As shown in Table 3.3 and Fig. 3.3, our method has advantages in the following aspects. First, our method works well for volumes with complex topology and structure. Second, our domain does not have any singularity and can control the type and number of ill-points (which is highly desirable for spline construction). Our domain construction does not require tedious design, even for very complex shape input. Meanwhile, we can also flexibly edit the cube domain to better approximate the shape interactively.

Table 3.1: Statistics of various test examples

| Model | Genus | Twisted | # T-shape | # Cuboid | # Ill-points |
|---|---|---|---|---|---|
| genus-3 | 3 | no | 4 | 16 | 8 |
| bucky | 31 | no | 60 | 240 | 120 |
| mobius | 1 | yes | 1 | 1 | 0 |
| twirl | 1 | yes | 2 | 6 | 4 |
| 4-sphere | 4 | no | 6 | 24 | 12 |
| eight | 2 | no | 2 | 6 | 4 |
| bimba | 0 | no | 1 | 1 | 0 |
| femur | 0 | no | 1 | 1 | 0 |
| wrench | 1 | no | 2 | 8 | 4 |
| dancer | 1 | no | 3 | 14 | 6 |
| david | 3 | no | 4 | 12 | 24 |
| greek | 4 | no | 6 | 19 | 12 |

Table 3.2: Statistics of various spline examples.

| Model | #. Surface vertices | #. Volume vertices | #. Control points | RMS error | Running time |
|---|---|---|---|---|---|
| Kitten | 12403 | 40000 | 3020 | 0.35% | 202s |
| Wrench | 7550 | 12000 | 2966 | 0.2% | 105s |
| Eight | 766 | 6400 | 448 | 0.16% | 24s |
| 4-sphere | 2042 | 22800 | 1088 | 0.2% | 47s |
| Genus-3 | 6632 | 51200 | 1280 | 0.17% | 162s |
| David body | 15572 | 81600 | 5956 | 0.37% | 890s |
| Greek body | 20109 | 91900 | 7265 | 0.4% | 1096s |

Table 3.3: Comparison with the existing approaches.

| Method | Tarini [141] | He [48] | Martin [86] | Zhang [166] | Lin [80] | Wang [152] | Hexahedron[13] | Ours |
|---|---|---|---|---|---|---|---|---|
| Primitives | cube | cube | cylinder | cylinder | cube | sphere | small cube<br>no parameter | cube |
| Topology | axis-aligned | axis-aligned | symmetric | long branch | reeb-graph | genus-0 | arbitrary | arbitrary |
| Twisted model | no | no | yes | yes | no | no | yes | yes |
| Singularity | no | no | center | center | no | center | large number | no |
| Ill-points | no control | large number | - | - | no control | - | large number | controllable |
| Domain construction | Artiest design | Axis scan | Simple | Simple | Simple | Sphere only | No domain existed | Simple |
| Editable domain | yes | no | no | no | no | no | no | yes |

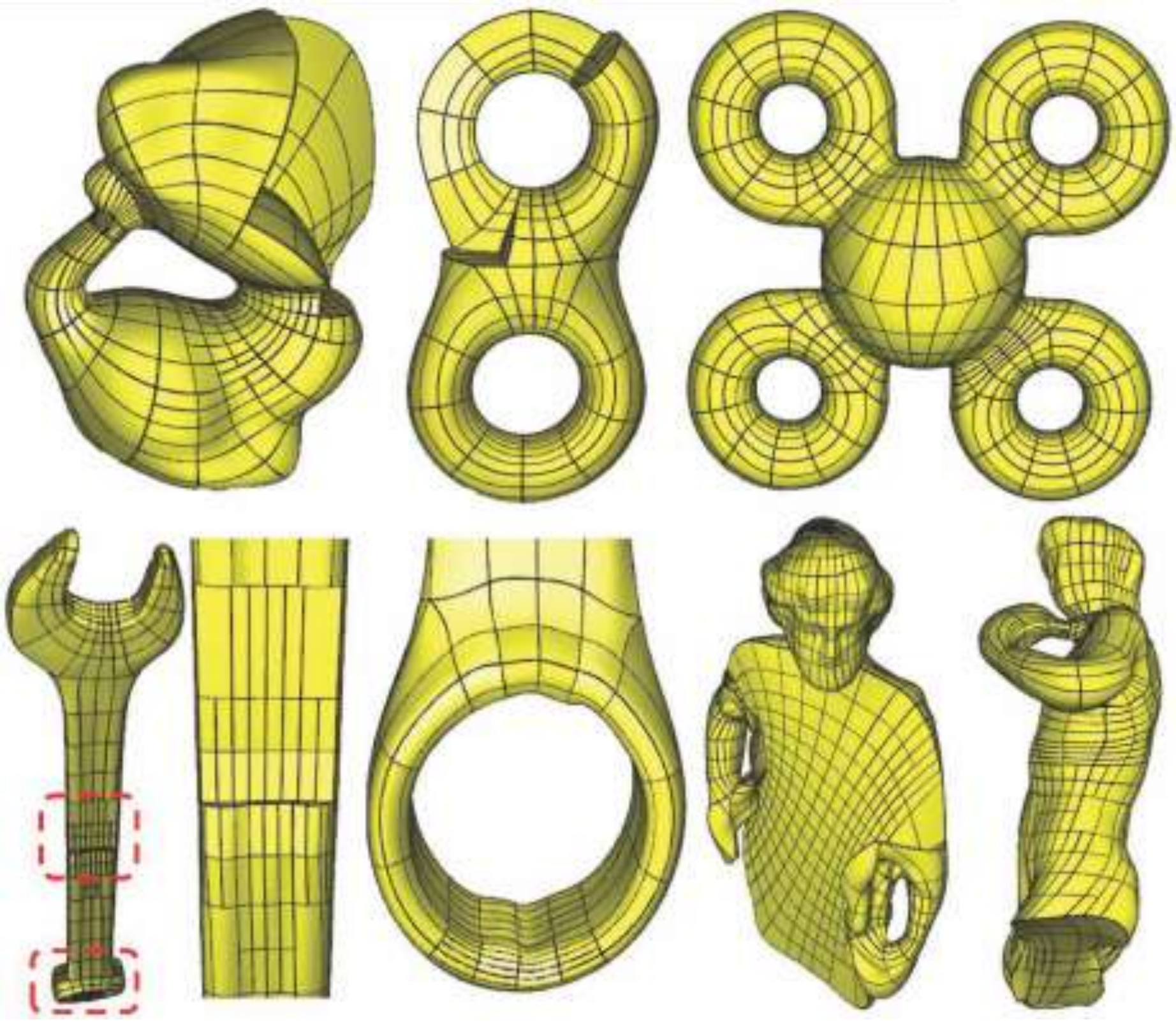

Figure 3.16: The volumetric spline approximation results.

## 3.7 Chapter Summary

In this chapter, we have presented a GPC spline framework for data transformation from surface meshes to continuous volumetric splines. The novelty in this chapter lies at the systematic handling of generalized poly-cube (GPC) parametric domain without any strong assumption. Compared with conventional poly-cube (CPC), GPC provides more generalized shape domain and better numerical stability to represent complicated models of arbitrary structure. We design a volumetric parameterization procedure based on GPC, which better handles solid objects with general topology and structure than existing volumetric parameterization techniques. We then devise a global "one-piece" volumetric spline based on GPC parameterization. The GPC construction enables a novel and desirable mechanism that facilitates the "one-piece" spline representation. Using local point-based strategy, global volumetric T-splines can be constructed on piece-wise GPC because transition functions can be effectively computed from the GPC's topological structure. The entire spline framework affords hierarchical refinement and level-of-detail control. Our GPC volumetric splines have great potential in various shape design and physically-based analysis applications. Our GPC is of great value to a wide range of geometry processing tasks, including volumetric isogeometric analysis [56], volume deformation, anisotropic material/texture synthesis.

# Chapter 4

# Component-aware Trivariate Splines

In the last chapter, we have proposed the technique to map the model into a generalized poly-cube domain. That means the integral model is decomposed into several cube components. Subsequently, it is natural to construct splines on each component and then glue them together. However in the previous chapter, we still use the global parameterization to approximate the global splines integrally for numerical reasons. Very naturally, this phenomena intrigues us to answer the question: "How to apply divide-and-conquer schemes onto decomposition-already inputs?" In this chapter, our primary goal is to develop efficient methods for arbitrary solids undergoing spline transformation, with local spline construction and global spline merging.

## 4.1 Motivation

To achieve this goal, we must address the following key challenges.

(1) **High genus.** An attractive spline representation must accommodate high-genus solid models with complicated shapes.

(2) **Local refinement and adaptive fitting.** For trivariate splines, both structurally-complicated shape models and feature-enriched models need local refinement. For example, a genus-0 solid bounded by 6 simple four-sided B-spline surfaces has originally $6 \times 1024^2$ control points (DOFs). The size of DOFs increases drastically to $1024^3$ or even larger when we naively convert it to a volumetric spline representation. This exponential increase during volumetric spline conversion poses a great challenge in terms of both storage and fitting costs. Therefore, it is advantageous to use high resolution to approximate boundary surface and low resolution for interior space.

(3) **Singularity free.** A *singular point in a volumetric domain* is a node with valence larger than four along one iso-parametric plane (Fig. 4.1(a)). Handling singularity with tenor-product splines is highly challenging in FEM, thus a singularity-free domain is highly desirable. Unfortunately, singularities commonly exist in many volumetric domains such as hexahedral meshes and cylinder (tube) domains.

(4) **Boundary restriction.** It is a basic requirement for a spline that all blending functions are completely confined within the parametric domain.

(5) **Semi-standardness.** A hierarchical spline is always formulated as Eq. 4.1. Semi-standardness, meaning that $\sum_{i=1}^{B} w_i B_i(u, v, w) \equiv 1$ always holds for all $(u, v, w)$, has a broader appeal to both theoreticians and practitioners.

Recently, much work has been attempted towards spline modeling of arbitrary topology shape while satisfying the aforementioned requirements, following a top-down fashion like Wang et al.[151]. They have proposed a theoretical trivariate spline scheme, being built upon volumetric poly-cube domains. Poly-cube is a shape composed of cuboids that abut with each other. All cuboids are glued in various merging types like Fig. 4.2, without any singular point (Note that the yellow dots are *not* singular points in the trivariate splines, even though they are singular for surface study). For example, a poly-cube parametric domain like Fig. 4.1(c) is designed to mimic shape geometry Fig. 4.1(b). Although their spline refinement guarantees the features such as semi-standardness and boundary restriction, this theoretical formulation encounters many difficulties. A global one-piece poly-cube domain, together with its 3D embedding, is not versatile enough to handle highly-twisted and high-genus solid datasets. Creating a poly-cube to mimic the input shape requires tedious user work. The boundary restriction procedure in the vicinity of gluing regions (Fig. 4.2, yellow dots/lines) is extremely complicated. Computationally speaking, the global fitting is very time consuming which is completely unsuitable for trivariate splines.

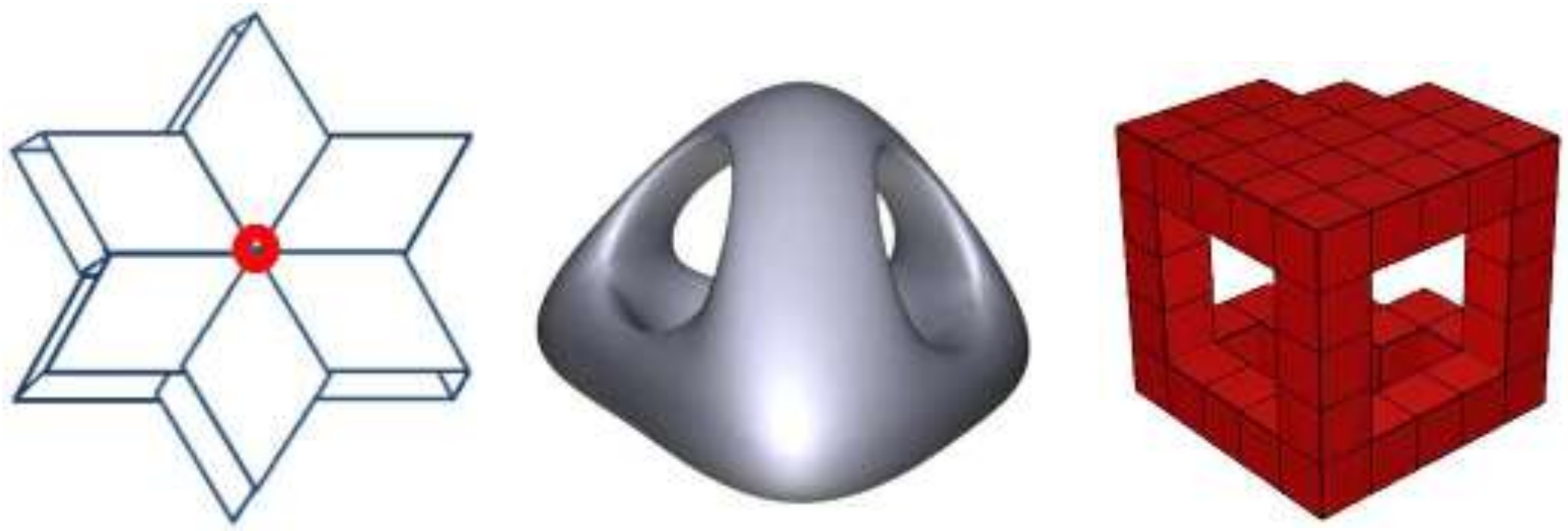

(a) (b) (c)

Figure 4.1: (a) The singular point in the volumetric domain. (b-c) A poly-cube domain can mimic the geometry of input and avoid such type of singular point.

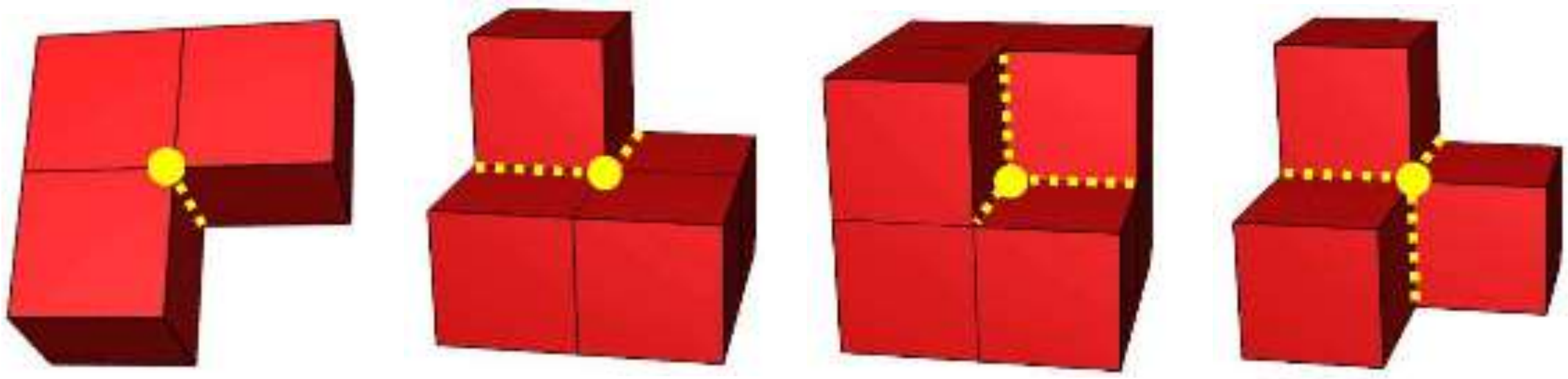

(a) (b) (c) (d)

Figure 4.2: All possible merging types in a poly-cube (“Type-1” to “Type-4”). To preserve both boundary restriction and semi-standardness, we add extra knots around the control points on the merging boundary (yellow lines and dots).

To ameliorate, our framework takes advantage of the bottom-up scheme. The global domain is divided into several components, with a controllable number and types of the cuboid merging. We build tensor-product trivariate splines separately for each component, and then glue them together. Compared with the top-down scheme, our divide-and-conquer method is more flexible and powerful to handle high-genus and complex shape. The interior space mapping and remeshing in each component is much easier. Compared with global fitting, our local fitting reduces both the computation time and space consumption significantly.

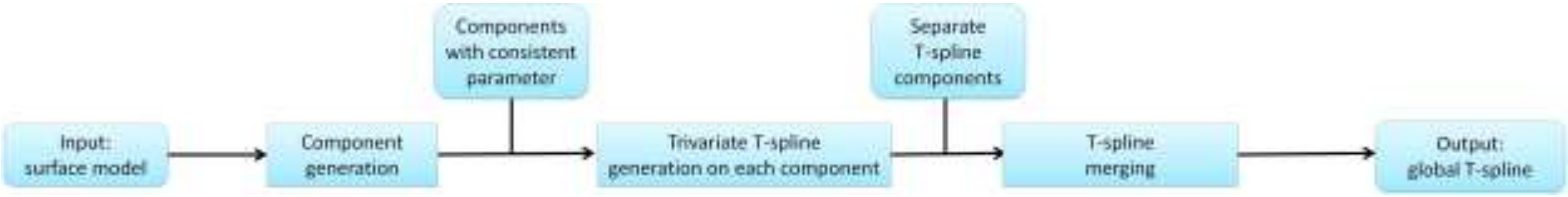

Figure 4.3: The divide-and-conquer scheme.

One key theoretical challenge in our divide-and-conquer scheme lies at designing merging strategies, so that the global spline after merging will still satisfy the semi-standardness and boundary restriction properties, especially around merging regions across adjacent cuboids. All possible cuboid merging types for a poly-cube are shown in Fig. 4.2. The traditional merging technique [118] only handles standard surface T-spline models defined over 2D domains without considering any 3D merging. In our framework, we have to design a new merging strategy, through adding extra knots and modifying weights of blending functions, to handle each merging case in Fig. 4.2, enforcing the semi-standardness and boundary restriction properties everywhere. Fig. 4.3 and 4.4 show the detailed, step-by-step procedure using a high-genus G3 model as an example. Specifically, it includes the following major phases:

(1) Construct a surface poly-cube mapping. To better support our divide-and-conquer scheme, we use the technique [74, 171-174] to decompose the entire surface model into several components. Each component is a part-aware surface patch and we map it to the boundary surface of a cuboid. We also guarantee in this step that separate cuboid mappings are globally aligned.

(2) Construct a local trivariate tensor-product T-spline on each cuboids (Section 4.3). Adaptive fitting is allowed for a better fitting result.

(3) Merge local cuboids into a single global spline (Section 4.4). Note that, the novelty of our merging strategy lies at its comprehensive and complete solution to guarantee the desirable properties: semi-standardness and boundary restriction.

Our new shape modeling framework has the following advantages:

1. Compared with prior top-down strategies, our new divide-and-conquer approach is more flexible and powerful to handle complex solids with arbitrary topology. Each component can be easily converted to a trivariate semi-standard regular spline, which is embraced by industry-standard CAD kernels and facilitates GPU computing like [93].

2. We develop the theory and algorithm to merge adjacent trivariate splines together. Through adding knots and modifying weights, our merging method can enforce semi-standardness and boundary restriction for all possible merging types, even after local adaptive refinement.

3. For solids with homogeneous material, we are capable of generating trivariate splines from poly-cube surface parameterization directly, thus we avoid complicated interior volumetric remeshing. Moreover, our divide-and-conquer strategy makes the modeling and analysis tasks scalable to large-scale volumetric data, in terms of computation time and space consumption during the fitting.

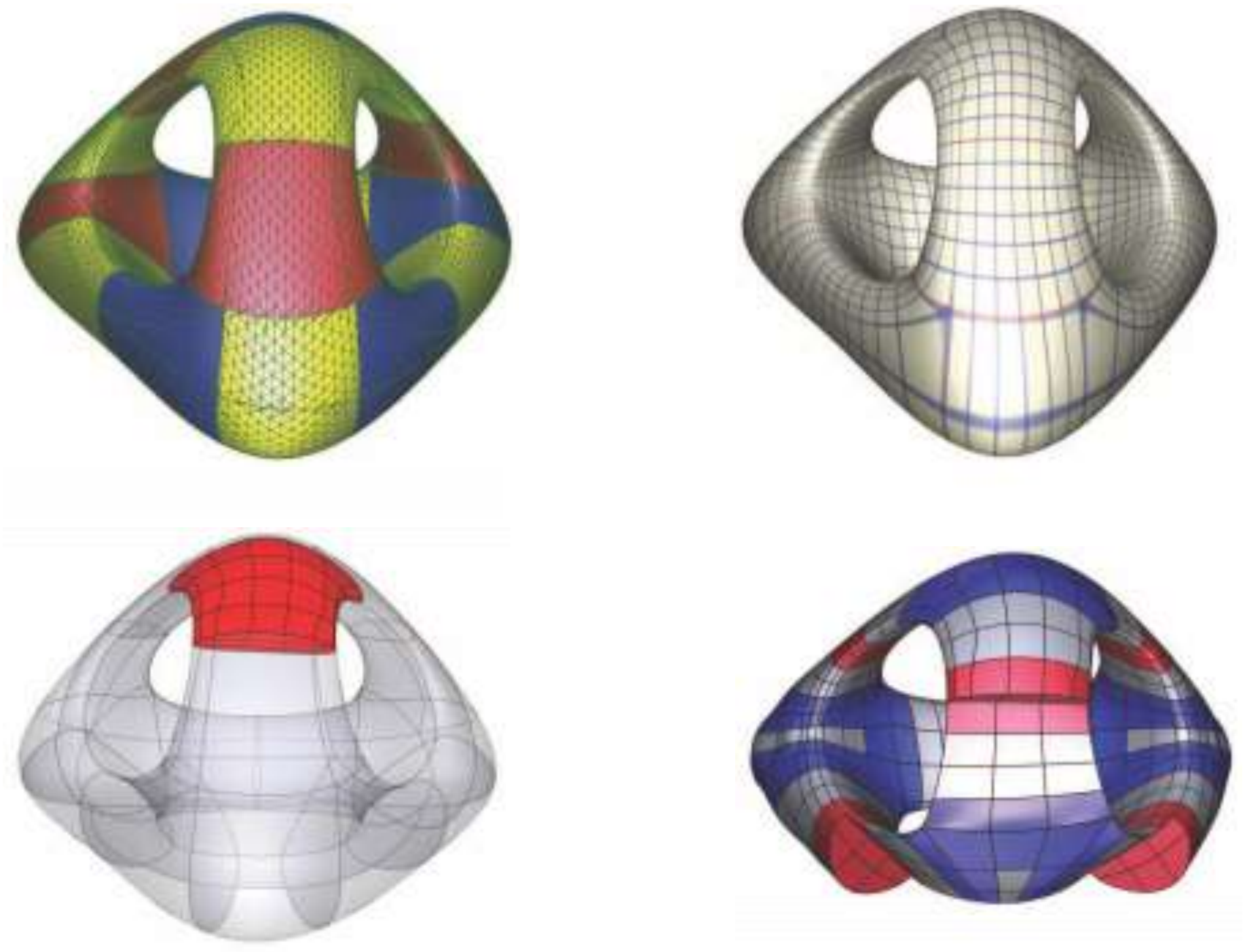

Figure 4.4: Steps to convert the G3 model into a trivariate T-spline solid.

## 4.2 Component Generation and T-splines

This section briefly reviews the required surface poly-cube generation algorithm. We also define the necessary notations for the rest of this chapter. In the interest of understanding, most illustrative figures about knots are simply shown in 2D layout, as their 3D generalizations are straightforward.

### 4.2.1 Component Generation

The starting point of our whole procedure is to decompose an input surface model into several component surface. Each component surface is part-aware and maps to a cuboid surface. The decomposition and mapping must follow the rule that parameters between neighboring components are consistent (i.e., we can glue their parameters together directly as a seamless aligned global poly-cube mapping). We remain agnostic as to which method should be used for such decomposition. However, in order to better promise these requirements, we utilize the algorithm introduced in Chapter 3. Compared with the conventional poly-cube mapping method like [141], our construction is specifically suitable for the divide-and-conquer strategy and spline construction. (1) The conventional poly-cube method always generates an integral poly-cube domain to mimic the whole shape at first. Then we have to decompose this integral domain into small pieces for applying the divide-and-conquer strategy. In contrast, our method directly uses a small set of connected local cuboids, each of which represents a geometrically meaningful patch (e.g., part-aware). This property is particularly suitable for highly-twisted/non-axis aligned/high-genus models (e.g., the g3 model). More importantly, we can use the divide-and-conquer technique directly on our resulting poly-cube without further decomposing anymore. (2) Our method can also reduce the number of cuboids, and control the merging types efficiently: It only generates "Two-cube" and *"Type-1"* (Fig. 4.2(a)) merging, thus it simplifies the merging requirement.

### 4.2.2 Trivariate T-spline

To better prepare readers for the better understanding of the following algorithm, we briefly define the volumetric T-spline representation (The surface T-spline formulation is detailed in [118]). Also we give the detailed explanation of *"Semi-standardness"* and *"Boundary Restriction"* as follows.

We use $T(\mathsf{V}, \mathsf{F}, \mathsf{C})$ (or simply $T$) to denote a control grid domain, where $\mathsf{V}$, $\mathsf{F}$, and $\mathsf{C}$ are sets of vertices, faces and cells, respectively. Given $T$, a trivariate T-spline can be formulated as:

$$F(u, v, w) = \frac{\sum_{i=1}^{\mathsf{B}} w_i p_i B_i(u, v, w)}{\sum_{i=1}^{\mathsf{B}} w_i B_i(u, v, w)}, \tag{4.1}$$

where $(u, v, w)$ denotes parametric coordinates, $p_i$ is a control point, $\mathsf{W}$ and $\mathsf{B}$ are the weight and blending function sets. Each pair of $< w_i B_i >$ is associated with a control point $p_i$. Each $B_i(u, v, w) \in \mathsf{B}$ is a blending function:

$$B_i(u, v, w) = N^3_{i0}(u) N^3_{i1}(v) N^3_{i2}(w), \tag{4.2}$$

where $N^3_{i0}(u)$, $N^3_{i1}(v)$ and $N^3_{i2}(w)$ are cubic B-spline basis functions along u, v, w, respectively.

In the case of cubic T-spline blending functions in Eq. 4.1, the univariate function $N^3_j$ for each blending function $B_i$ is constructed upon knot vector $R^j = [r^j_{-2}, r^j_{-1}, r^j_0, r^j_1, r^j_2]$, where $R^j$ is a tracing ray parallel to the control grid (See Fig. 4.5(b)): Starting from a knot $k = r^0_0, r^1_0, r^2_0$, we can trace to $r^0_1$ and $r^0_{-1}$, which are the very first intersections when the ray $R(t) = (r^0_0 \pm t, r^1_0, r^2_0)$ comes across one cell face. Naturally, we define the parameter of a control point as the central knot of the knot sequence for the control point.

To support downstream CAE applications, our spline framework has the following requirements:

**Semi-standardness.** $\sum_{i=1}^{\mathsf{B}} w_i B_i(u, v, w) \equiv 1$ holds for all $(u, v, w)$ in Eq. 4.1, so that the evaluation of spline functions and their derivatives is both efficient and stable. Eq. 4.1 can be rewritten as:

$$F(u, v, w) = \sum_{i=1}^{\mathsf{B}} w_i p_i B_i(u, v, w), \tag{4.3}$$

**Boundary restriction.** We require that blending functions of all control points are strictly confined within parametric domain boundaries. Unfortunately, achieving this requirement is not trivial, especially around the cuboid merging regions. Fig. 4.5 shows a counter-example. A standard control point's blending function (green box), without confinement procedure, tends to intersect with the boundary. In CAE-based force analysis, it means the strain energy "escapes the border", which might lead to an abrupt bend, twist, and flip-over phenomena in experiments. In the follow sections, we usually use *"central points"* for the control point/knot with an unconfined blending function, since the confinement procedure is mainly through adding extra knots/control points around the central point. However, even we design the additional knots carefully and successfully confine the blending function, we still have to recompute all control points' weights around the knots-adding region, otherwise we will break the semi-standardness around this local region.

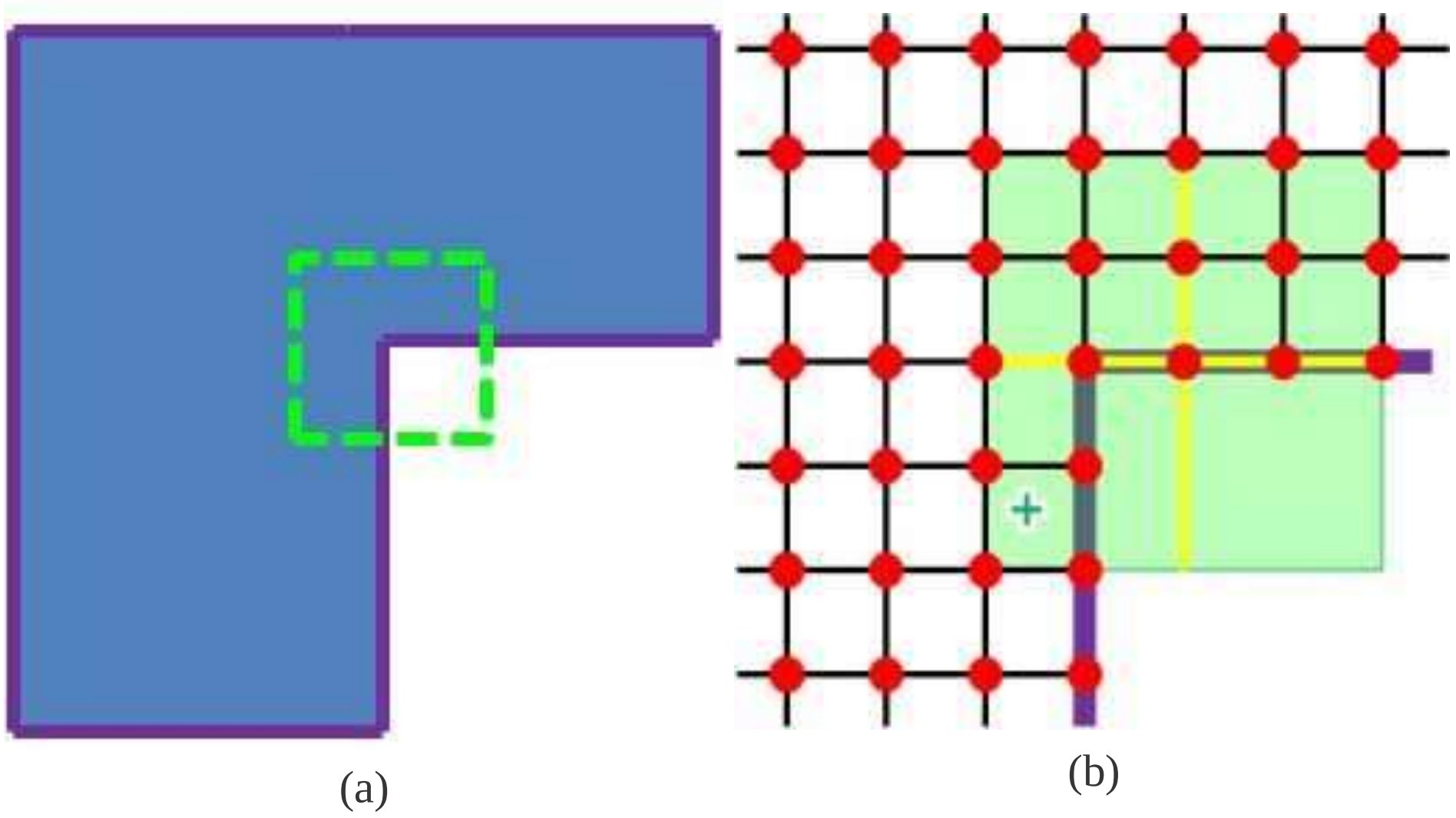

Figure 4.5: Counter-example of boundary restriction. (a) A "Type-1" merging in a 2D layout. (b) The blending function's supporting region (green box) crosses the boundary. The supporting region is determined by tracing rays (yellow lines).

## 4.3 T-spline Construction for Each Component

The construction of trivariate splines on each component is very critical in our divide-and-conquer method. Two major goals are involved in this step. Besides constructing T-splines preserving desirable features, we have to satisfy the necessary requirement in each component in anticipation for merging. We propose the following procedure to satisfy both goals:

1. Construct a boundary restricted control grid.
2. Perform the meshless fitting to determine locations of all control points.
3. Subdivide the control grid via local refinement iteratively. Perform fitting again in each iteration for a better fitting result.
4. Modify the control grid around merging boundary after each subdivision iteration in anticipation for merging.

### 4.3.1 Boundary Restricted Control Grid

In order to construct a control grid, we first divide the cuboid block into cells by grid coordinates. The grid coordinates along k-axis are denoted as:

$$S_k = [s^k_1, s^k_2, \ldots, s^k_{n_k}], k = 1, 2, 3,$$

where $n_k$ is the resolution of rectilinear grid along k-axis and each value in $S_k$ is the normal subdivision of cuboid parameter along k-axis. The tensor product of $S_1, S_2, S_3$ divides the block into $(n_1 - 1) \times (n_2 - 1) \times (n_3 - 1)$ cells and gives rise to a point-based spline on $n_1 \times n_2 \times n_3$ control points.

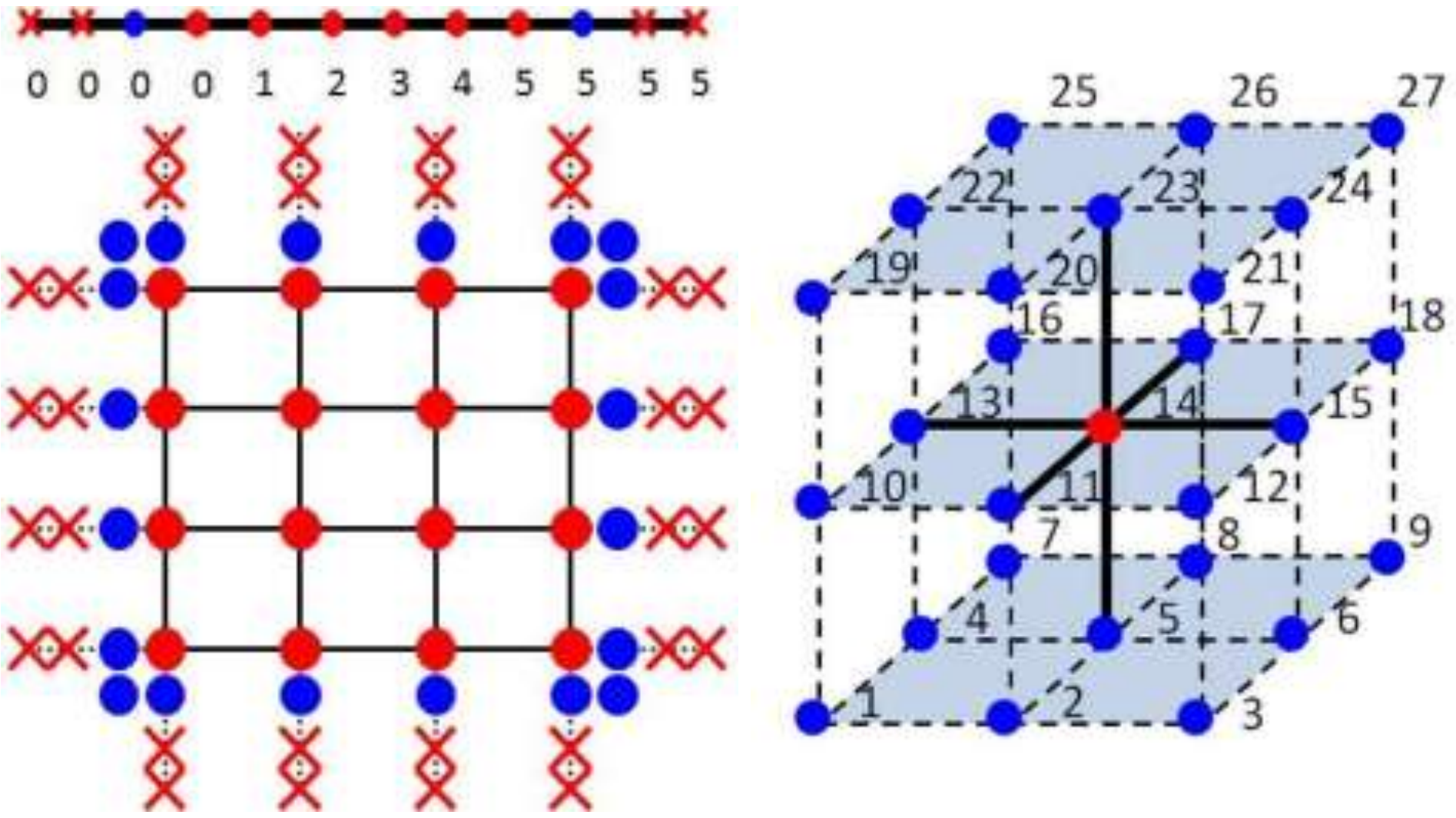

Figure 4.6: (a) Top: Boundary restriction is illustrated on a 1D domain, with 6 “boundary knots” (or called “bd-knots”, [0, 0, 0] and [5, 5, 5]) and two “boundary control points” (or called “bd-control-points”, blue dots) inserted. (a) Bottom: Boundary restricted control grid in a 2D layout. (b) All possible bd-control-points around one central point.

Table 4.1: Refining $N_R$ by inserting k into knot vector $[r_0, r_1, r_2, r_3, r_4]$ generates two basis functions $N_{R_1}$ and $N_{R_2}$.

| k | $R_1$ | $R_2$ |
|---|---|---|
| $r_0 \leq k < r_1$ | $[r_0, k, r_1, r_2, r_3]$ | $[k, r_1, r_2, r_3, r_4]$ |
| $r_1 \leq k < r_2$ | $[r_0, r_1, k, r_2, r_3]$ | $[r_1, k, r_2, r_3, r_4]$ |
| $r_2 \leq k < r_3$ | $[r_0, r_1, r_2, k, r_3]$ | $[r_1, r_2, k, r_3, r_4]$ |
| $r_3 \leq k \leq r_4$ | $[r_0, r_1, r_2, r_3, k]$ | $[r_1, r_2, r_3, k, r_4]$ |

However, this naive spline construction is open boundary and violates the requirement of boundary restriction. To improve, we replicate the non-uniform knots at both ends of $S_k$ to restrict the blending functions within the domain (See Fig. 4.6(a)Top): We add 3 extra knots, called *boundary knots (bd-knots)*, at the end of domain to restrict the boundary. The knot set is expanded:

$$S_k = [s_1^k, s_1^k, s_1^k, s_1^k, s_2^k, \ldots, s_{n_k}^k, s_{n_k}^k, s_{n_k}^k, s_{n_k}^k].$$

We also add 1 extra *boundary control point (bd-control-point)* (blue dots), on the bd-knot outside of the last control point on the boundary. Fig. 4.6(a)Bottom extends it to a 2D domain, and its extension to the 3D domain is in the same pattern. Our spline definition achieves: (1) Every blending function in each domain is confined within the domain boundary; (2) Only bd-control-points’ blending functions influence the cuboid boundary, so our following fitting method can rely on this usable property.

In order to represent the bd-control-points conveniently, we can arrange them into a $3 \times 3 \times 3$ grid around the central point as Fig. 4.6(b) (Recall that the central point is the control point with an unconfined blending function). These 27 possible knots share the same parameters as the central point. It is only designed to explicitly record topological relations of these control points in preparation for efficient spline merging. After adding bd-control-points to the 3D control grid, each central point on the corner/edge/face has 8/4/2 control points, respectively (Fig. 4.7). This special bd-control-point representation is uniquely suitable for merging processing as shown in Section 4.4.

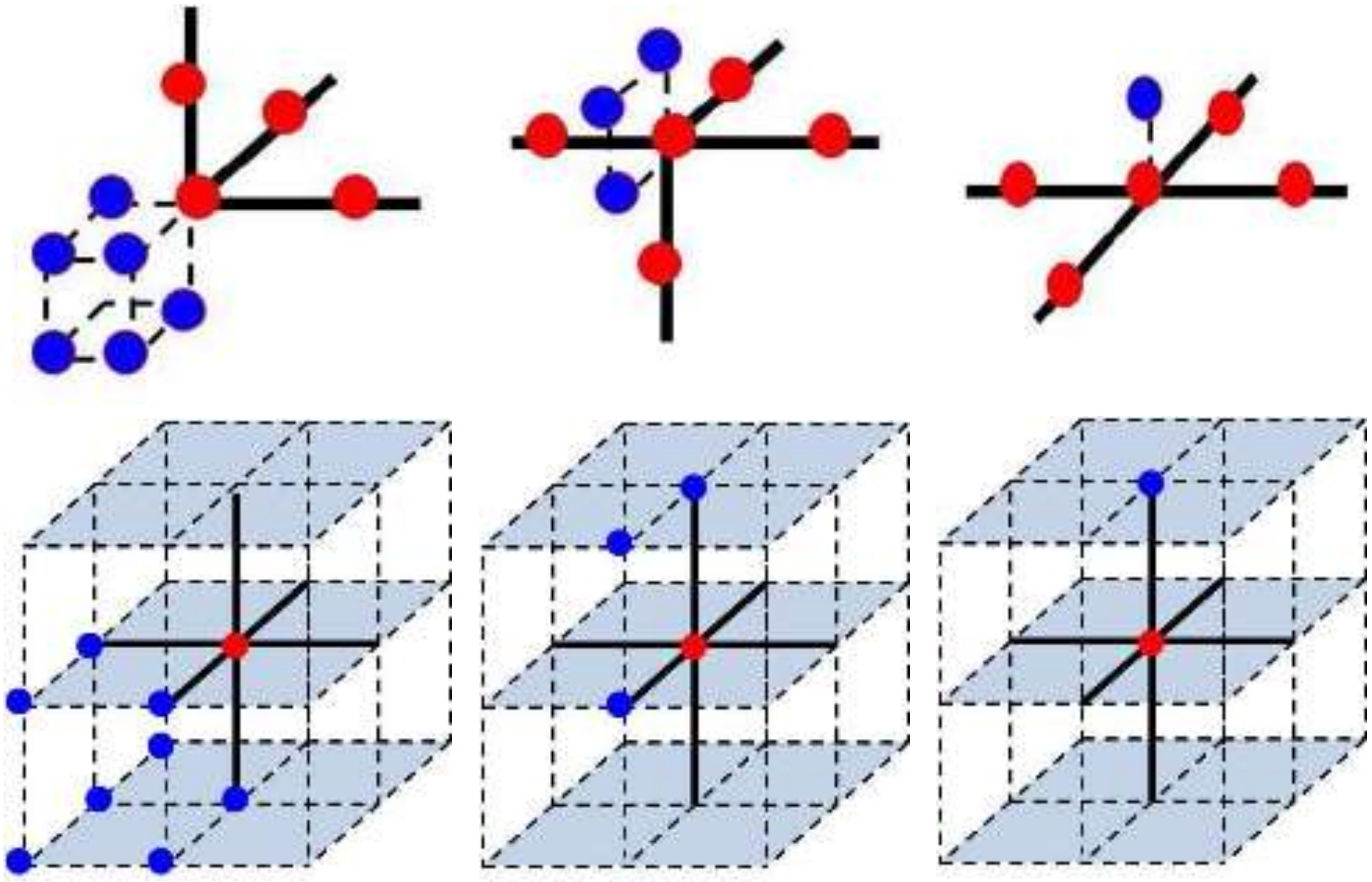

Figure 4.7: The bd-control-point distributions around a central point on the corner/edge/face vertex, respectively.

### 4.3.2 Meshless Fitting

Our input only includes a control grid and a group of surface sample points extracted from the surface patch (already mapped to a cuboid domain surface). The challenge consists in designing a fitting method for solids without interior volumetric parameterization or remeshing.

1. **Boundary fitting.** We first determine the positions of bd-control-points only. Recall that only bd-control-points $p_i^b$ influence the cuboid surface sample points. Therefore, we can determine their positions by minimizing Eq. 4.4 w.r.t. to surface sample point $v_j^b$:

$$\text{argmin}\left(\sum_{j=1} ||F(f^{-1}(v_j^b)) - v_j^b||\right) \tag{4.4}$$

$$\Rightarrow \frac{\partial}{\partial p_i^b} \sum_{j=1} (F(f^{-1}(v_j^b)) - v_j^b)^2$$

where F denotes the spline function as Eq. 4.1 and $f^{-1}(v_j^b)$ the parameters of $v_j^b$ in the cuboid. The above equation can be rewritten in matrix format as in the least square method:

$$\frac{1}{2} P^T B^T BP - V^T BP = 0, \tag{4.5}$$

where B is the matrix of blending functions $B_{ij} = I_{3\times3}B_i(f^{-1}(v_j^b))$, V and P denote the vectors of surface sample points $v_j^b$ and bd-control-points $p_i^b$, respectively. This equation determines bd-control-points and they serve as the constraint in the next interior fitting step.

2. **Interior fitting.** Let u in the set U be the interior parametric value. Each $u_i = (u, v, w)$ is the interior parameter triplet in the tensor-product parametric grid $(u_0, u_1, \ldots, u_{n_0}) \times (v_0, v_1, \ldots, v_{n_1}) \times (w_0, w_1, \ldots, w_{n_1})$. Theoretically, we have the following harmonic equation w.r.t. interior control points $p_j^{in}$:

$$\operatorname{argmin}\left(\sum_{i=1}^{m}\int_{\Omega_i} ||\nabla\cdot\nabla F(u_i)||du\right) \tag{4.6}$$

$$\Rightarrow \frac{\partial}{\partial p_j^{in}}\sum_{i=1}^{m}\int_{\Omega_i}(\triangle F(u_i))^2 du = 0,$$

where $\Omega_i$ is an infinitesimal parametric volume around $u_i$. Similar as [168], the above minimized energy $\int_{\Omega_i} ||\triangle F(u_i)||$ can be approximated by the following formulation:

$$\sum_{j=0}^{m} w_{ij}F(u_j) = 0, w_{ij} = \begin{cases} 1 & i = j, \\ -\frac{1}{6} & u_j \in Nbr(u_i) \\ 0 & \text{others} \end{cases} \tag{4.7}$$

where Nbr includes 6 immediate neighbors of $u_i$ in the tensor-product parametric grid. We substitute Eq. 4.7 into Eq. 4.6, which can be solved by the least square method similar to Eq. 4.4. During computing we set already-known $p_i^b$ as constraints and get all other control point positions.

**Global alignment.** Although we execute volumetric fitting separately on every cuboid, our fitting technique still guarantees global alignment of interior fitting results. Recall that we already obtain the identical surface parameters between cuboids before fitting, since we generate aligned poly-edges (i.e., cuboid edges). Therefore, two cuboids minimize precisely the same energy in Eq. 4.4 and Eq. 4.6 on the boundary, leading to the equivalent fitting results.

### 4.3.3 Cell Subdivision and Local Refinement

If the fitting results do not meet certain criteria on each cuboid, we can always perform subdivision over cells in the control grid with large fitting errors and then conduct the volumetric fitting. Each cell is split along 3-axis and divided into eight sub-cells naturally.

The challenge is how to preserve the semi-standardness during subdivision. Sederberg et al. [118] have proposed a feasible approach to refine blending functions on surface patch. We generalize this technique onto our 3D control grid. Let $R = [r_0, r_1, r_2, r_3, r_4]$ be a ray-tracing knot vector and $N_R(u)$ denotes the corresponding cubic B-spline basis function. If there is an additional knot $k \in [r_0, r_4]$ inserted into R, N can be written as a linear combination of two B-spline functions:

$$N_R(u) = c_1N_{R_1}(u) + c_2N_{R_2}(u). \tag{4.8}$$

Two knot vectors $R_1, R_2$ are shown in Table 4.1, $c_1$ and $c_2$ are 2 weights that can not exceed 1:

$$c_1 = \min\left(\frac{k - r_0}{r_3 - r_0}, 1\right), c_2 = \min\left(\frac{r_4 - k}{r_4 - r_1}, 1\right).$$

Since the blending function of B is the tensor product of N along 3-axis, we can also formulate the refined blending functions along one axis:

$$B_i \equiv c_1 B_{i1} + c_2 B_{i2}. \tag{4.9}$$

The procedure of our 3D subdivision and local refinement consists of following steps. The input is a queue of cell $Q_c$.

1. Subdivide cells in $Q_c$ and insert the new vertices into the domain $T$, and update $T$ to $T^*$
2. For all pairs of blending functions $< w_i B_i >$, $w_i \in W, B_i \in B$, compute its new knot vector $R^*$ (See Section 4.2). Then,
   - If the $R^*$ includes the knot which does not exist in $T^*$, insert a new vertex on that knot into the domain $T^*$.
   - If the $R^*$ is more refined than R, compute the refinement $B_i = c_1 \times B_{i1} + c_2 \times B_{i2}$. Insert the new blending functions $< w_i \times c_1 B_{i1} >$ and $< w_i \times c_2 B_{i2} >$ into the control grid. Delete the
old pair $< w_i B_i >$.
3. Repeat the last step until no new knot vector in $R^*$. Collect all blending functions on the same control point and use the total weight as its new weight.

The above procedure can handle refinement and knot extraction on a complicated 3D control grid. It also determines new required control points automatically to guarantee the semi-standardness. Note that unlike [118], we perform spline fitting again after each refinement iteration to update control point positions. This is mainly because our goal of refinement is to seek for more accurate fitting result. In contrast, the refinement in [118] aims to keep the shape unchanged.

### 4.3.4 Boundary Modification

Boundary modification is necessary for our semi-standard T-spline component, because of the fundamental difference between standard B-spline and our merging strategies. Fig. 4.12 intends to visually show the difference between them. It illustrates the 1D merging method introduced in [120] on our boundary restricted grid. For a $C^2$ merging, 3 control points on one component will be merged with 3 control points on the other component to form a joint new spline. However, the procedure does not take the associated weights into consideration. In standard B-spline, all the weights are uniform. However, in semi-standard T-spline, it is possible that two corresponding soon-to-be-joined control points have different weights. As a result, the semi-standardness around the merged regions will break down. Therefore, we have to add extra requirement about weights to make these control points be capable of merging.

**To-be-merged control point (Definition 1).** *For a control point, if its blending function includes bd-knots around merging boundary, we say this control point is a "To-be-merged" control point* (For example, Fig. 4.11(a-b) in a 2D layout).

**Modification zone (Definition 2).** *For any cell in the control grid, if one of its* 8 *vertices is "To-be-merged" control point for one boundary, we say this cell is in the "Modification zone".*

A merging-ready spline must have the following properties:

**Boundary requirement (Proposition 1).** *The weights of all "To-be-merged" control points on this boundary must equal to one, such that we can merge two splines and the resulting spline still preserves semi-standardness.*

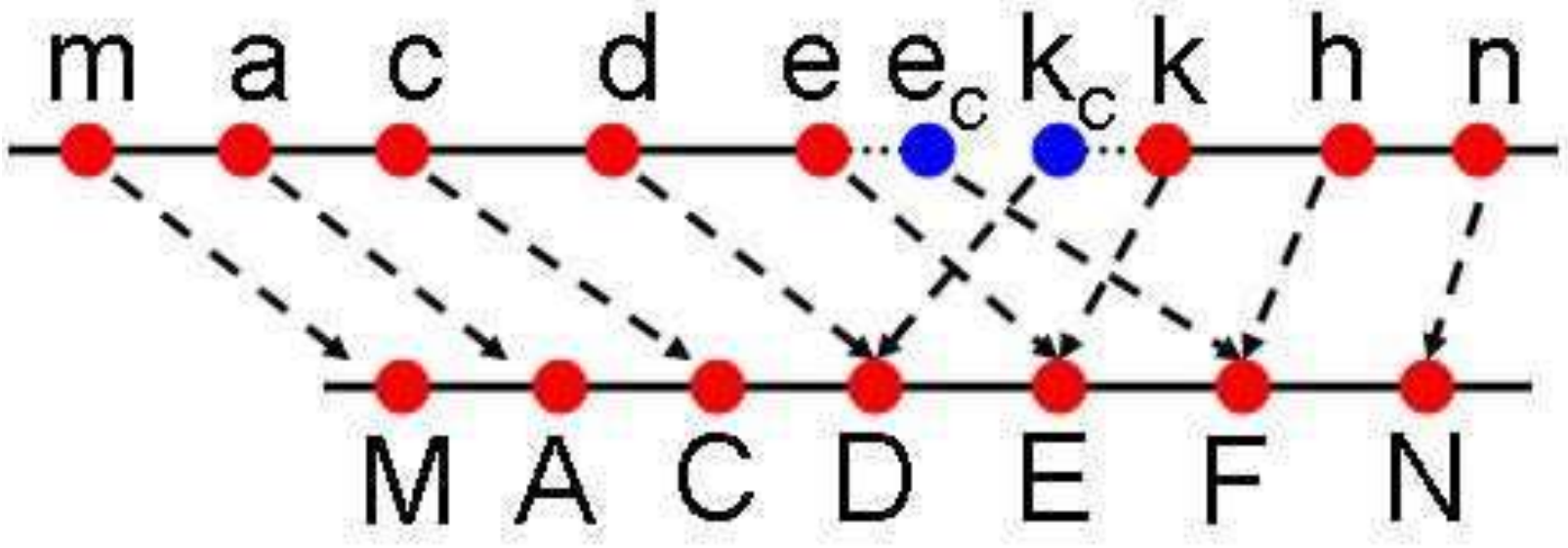

Figure 4.8: Proof of Observation 1.

**Proof:** As shown in Figure 4.8, two 1D splines are merged. Because of the symmetry, we only analyze the left side of merging boundary (line segment $[M, E]$). During merging, only control points D, E, F update blending functions and weights. The influenced region thus narrows down to the line segment $[A, E]$.

1) Line segment $[A, C]$: The semi-standardness preserves before merging, thus $w_m B_m + w_a B_a + w_c B_c + w_d B_d \equiv 1$ on line segment [a, c], $B_i$ is the blending function at i. After merging, $w_m = w_M$, $w_a = w_A$, $w_c = w_C$, $w_d = w_D = 1$, $B_m = B_M$, $B_a = B_A$, $B_c = B_C$, thus we only need to prove that $B_d = B_D$ between $[A, C]$. The knot vector of $B_d$ is [a, c, d, e, e]. The knot vector of $B_D$ is [a, c, d, e, f]. According to Eq. 8,

$$B_D = B_d + \frac{f - e}{f - c} B'$$

The knot vector of $B'$ is [c, d, e, e, f], which does not influence the line segment $[A, C]$. Therefore, we get $B_D = B_d$ on $[A, C]$.

2) Line segment $[C, E]$: Similar to $[A, E]$, we only need to prove $B_d + B_e + B_{e_c} = B_D + B_E + B_F$. Our subdivision procedure under "boundary requirement" generates a local trivariate B-spline on line segment $[D, F]$ along this direction. According to B-spline merging, $B_d + B_e + B_{e_c} = B_D + B_E + B_F$ on $[C, E]$.

To guarantee that all "To-be-merged" control points' weights equal to one, we need to be able to recognize if a local subdivision breaks the above rule or not:

**Proposition 2.** *Any knot insertion outside of "Modification zone" does not violate boundary requirement (i.e., weights of bd-control-points equal to one).*

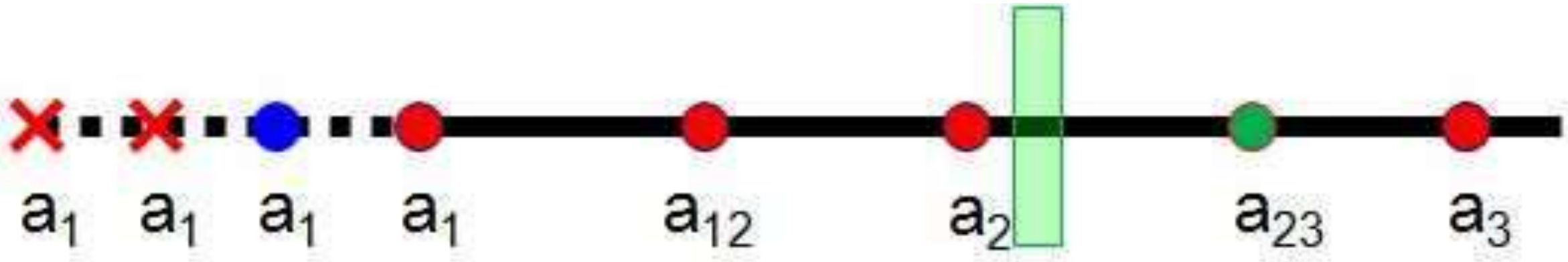

Figure 4.9: Proof of Proposition 2. "Modification zone" is the left of "Green Bar". Three nodes on $a_1$, $a_{12}$ represent "to-be-merged" control points.

**Proof:** The point on $a_{23}$ is the nearest newly-inserted control point outside "Modification Zone". According to Table 1, none of refined blending functions takes $a_1$ as the center of knot vector. Thus the weights of two “to-be-merged” control points on $a_1$ are unchanged. For “to-be-merged” control point $a_{12}$, its refined blending function is only subdivided from the original blending function $B_{a_{12}}$, located at $a_{12}$. According to Eq. 8, $B_{a_{12}} = c_1 \times B^{\prime}_{a_{12}} + c_2 \times B_{a_2}$. The new weight of $a_{12}$ is $c_1 = \min(\frac{a_{23} - a_1}{a_2 - a_1}, 1) = 1$. Therefore, the new weight on $a_{12}$ still equals to one.

After detecting the potential violation, we can properly handle it using the following proposition:

**Proposition 3.** *If we subdivide all boundary cells around merging region at the same time, the new “To-be-merged” control points still guarantee “Boundary requirement” and their wights all equal to one.*

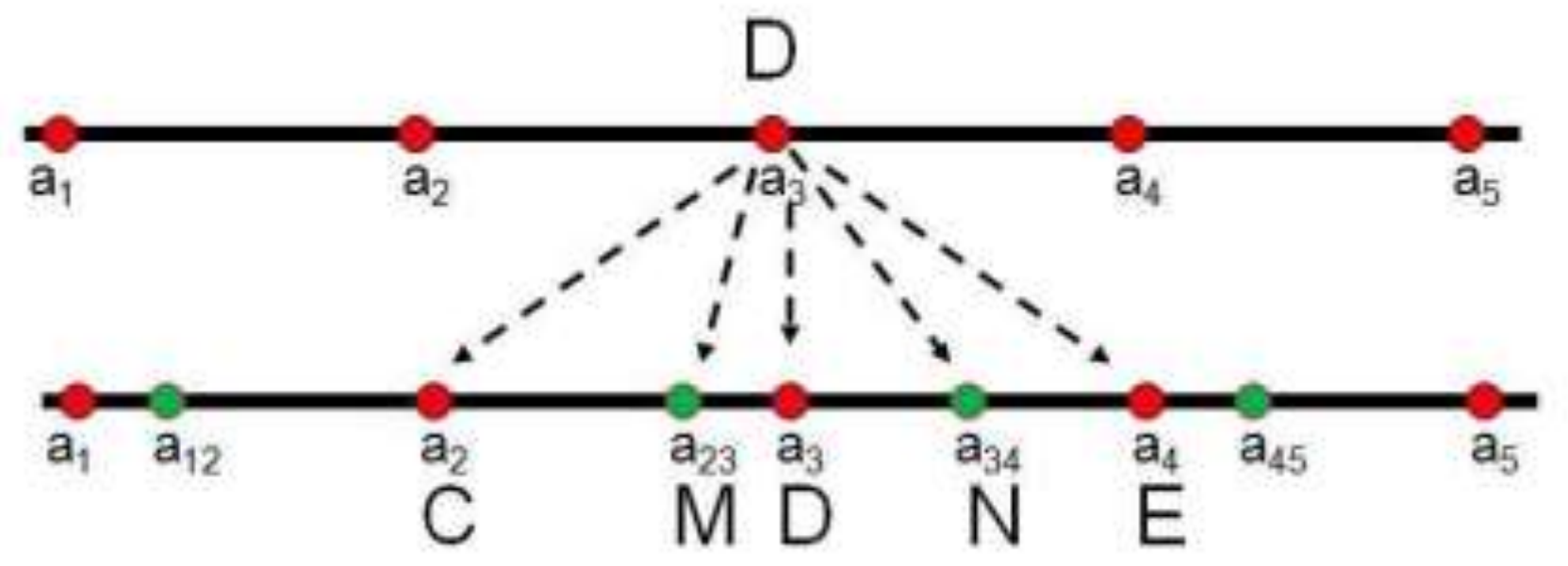

(a)

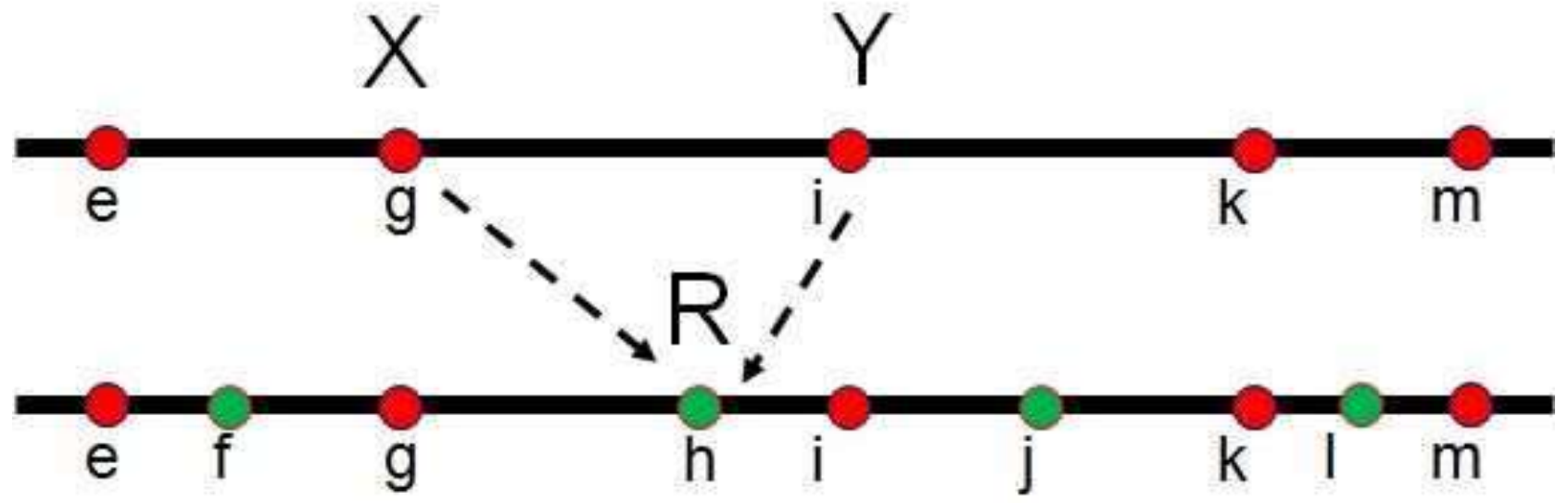

(b)

Figure 4.10: Proof of Proposition 3. (a) blending function on $D$ subdivides and generates new blending functions at C, M, D, N, E. (b) updated blending functions on $R$ only result from subdivision of $X$, $Y$.

**Proof:** After subdivision, each blending function is subdivided to several sub-blending functions pairs $< w_i B_i >$. These pairs are distributed to other knots: For example, subdivision of blending function located at $D$ generates new pairs on C, M, D, N, E and the weights on each node can be computed by Eq. 8:

$$D = \frac{(a_{12} - a_1)(a_4 - a_{23})(a_4 - a_{34})}{(a_4 - a_1)(a_4 - a_{12})(a_4 - a_2)} + \frac{(a_{23} - a_{12})(a_4 - a_{34})}{(a_4 - a_{12})(a_4 - a_2)} + \frac{(a_5 - a_{23})(a_{34} - a_2)}{(a_5 - a_2)(a_4 - a_2)}$$

$$M = \frac{(a_{12} - a_1)(a_4 - a_{23})}{(a_4 - a_1)(a_4 - a_{12})} + \frac{a_{23} - a_{12}}{a_4 - a_{12}}, N = \frac{a_5 - a_{34}}{a_5 - a_2}$$

$$C = \frac{(a_{12} - a_1)(a_{23} - a_1)}{(a_4 - a_1)(a_3 - a_1)}, E = \frac{(a_5 - a_{34})(a_5 - a_{45})}{(a_5 - a_2)(a_5 - a_3)}$$

The weight of refined blending function is the summation of subdivided weights. Considering $R$ as an example, the refined blending function on $R$ is only derived from unrefined blending functions on $X$ and $Y$.

According to the above equations, we can compute the weights from $X$ and $Y$:

$$X \Rightarrow R : \frac{k-h}{k-e}, Y \Rightarrow R : \frac{(f-e)(k-h)}{(k-e)(k-f)} + \frac{h-f}{k-f}.$$

The summation of weights on R is

$$\frac{f-e}{k-e} \times \frac{k-h}{k-f} + \frac{h-f}{k-f} + \frac{k-h}{k-e} \equiv 1.$$

Based on the above propositions, we propose our modification procedure as follows. The input is the newly refined control grid with new subdivided cell set $C_{new}$.

1. For each boundary, assign the cell set $C_T$ as "Modification zone". For any cell with one vertex as a "To-be-merged" control point, we add this cell into $C_T$.

2. For each boundary, detect if there is any new subdivided cell in the "Modification zone":
   - $C_{new} \bigcap C_T = \emptyset$. According to Proposition 2, the refined grid preserves the standardness on the boundary, so no further processing.
   - $C_{new} \bigcap C_T = \emptyset$. Modify the boundary according to Proposition 3: Subdivide all cells on the boundary to satisfy "Boundary requirement".

3. Update control point positions. Instead of fitting again like in Section 4.3.3, we use the same method as in [118] because we seek for keeping spline shape unchanged in this step.

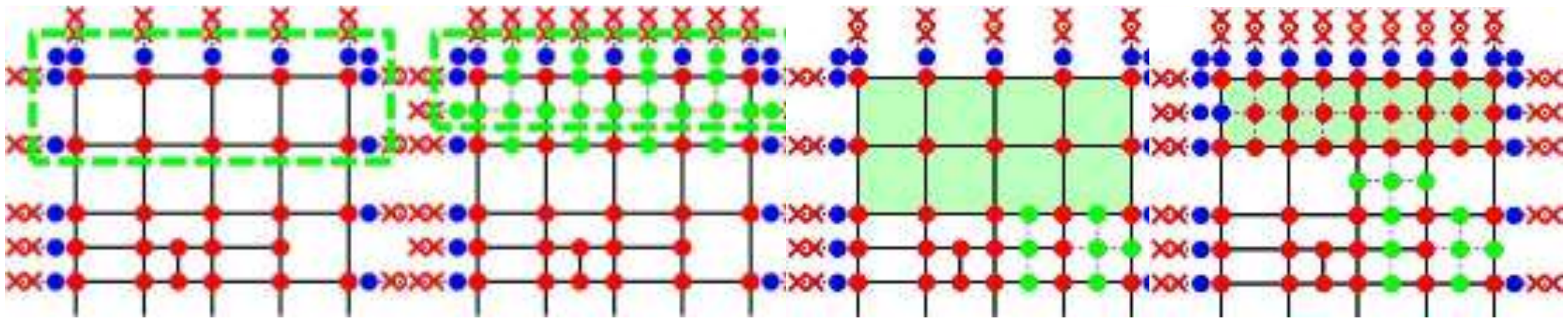

(a) (b) (c) (d)

Figure 4.11: Boundary modification. (a) Original "To-be-merged" control points (in the green box). (b) Subdivision all cells along the boundary, according to Proposition 3. The green box covers updated "To-be-merged" control points. (c) and (d) "Modification zone" (green box) of (a) and (b). According to Proposition 2, cell subdivision (by green dots) outside "Modification zone" does not violate "Boundary requirement" (Proposition 1).

## 4.4 Global Merging Strategies

In our framework, the decomposed components can be merged in various different merging types. We develop algorithms to handle different types of merging in this section. As we discussed in Section 4.2.1, our domain only includes "Two-cube" merging (Section 4.4.1) and "Type-1" merging (Section 4.4.2). Also, we seek to handle more complicated conventional poly-cube domains, including all other types of merging in Fig. 4.2 (Section 4.4.3).

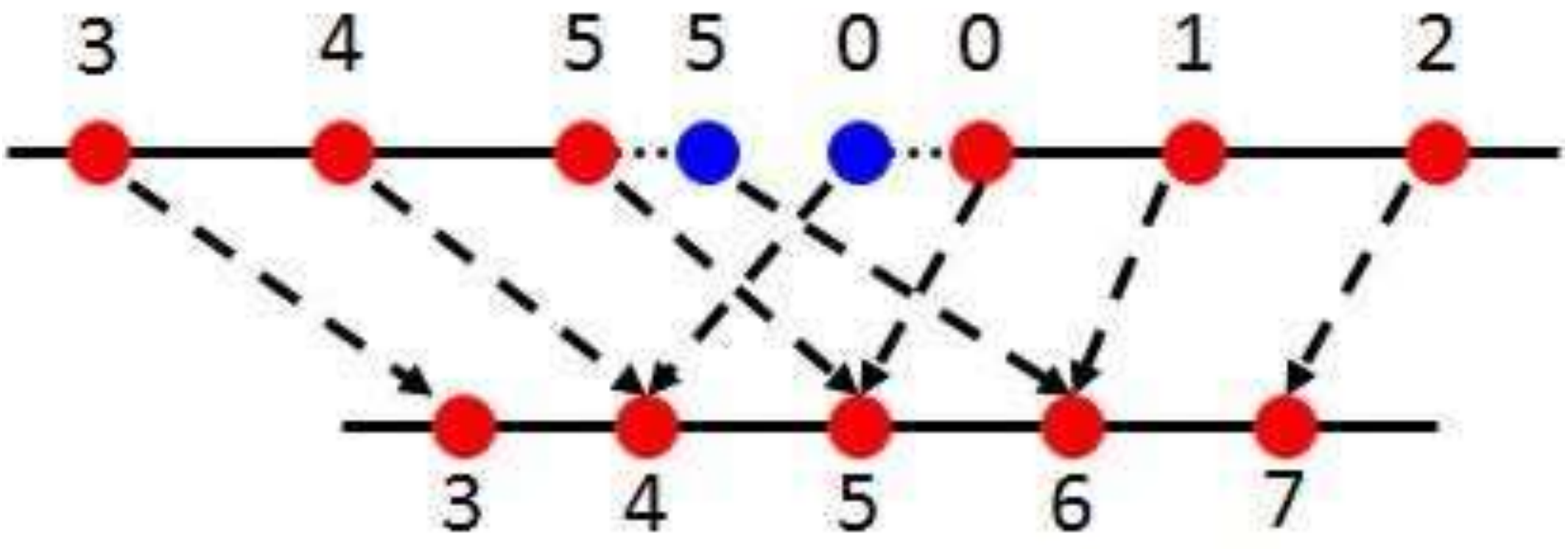

Figure 4.12: Two cube merging in 1D layout. Two control points are combined to form one new control point (4, 5 and 6).

### 4.4.1 "Two-cube" Merging

Merging of 3D components can be simply illustrated by 1D merging. In 1D merging, each boundary parameter corresponds to a new position after merging. For example, in Fig. 4.12, the bd-control-point with parameter 5 corresponds to a new parameter 6. The control point corresponding to $n (n \geq 2)$ original control points simply takes the average position as its new position. Similarly, the merging of two cuboids includes the following steps.

1. Boundary modification. If bd-knot intervals of two components are different, subdivide the cube boundary using the procedure in Section 4.3.4 iteratively until they share the same knot interval (Fig. 4.13(a)).

2. Merging control points. Correspond the original control point to the new control grid. As shown in Fig. 4.13(a)Right, we merge each column along the merging direction as 1D case.

3. Computing control point positions. Each new control point $p'$ corresponds to $n (n \leq 2)$ original control points $p_i$. The new control point position is computed by $p' = \frac{\sum_i^n p_i}{n}$.

### 4.4.2 "Type-1" Merging

The goal of this merging type is to merge 3 cuboids into one control grid, like Fig. 4.2(a). We can still use the "Two-cube" merging technique to treat most merging regions. But we have to design special confinement method to handle the central points on the yellow dot/lines. Fig. 4.14(b) shows the extra bd-control-points we add around the central point on the yellow lines. For the yellow dot, we add additional 8 bd-control-points around it to confine it into the surface boundary, as shown in Fig. 4.14(c).

Fig. 4.15 illustrates the confinement effect Fig. 4.15(a) shows a confined 2D control grid in 2D layout. The extra bd-control-points (blue dots) are inserted around the central point. Fig. 4.15(b-d) showcase its advantage: unlike Fig. 4.5, for any chosen parametric position, none of its control points penetrates the boundary to influence the chosen position.

**Preserving semi-standardness.** Now we still have another challenge. Simply adding these extra control points would violate the semi-standardness property. To preserve semi-standardness, we also modify weights in this newly-merged control grid structure. The weight can be computed as follows (See

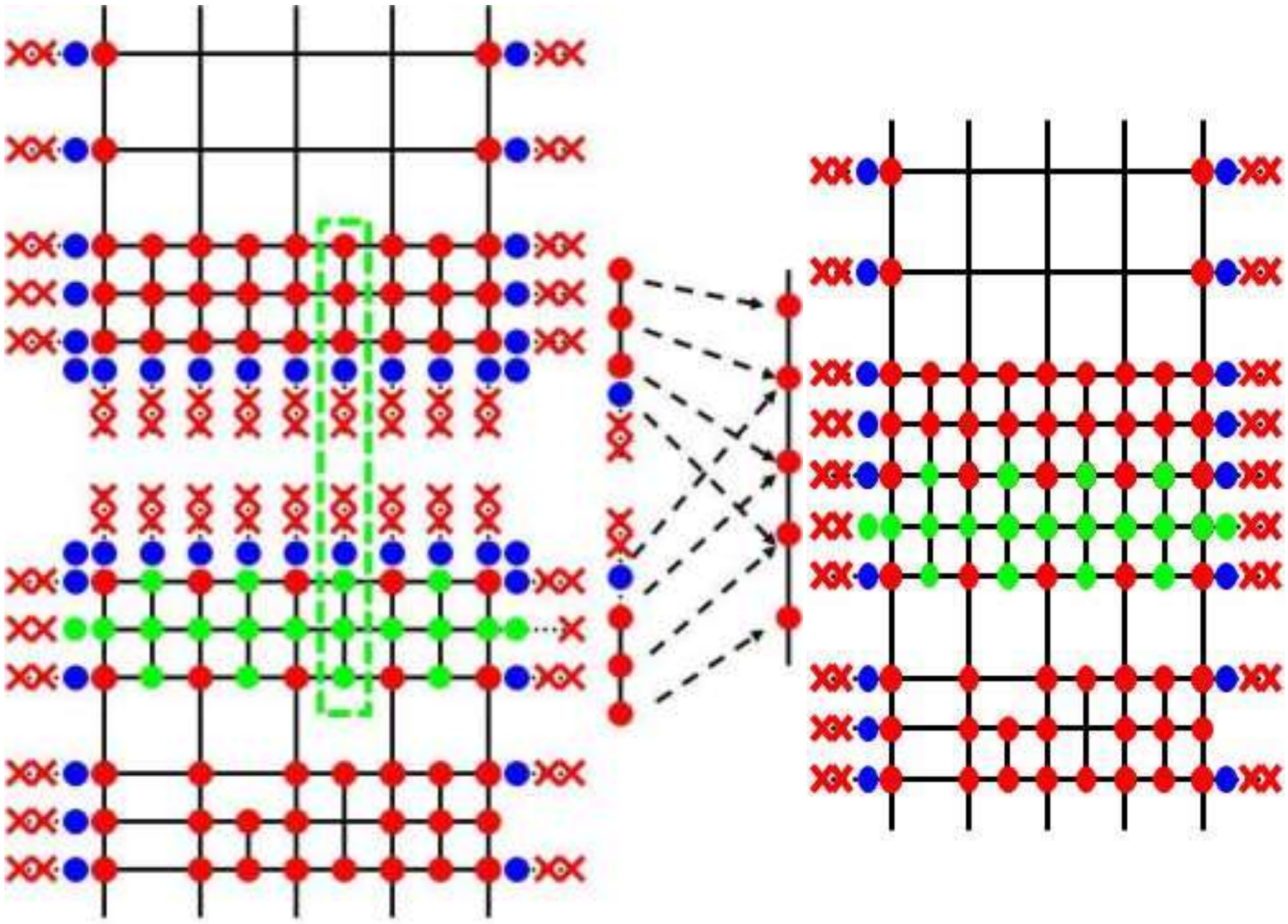

(a) (b)

Figure 4.13: "Two-cube" merging. (a) Left: Subdivide the bottom cuboid and insert new control points (green dots) to keep the same knot intervals. Right: Merging along merging direction. (b) The merged control grid.

Fig. 4.14(a)): (1) Before adding bd-knots around the central point, we add an auxiliary control point (green dot) at the corner. Now we locally have a standard rectangular control grid with weights all equal to one initially; (2) Insert the designed bd-knots (blue knots and red crosses in Fig. 4.14(a)) to the grid; (3) Inserting knots triggers the local refinement procedure to recompute the weight of each control points. Note that after refinement, the auxiliary point does not affect inside boundary anymore. Therefore, it is "transparent" and free to be deleted from the spline representation.

Besides preserving semi-standardness, our weight modification technique also has advantage for pre-computation. The weight computation only depends on the initial knot interval of merged control grid. Thus, we can pre-compute this step and build a look-up table for speedup. Table 4.2(a) shows the indices of control points around the central point (the same as indices in Fig. 4.6(b)). Table 4.2(b) shows the corresponding weights for all control points in Fig. 4.14(a) (Numbers in parentheses correspond to additional control points in Fig. 4.14(b)).

To summarize, "Type-1" merging includes the following steps. The first 3 steps are the same as "Two-cube" merging.

**Step 1** Modify boundary;

**Step 2** Merge control points;

**Step 3** Compute control point positions;

**Step 4** Insert extra bd-control-points as shown in Fig. 4.14(a-b) (We assign the position of the control point on the central point to these new inserted control points);

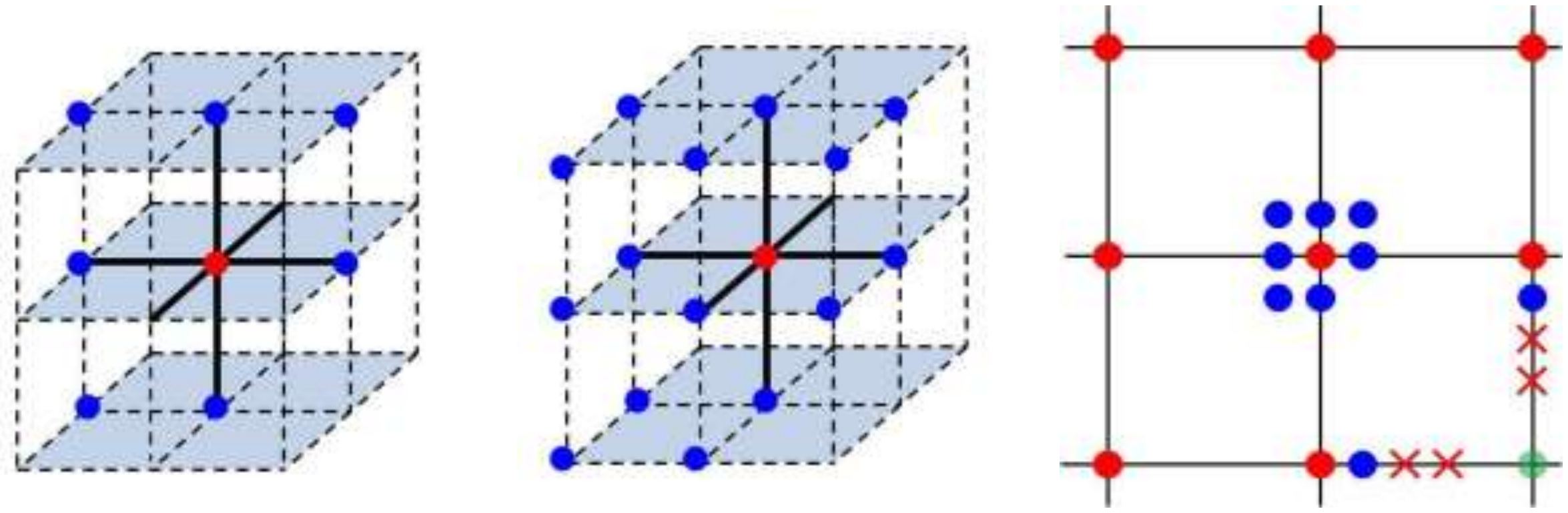

(a) (b) (c)

Figure 4.14: “Type-1” merging: (a-b) The 3-D distribution of bd-control-points around the central point on the yellow lines/dot in Fig. 4.2(a). (c) To preserve semi-standardness, bd-knots (blue dots and red crosses) and an auxiliary knot (green) are added. Then we can use local refinement algorithm to compute new control points’ weights.

**Step 5** Modify weight (Change the weight of these bd-control-points by checking the look-up table, as shown in Table 4.2(a)).

### 4.4.3 “Type-2,3,4” Merging

The above two merging algorithms (Sections 4.4.1, 4.4.2) are already functionally sound when handling the merging of all components in our divide-and-conquer framework, because these are the only two merging types in our T-shape based poly-cube. Not just limited to that, Our ambitious goal is to handle any shape of poly-cube domains. Therefore, we offer several more powerful merging operations, which are designed to merge the components like “Type-2,3,4” in Fig. 4.2(b-d). Once again, in order to enforce the boundary restriction, we need to insert extra bd-control-points. For the central points on all yellow line in Fig. 4.2(b-d), they are just “Type-1” merging, so we use the same merging method as as shown Fig. 4.14(a). For the central points on 3 yellow dots, we design the extra bd-control-points, as shown in Fig. 4.16, to preserve boundary restriction.

To guarantee semi-standardness, we recompute the weight using the same method in Section 4.4.2 as follows. First, we add auxiliary control points, expanding given control grid around the central point to a complete cube-like grid. Second, we insert the designed bd-control-points and perform local refinement to compute the new weight for each control point. Their look-up tables are shown in Table 4.2.

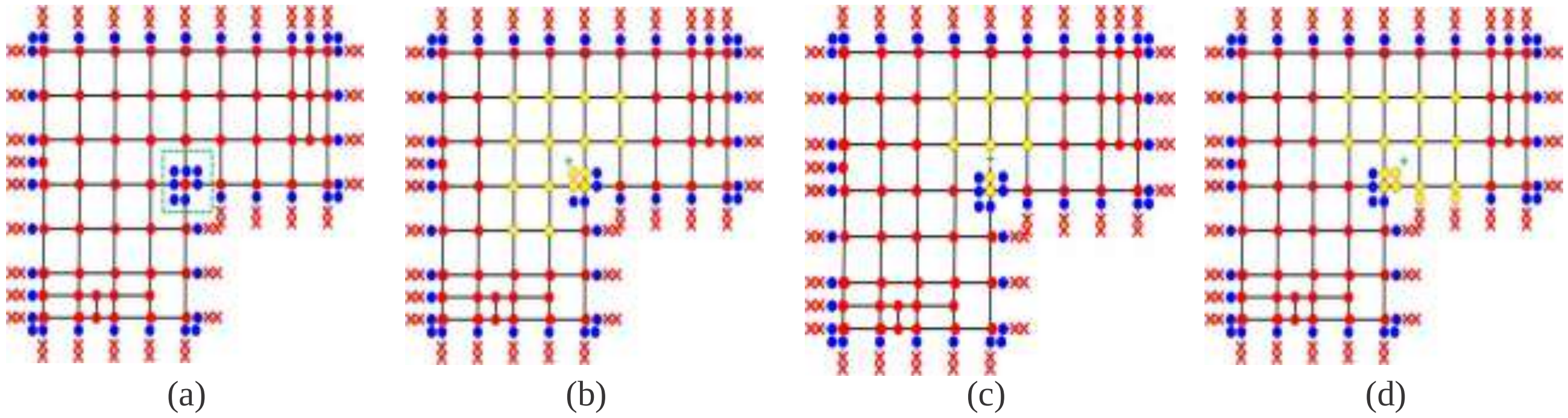

Figure 4.15: Confinement effect of "Type-1" merging. (a) The 2D layout of a refined control grid, with added bd-control-points (blue dots) around the central point (green box). (b-d) For each parameters (green cross), we highlight all control points (yellow points) that influence this parameter. The violation like Fig. 4.5 is completely eliminated.

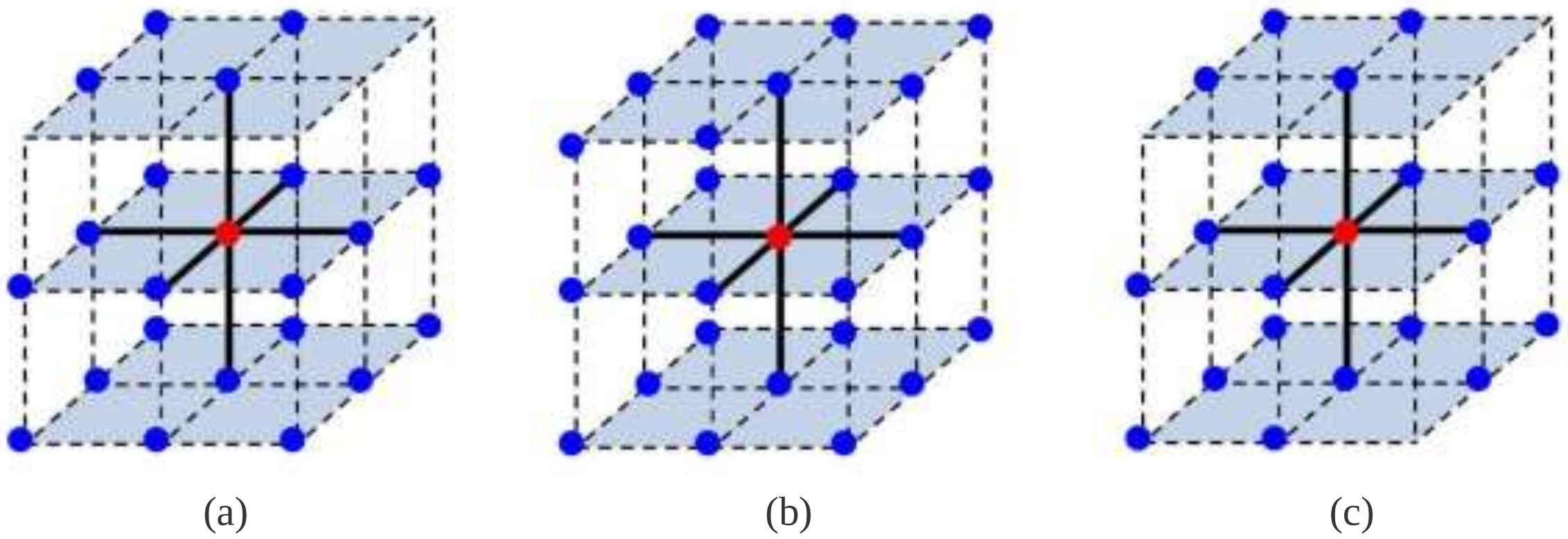

Figure 4.16: The 3D distribution of bd-control-points in "Type-2,3,4" merging. The central points are on the yellow dots in Fig. 4.2(b-d).

Table 4.2: Look-up tables. Row 1: an index table for 27 possible br-control-points in Fig. 4.6. Row 2: weights for "Type-1" merging in Fig. 4.15(b) (weights in parentheses correspond to additional 8 control points in Fig. 4.15(c)). Row (3-5): weights for "Type-2,3,4" merging.

Indices

| | | | | | | | | |
|---|---|---|---|---|---|---|---|---|
| 7 | 8 | 9 | 16 | 17 | 18 | 25 | 26 | 27 |
| 4 | 5 | 6 | 13 | 14 | 15 | 22 | 23 | 24 |
| 1 | 2 | 3 | 10 | 11 | 12 | 19 | 20 | 21 |

Type-1

| | | | | | | | | |
|---|---|---|---|---|---|---|---|---|
| - | - | - | - | - | - | - | - | - |
| 1 | 1 | - | $\frac{17}{18}$ | $\frac{35}{36}$ | 1 | $\frac{8}{9}$ | $\frac{17}{18}$ | 1 |
| (1) | (1) | - | ($\frac{17}{18}$) | ($\frac{35}{36}$) | (1) | ($\frac{8}{9}$) | ($\frac{17}{18}$) | (1) |

Type-2

| | | | | | | | | |
|---|---|---|---|---|---|---|---|---|
| $\frac{26}{27}$ | $\frac{53}{54}$ | 1 | $\frac{53}{54}$ | $\frac{107}{108}$ | 1 | 1 | 1 | - |
| $\frac{53}{54}$ | $\frac{107}{108}$ | 1 | $\frac{107}{108}$ | $\frac{209}{216}$ | $\frac{17}{18}$ | 1 | 1 | - |
| 1 | 1 | 1 | 1 | $\frac{17}{18}$ | $\frac{8}{9}$ | - | - | - |

Type-3

| | | | | | | | | |
|---|---|---|---|---|---|---|---|---|
| $\frac{20}{27}$ | $\frac{22}{27}$ | $\frac{8}{9}$ | $\frac{22}{27}$ | $\frac{95}{108}$ | $\frac{17}{18}$ | $\frac{8}{9}$ | $\frac{17}{18}$ | 1 |
| $\frac{22}{27}$ | $\frac{95}{108}$ | $\frac{17}{18}$ | $\frac{95}{108}$ | $\frac{25}{27}$ | $\frac{35}{36}$ | $\frac{17}{18}$ | $\frac{35}{36}$ | 1 |
| $\frac{8}{9}$ | $\frac{17}{18}$ | 1 | $\frac{17}{18}$ | $\frac{35}{36}$ | 1 | 1 | 1 | - |

Type-4

| | | | | | | | | |
|---|---|---|---|---|---|---|---|---|
| $\frac{26}{27}$ | $\frac{53}{54}$ | 1 | $\frac{53}{54}$ | $\frac{107}{108}$ | 1 | 1 | 1 | - |
| $\frac{53}{54}$ | $\frac{107}{108}$ | 1 | $\frac{107}{108}$ | $\frac{215}{216}$ | 1 | 1 | 1 | - |
| 1 | 1 | - | 1 | 1 | 1 | - | - | - |

## 4.5 Implementation Issues and Experimental Results

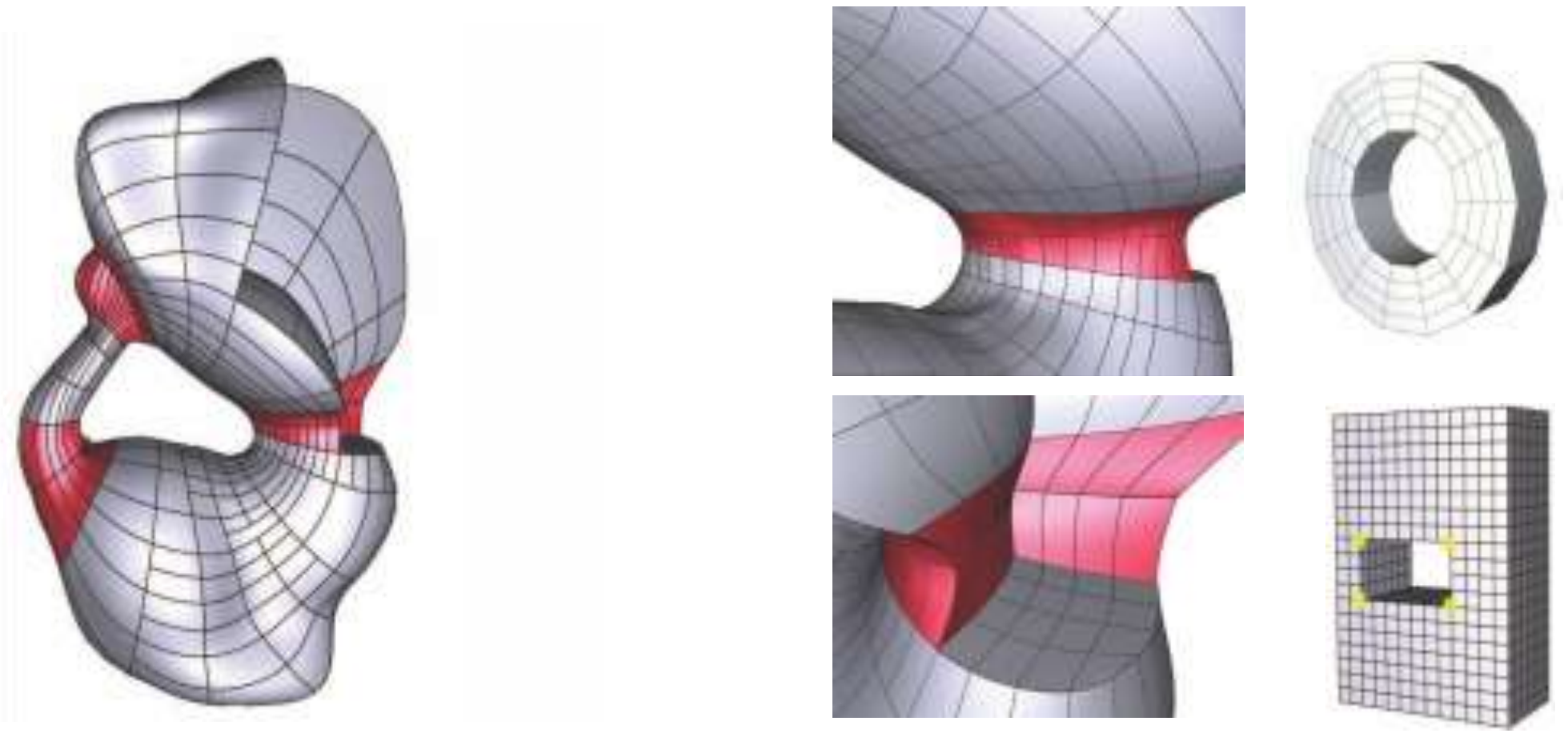

Figure 4.17: “Two-cube” merging for the kitten model.

Our experimental results are implemented on a 3 GHz Pentium-IV PC with 4 Giga RAM. Our first experimental results (Fig. 4.17, and Fig. 4.18) show the application of “Two-cube” merging by considering the kitten and beethoven model as the test datasets. These are the only merging types that exist in our component generation framework. For “Type-2,3,4” merging types that do not exist in our framework, we design a special screwdriver model and domain to demonstrate the power of “Type-2” merging (Fig. 4.19). In terms of poly-cube construction, we recognize that “Type-2” merging is very popular to handle the input with long branches. Yet, “Type-3,4” merging cases rarely exist even in the most conventional poly-cube domains. Geometrically speaking, they are more suitable to mimic highly concave shapes. We use the dark T-junction lines to show control grid knots and use different colors to represent different merging types. Red/Blue/Yellow marks all “to-be-merged” control point knots in 3 merging cases, respectively. We also have a close-up view to show the interior fitting result, demonstrating smoothness around the merging region. The yellow marks on the control grid highlight the ill-points.

In the second group of experimental results (Fig. 4.21, Fig 4.4, Fig. 4.22, and Fig. 4.23), we integrate all merging types together to handle the models with high-genus and complex bifurcations, including the eight (genus 2), g3 (genus 3), rockarm, and wrench (genus 1 with bifurcations) model. We first display their component generation results. Then we show a spline model for one local component and the final spline results with a close-up view to highlight the interior fitting and merging regions. Fig. 4.20 also visualizes components’ T-shape/poly-cube structures in a more efficient way. We use the same color cuboid to represent one component and the edges to show the cuboid connections. Each green box covers cuboids from the same T-shape. This structure clearly demonstrates that only “Two-cube” and “Type-1” merging are functionally sound in our framework.

In Table 4.3, we document numbers of control points and fitting error. The T-spline scheme can significantly reduce the number of control points. The fitting results are measured by RMS errors which are normalized to the dimension of corresponding solid models. Meanwhile, we demonstrate the interior fitting quality in a close-up view of each model. Also, the table illustrates that adaptive refinement is necessary for trivariate splines, even on a simple surface input model. It is desirable to use high resolution with more DOFs to approximate boundary surface and low resolution with fewer DOFs for volume interior. For exam-ple, in the kitten/beethoven model, if we naively use B-spline scheme with hierarchical refinement inside the volume, their control points will increase to 3718/4850, respectively. In the last experiment (Figure 4.24), we apply our technique to convert the fertility model, with the noisy surface, into a trivariate spline and remesh it into a smooth result. The poly-edges (gray-lines) decompose the fertility model into components. Note that poly-edges are aligned everywhere so our local parameters are consistent globally.

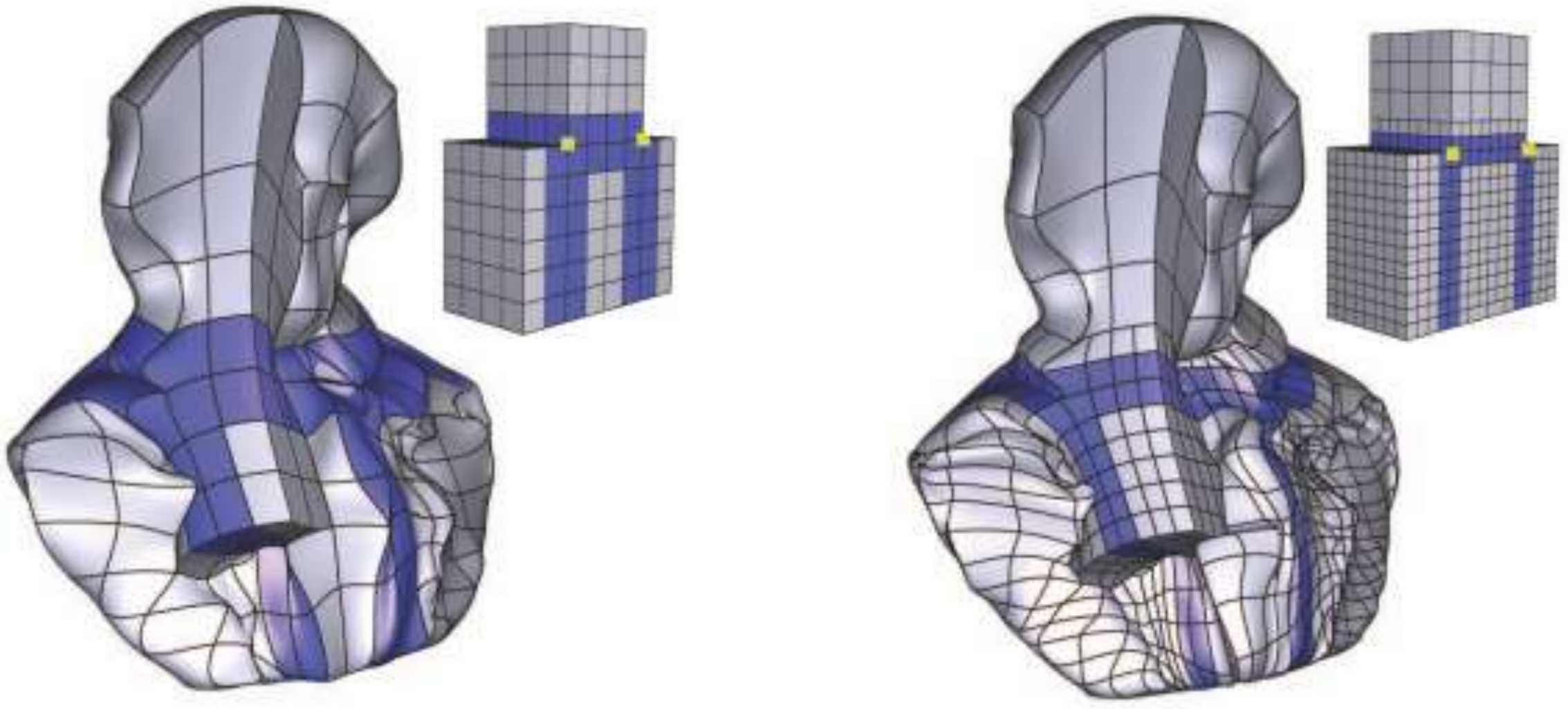

Figure 4.18: “Type-1” merging for the beethoven model.

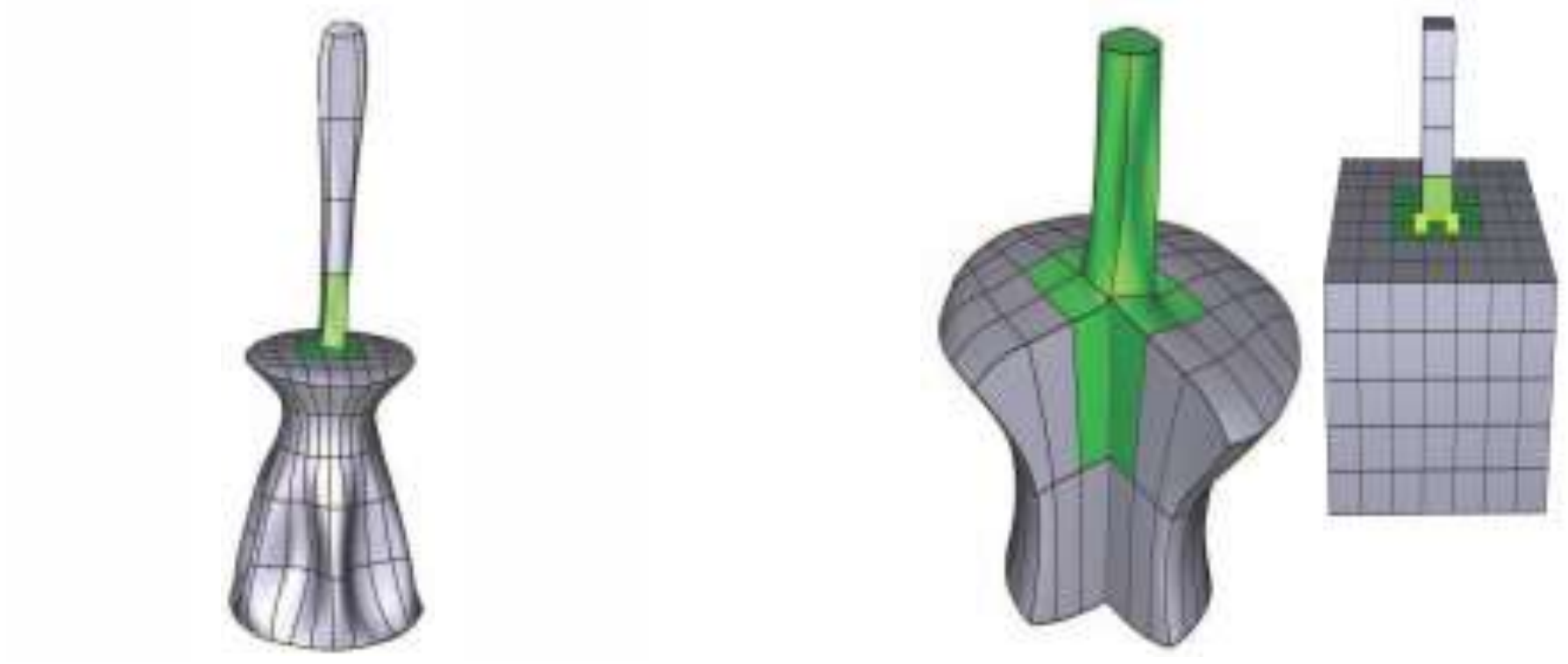

Figure 4.19: “Type-2” merging for the screwdriver model.

**Top-down vs. Divide-and-conquer schemes.** In Table 4.4, we compare the performance between our divide-and-conquer framework with general T-splines using single integral domain in a traditional top-down approach. The most prestigious advantage of divide-and-conquer framework is to easily handle models with bifurcations/highly twisted shape/high-genus. For example, a poly-cube like Fig. 4.1 designed using a top-down scheme is very complicated, with 46 cuboids and they are connected in various types, to mimic the shape of the g3 model. The poly-cube construction also requires tedious manual design. By comparison, its divide-and-conquer domain (Fig. 4.20) includes only 16 cuboids with two certain merging types. Second, we also compare the required spatial consumption between our divide-and-conquer scheme with the top-down scheme. In general, our memory cost is reduced to $\frac{1}{n_s}$, where $n_s$ is the number of cuboids. Third, we compare the computation of $B^0$ between semi-standard T-spline and rational T-spline. We record the computation time on $10^4$ samples for each model. The result shows that our method is at least twice as fast as rational T-splines. This is because the computation avoids division operation completely (See the difference between Eq. 4.1 and 4.3).

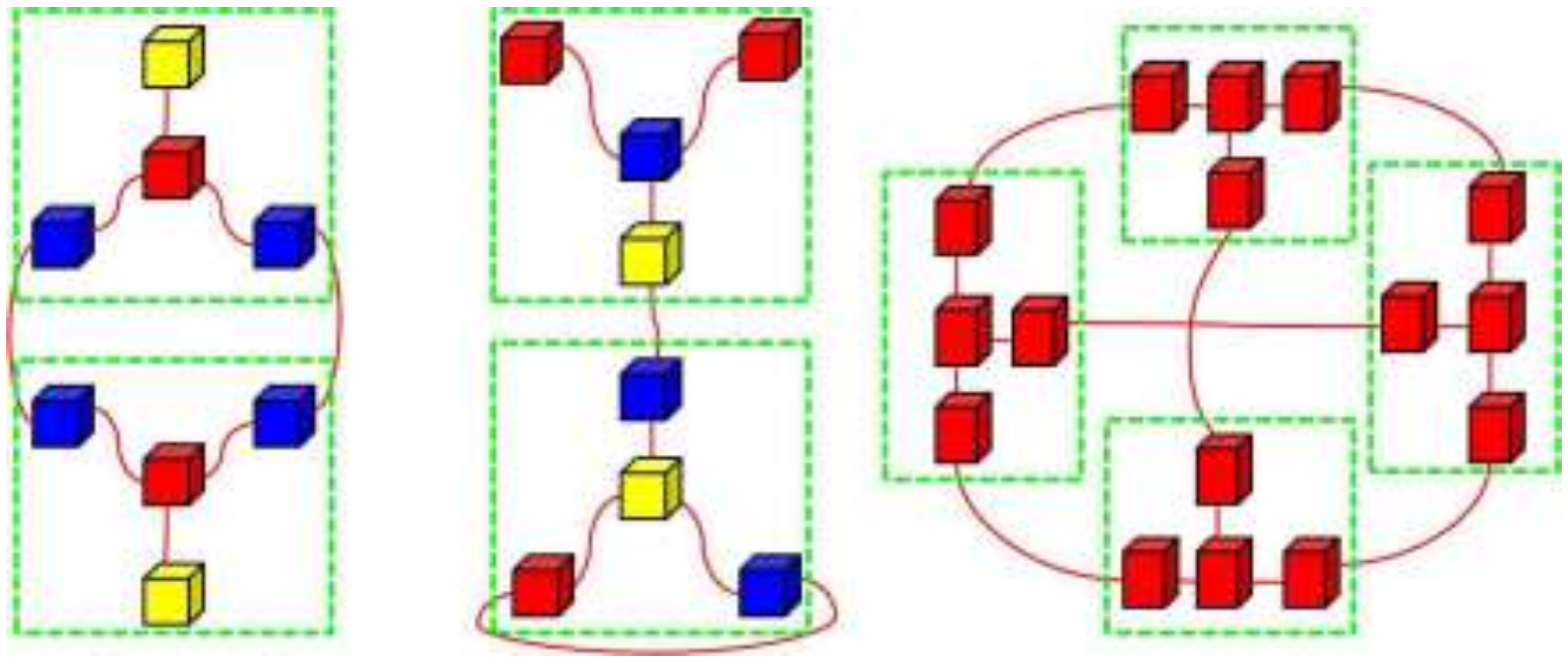

Figure 4.20: The divide-and-conquer structures of the rockarm/wrench/g3 model.

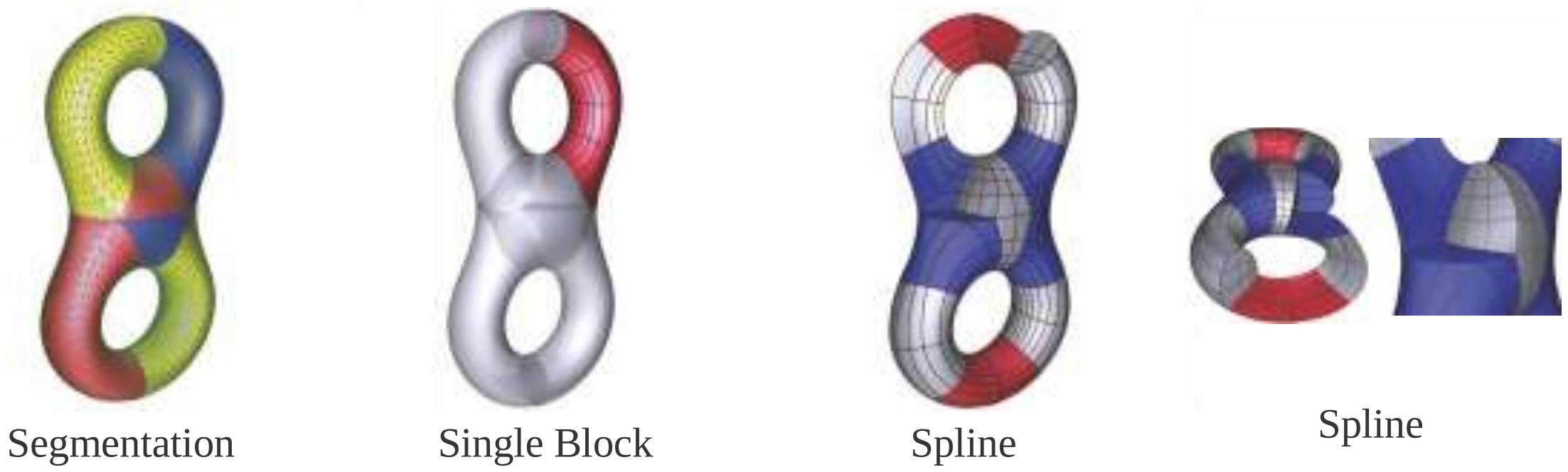

Figure 4.21: The eight model.

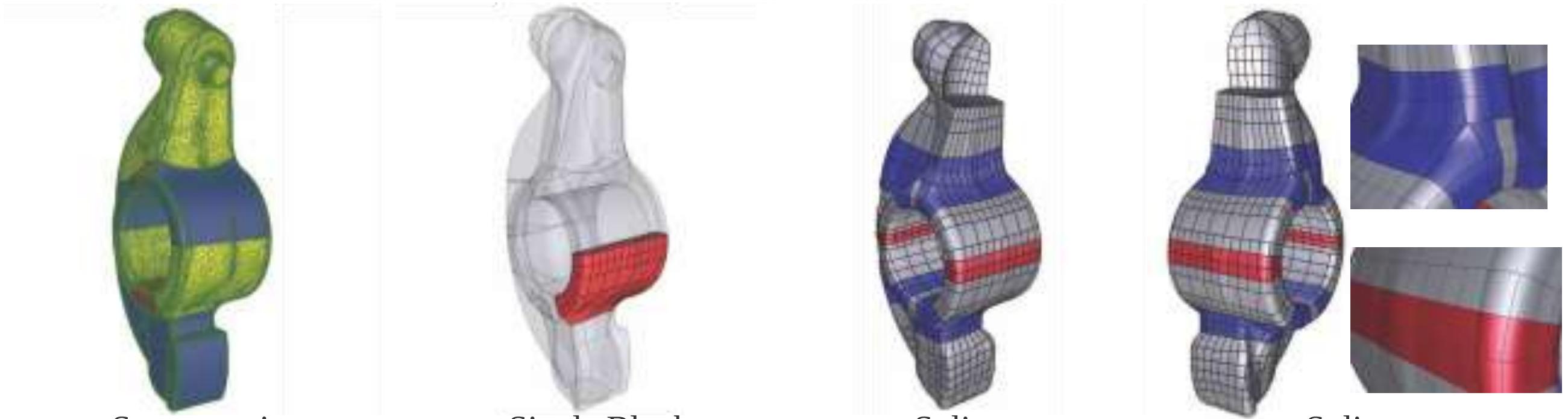

Segmentation Single Block Spline Spline

Figure 4.22: The rockarm model.

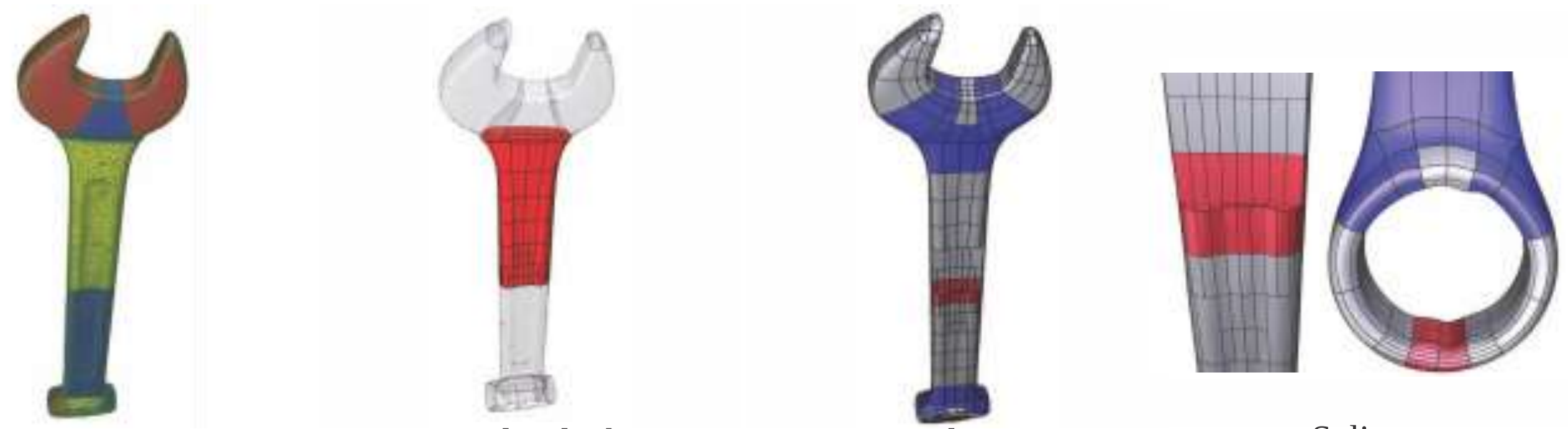

Segmentation Single Block Spline Spline

Figure 4.23: The wrench model.

## 4.6 Chapter Summary

Table 4.3: Statistics of various test examples: $N_c$, # of control points; RMS, root-mean-square fitting error ($10^{-3}$). “bv1”, “bv2”, “ra” and “sd” represent the beethoven (low and high resolution), rockarm, and screwdriver models.

| Model | $N_c$ | RMS | Model | $N_c$ | RMS |
|---|---|---|---|---|---|
| eight | 2058 | 1.63 | wrench | 3756 | 2.3 |
| kitten | 2840 | 3.32 | g3 | 2976 | 1.74 |
| bv1 | 1001 | 1.8 | bv2 | 3273 | 1.36 |
| ra | 4582 | 3.75 | sd | 1261 | 1.65 |

In this chapter, we have presented a novel framework to construct trivariate T-splines with arbitrary topology. Because of the flexible and versatile divide-and-conquer scheme, our framework can naturally handle solid objects with high genus and complex bifurcations. We decompose the input surface model into several part-aware components so that we can fit each component without the need of volumetric parameterization. The proposed spline scheme supports local refinement hierarchically, and the global trivariate T-splines satisfy the attractive properties of semi-standardness and boundary restriction. These novel contributions have a broad appeal to both theoreticians and engineers working in the shape modeling and its application areas.

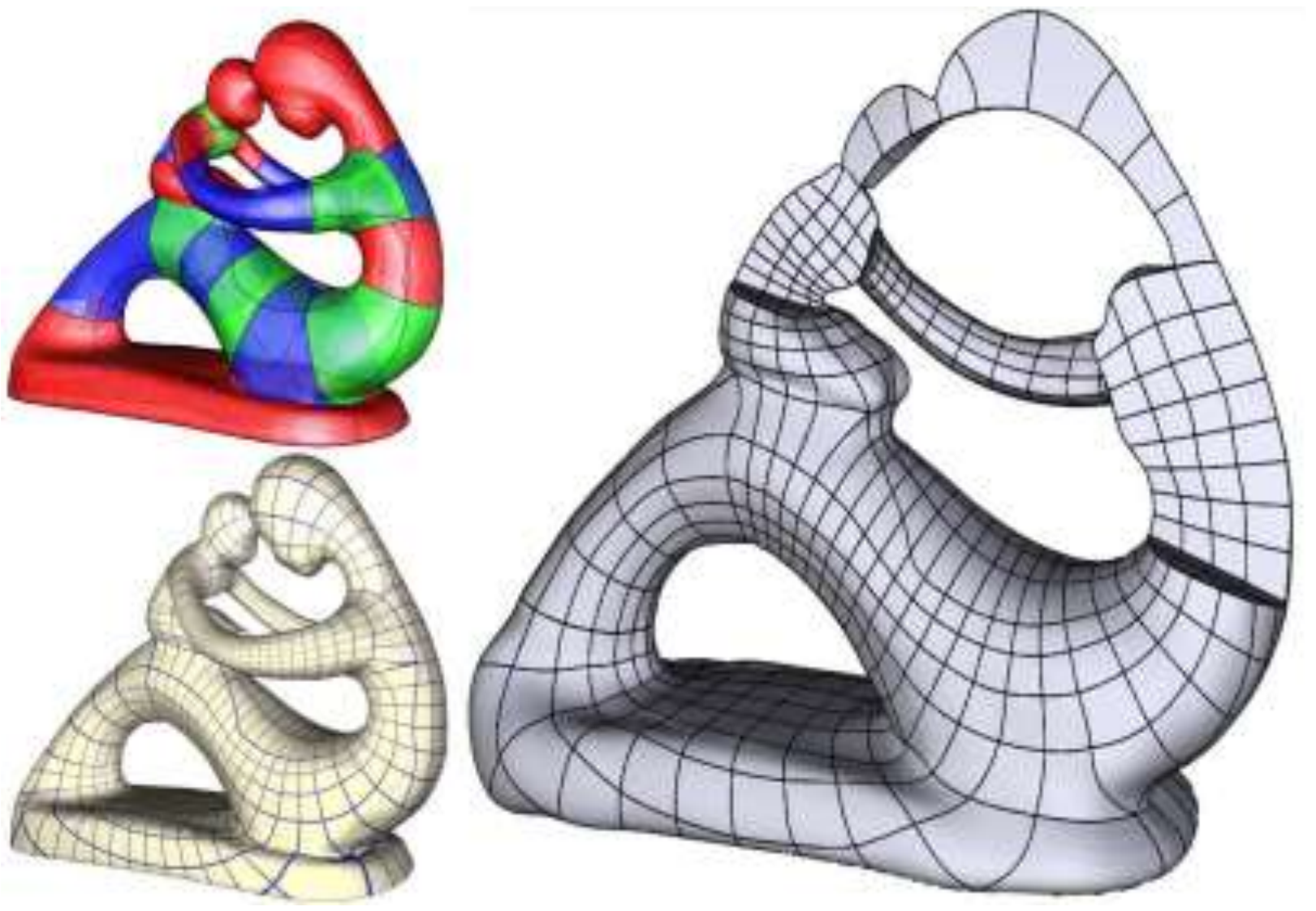

Figure 4.24: Mesh smoothing: We convert the fertility model to a trivariate spline and remesh it into a smooth result. Three figures show the components (with poly-edges), the globally aligned parameters, the remeshing result (with interior cut views), respectively.

Table 4.4: Comparison between our splines and general splines: Space required by fitting; Time to compute derivatives of basis functions; $N_c$, and Number of cuboids.

| Model | Our Method | | | General Method | | |
|---|---|---|---|---|---|---|
| | Space | $B'$ | $N_c$ | Space | $B'$ | $N_c$ |
| kitten | 116802 | 2.38s | 1 | 300688 | 4.53s | 8 |
| eight | 24714 | 2.25s | 6 | 174124 | 4.35s | 15 |
| g3 | 18952 | 2.17s | 16 | 314832 | 4.23s | 46 |

# Chapter 5

# Spline-based Volume Reconstruction

In Chapter 3 and 4, we have introduced the techniques to construct trivariate splines from surface 3D model. In this chapter, we adapt relevant trivariate splines to volumetric data reconstruction: we attempt to apply trivariate splines to represent 3D volume images.

## 5.1 Motivation

For volumetric scalar fields defined over a set of discrete samples, the reconstruction of the data is a fundamental problem with very significant applications. For instance in visualization, the size of volume data we have been dealing with increases dramatically to $1024^3$ voxels commonly or even larger. This trend of ever-increasing data size poses a great challenge in terms of both storage and rendering costs and thus requires reconstruction.

An ideal model would provide an accurate and efficient approximation for huge data sets, as well as the exact evaluation of function values and gradients which are required for high-quality visualization and physical simulation. An appropriate reconstruction involves following common quality requirements:

**Accuracy.** The reconstructed model should faithfully preserve the density function.

**Feature-alignment.** In regions with well-pronounced feature directions, parametric lines should guide and follow the shape feature.

**Compactness.** The number of patch layout as well as the degree of freedom for each patch should be as few as possible.

**Structured regularity.** Locally, each 3D patch is a subdivided cube-structured domain; Globally, the gluing between patches should avoid singularity.

**As-homogenous-as-possible.** The density distribution in one single patch should be narrowed in favor of approximation accuracy.

**Continuity.** A continuous representation supports high-order derivatives for high quality visualization.

An ideal reconstruction framework should optimize the output simultaneously with respect to all above criteria. However, existing techniques typically prefer offering a tradeoff between above conflicting requirements. The major reconstruction strategy is through multi-resolution data hierarchy to compress the data representation. Many algorithms have been developed to support hierarchical data reconstruction, including multi-dimensional trees [157] and octree-based hierarchies [7], [69]. However, these methods tend to produce an extreme large set of sub-blocks and require extra effort to pack them into a single structure, which undoubtedly violates the aforementioned compactness requirement. Moreover, the shape of pro-duced block is limited as the axis-aligned texture/cube (i.e.,"flat block"). In contrast, an ideal candidate for feature-driven applications should utilize feature-aligned texture/cube. Other reconstruction methods seek to generate a continuous spline representation to approximate the data. In general, spline based reconstruction can be divided into non-regular and regular splines. Rossl et al. [112] have developed quadratic super splines to reconstruct and visualize non-discrete models from discrete samples. Finkbeiner et al. [31] have demonstrated that box splines deployed on body-centered cubic lattices in the input data are also feasible models for fast evaluation and GPU-acceleration. Tan et al. [140] have presented a reconstruction algorithm for medical images taking advantages of trivariate simplex splines. Meanwhile, compared to non-regular splines, many types of techniques (e.g., volume rendering [88]) and applications (e.g., iso-geometric analysis [56]) have a preference for regular-structured schemes. However, the major challenge lies at they rely heavily on spatial parameterizations and for arbitrary 3D objects such parameterizations become a rather non-trivial task. The goal of vectorization is to convert a raster object (2D or 3D image) into a vector graphics that is compact, scalable, editable and easy to animate, which is very similar to our research goal. In object-based vectorization [106], the whole image is segmented into a few objects. The color of each object then is approximated by spline patch. Recently, gradient meshes ([139]) serve as very powerful tools on 2D image representations and have been studied in depth. In a gradient mesh, position and pixel vary according to the specified gradients. However, it is not easy to directly update it to 3D volumetric image application because of its inefficiency of handling complex topology.

In order to achieve all above requirements, we propose a novel reconstruction approach that converts the discrete data to a small number of volumetric patch layouts. Each patch is a regular tensor-product cube grid while maintaining shape features. The voxels in every single patch have the almost homogenous density values in favor of accurate approximation for each patch.

In this chapter, we provide a novel framework to help a user to reconstruct a discrete volume data into regular patches and spline representations. Our representation has significant advantages: Each patch has regular structure while maintaining the shape features. The whole data is compactly represented by a very small number of patches. The density in each patch is as-homogenous-as-possible thus both the shape and density function can be accurately approximated by a high-order spline representation.

In order to achieve these advantages, our approach consists of the following major steps:

1. Starting with the computing of local tensors and principal curvatures, we generate an optimized frame field to respect the shape feature.

2. A regular structured parametrization of $(u, v, w)$ is generated, whose gradients align the above field everywhere. Then we produce a set of volumetric patches based on the parametrization result.

3. We construct on each patch a trivariate T-spline to approximate the function $F(u, v, w)$ using as-few-as-possible control points.

The remainder of this chapter is organized as follows. Section 5.2 is the frame field generation stage and Section 5.3 involves the volumetric parametrization and patch remeshing. We discuss the spline approximation, implementation details, and demonstrate experimental results in Section 5.4. We conclude in Section 5.5.

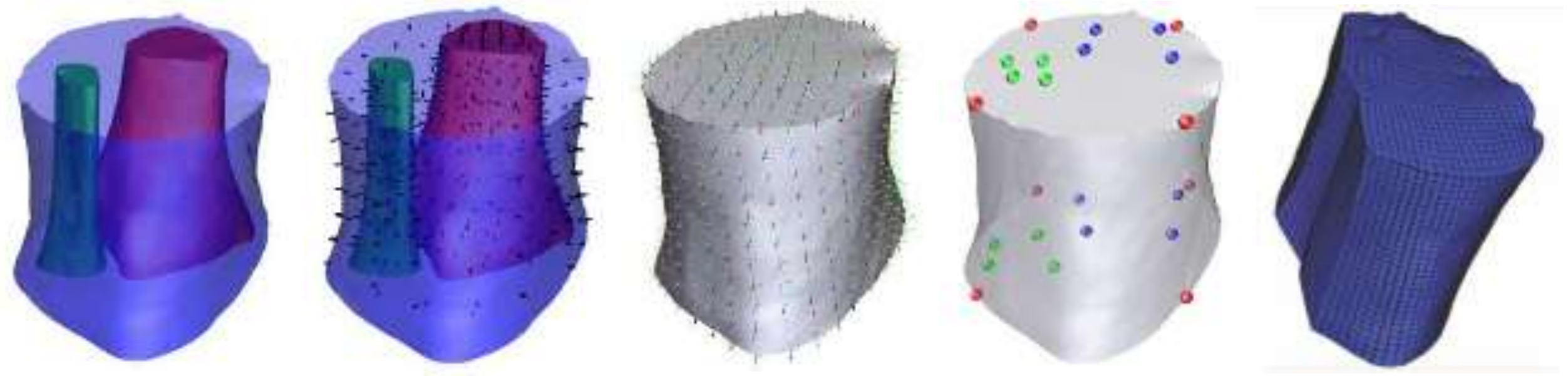

(a) (b) (c) (d) (e)

Figure 5.1: Main steps of the reconstruction. (a) Input model with material-aware boundary surfaces. (b) The tensor and principal direction field is computed on each voxel in the input data. The major principal directions on the boundary surfaces serve as the constraints for the next step. (c) In a frame field optimization procedure, an as-smooth-as-possible frame field is generated while maintaining the given constraints. (d) Corner points are selected to determine the domain structure. Additional constraints are added into next step of parametrization computing. (e) A volumetric parametrization.

## 5.2 Frame Field

In this section we mainly focus on frame field generation. We start from designing an operator of tensor to describe the local feature ( Section 5.2.1). Then we discuss the optimization of a 3-direction frame field in Section 5.2.2.

### 5.2.1 Tensor and Principal Curvature

Traditionally, curvature has be used as a shape feature descriptor widely. Theoretically this differential property characterizes only an infinitesimal neighborhood. Therefore it is desirable to design a numerically adaptable operator for the discrete data to compute this property. Although much work deals with this task on the surface (see [37] for an overview), we still need a new curvature operator for the discrete 3D hyper-volume data. Our operator captures statistically the shape of a neighborhood around a central point by fitting a continuous function, and thus mimics the 3D differential curvature and encodes anisotropy along 3 orthogonal directions. To summarize this shape, we use a cubic polynomial function $\mathbf{I}^H(u, v, w)$ to approximate the local density function, because they are the simplest form that can sufficiently express the shape variability we need to encode in a continuous manner.

Specifically, the given volumetric data set is represented using a uniform grid $G = (V, E, C)$, where $V = v_0, v_1, \ldots, v_n$ denotes the voxels and $E$, $C$ denote the set of edges and cubes in the grid, respectively. Each grid voxel $\mathbf{v}_i = (x_i, y_i, z_i, \mathbf{I}_i^D)$ includes 4 components: geometric position in the grid $(x_i, y_i, z_i)$ and the discrete density value $\mathbf{I}_i^D$.

In order to get a local polynomial function $\mathbf{I}^H(u, v, w)$ around center voxel $\mathbf{v}_i$, we assign a local parameter value $(u_0, v_0, w_0)$ to $\mathbf{v}_i$. For each of its adjacent k-ring neighbor voxels $\mathbf{v}_j \in N(\mathbf{v}_i)$, the local parameter is $(u_j, v_j, w_j) = (u_0 + x_j - x_i, v_0 + y_j - y_i, w_0 + z_j - z_i)$. Then our fitting cubic polynomial
can be formulated as:

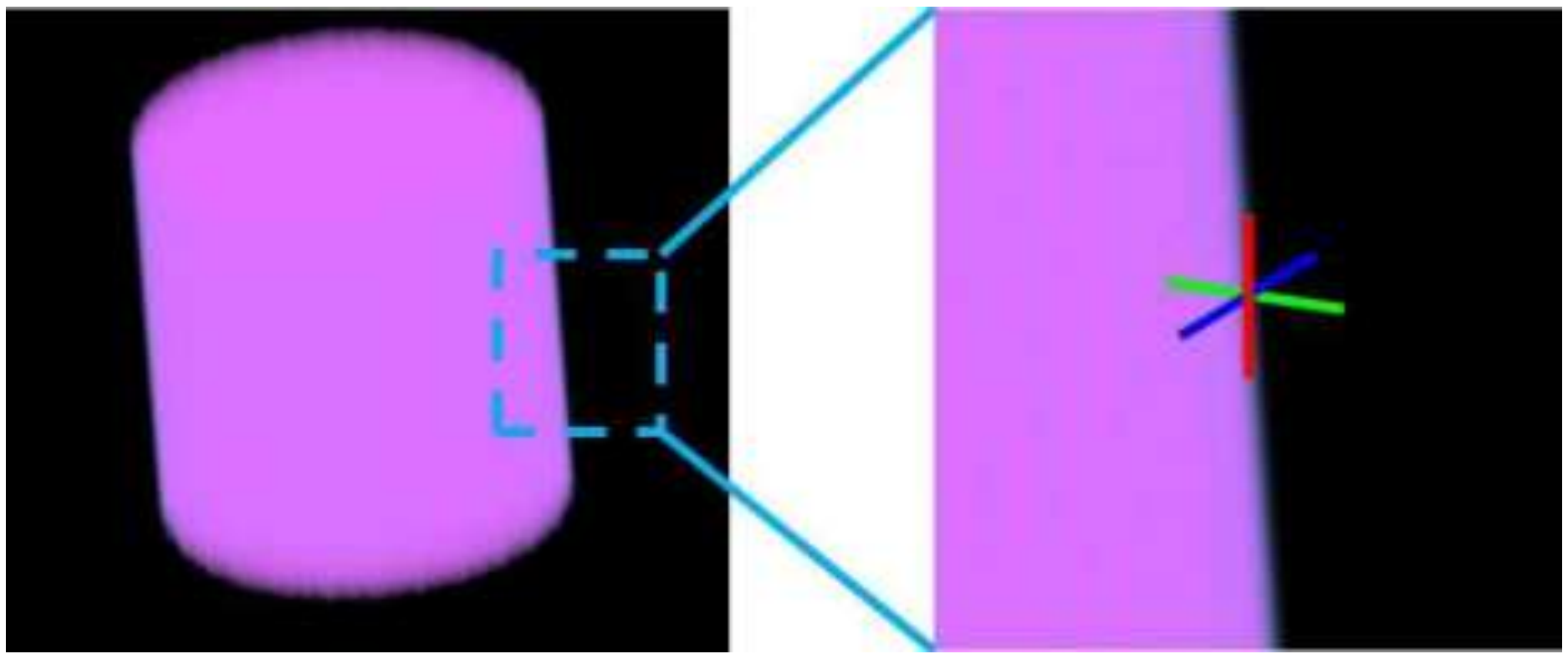

Figure 5.2: Left: The input local data around a voxel. Right: The approximated result and 3 principal directions.

$$I^H(u,v,w) = \sum_{i,j,k\geq 0}^{i+j+k\leq 3} c_m u^i v^j w^k = P(u,v,w)C^T, \tag{5.1}$$

where $C$ denotes the vector of unknown coefficients $c_m$. $\mathbf{P}$ is the vector of $u^i v^j w^k$. Similarly, we can also describe derivatives of $u, v, w$. For instance,

$$I_u^H(u,v,w) = \sum_{i,j,k\geq 0}^{i+j+k\leq 3} c_m \times i \times u^{i-1} v^j w^k = \mathbf{P}_u(u,v,w)C^T, \tag{5.2}$$

where $\mathbf{P}_u$ is the vector of $i \times u^{i-1} v^j w^k$ (we set $u^m = 0$ if $m < 0$). In the same way, we can also describe other derivatives $I_v^H$, $I_w^H$, $I_{uu}^H$, $I_{vv}^H$, $I_{ww}^H$, $I_{uv}^H$, $I_{uw}^H$, $I_{vw}^H$ by determining $\mathbf{P}_v$, $\mathbf{P}_w$, $\mathbf{P}_{uu}$, $\mathbf{P}_{vv}$, $\mathbf{P}_{ww}$, $\mathbf{P}_{uv}$, $\mathbf{P}_{uw}$, $\mathbf{P}_{vw}$.

In order to describe the currently unknown coefficients $C$, we construct a fitting equation:

$$QC^T = I^D, \tag{5.3}$$

where $Q$ is the fitting matrix. Each row $Q_{j:}$ in the matrix depends on a voxel $Q_{j:} = P(u_j, v_j, w_j)$, $j \in i \bigcup N(i)$. $I^D$ is the vector of discrete value $I_j^D$ on each voxel. Because the size of unknown variables is very small, we can solve this linear least-square problem through multiplying the matrix $Q$ by its transpose:

$$C = (Q^T Q)^{-1} Q^T I^D. \tag{5.4}$$

We notice that $(Q^T Q)^{-1} Q^T$ is constant for every local function if we choose the same $k$ for k-ring neighbors of each voxel.

**Tensor and Principal Curvature.** After the above calculations, we now can represent the tensor as the following matrix:

$$\mathbf{T} = \begin{bmatrix} I_{uu}^H & I_{uv}^H & I_{uw}^H \\ I_{uv}^H & I_{vv}^H & I_{vw}^H \\ I_{uw}^H & I_{vw}^H & I_{ww}^H \end{bmatrix}. \tag{5.5}$$

This matrix is equal to the second fundamental form of our hyper-volume representation. Therefore, we can compute 3 eigenvectors of the local tensor matrix $\mathbf{T}$ and thus get 3 directions. We use them to describe the feature on each voxel. Compared to the conventional texture-gradient based feature, our tensor feature has very obvious advantages: it produces 3 directions rather than one; all local 3 direction fields follow the shape anisotropy thus global fields are already almost smooth. As a result it simplifies the complexity and time consumption of the following optimization step.

### 5.2.2 Field Smoothing

Although we can use initial principal directions to compute the parametrization without optimization, it will stuck in a local minimum. To overcome this problem, we propose an optimization method which respects only the most dominant directions. First, we extract iso-surfaces of interest and take them as constraints to respect the shape. Second, the frame field in each iso-surfaces is iteratively optimized.

**Iso-surface extraction.** It is natural to take feature on iso-surfaces as constraints, because the final parametrization result and patch must respect the shape of iso-surfaces. Moreover, each sub-space in an iso-surface always tends to be as-homogenous-as-possible, which is an ideal property for final shape and density approximation.

Frequently, input datasets contain multiple structures and iso-surfaces that need to be differentiated. However, if those features have the same density and gradient values, existing clustering methods are limited at effectively classifying those similar features accurately. Thus, we apply the texture-based classification method for the iso-surface extraction. In the first step, we simply remove the background voxels. It does not influence the information of the feature of interest while significantly decreasing the computational time and operation complexity. After the background elimination, sixteen statistical attributes (angular second moment, contrast, correlation, variance, inverse difference moment, individual entropy, sum average, sum variance, sum entropy, skewness, kurtosis, correlation information measurements, intensity, gradient and second order derivative) can be extracted following the feature equations defined in [12] and [46]. For the sake of fast computation and easy programming, we use k-mean clustering in the high- dimension parameter space to automatically detect various features. One or more features can be selected with respect to the user's requirement. The boundary of each cluster finally becomes one of our iso-surfaces.

The constraints are added towards voxels on iso-surfaces, automatically or manually. In practice, to efficiently describe the feature of iso-surfaces, we set only one of 3 principal directions as the constraint, one of which follows the normal direction of the iso-surface. As shown in the following sections, only-one-direction constraints are functionally sound to preserve the feature and have extra flexibility when handling smoothing and parametrization.

After this preprocessing step, the input is decomposed to an independent subspace $V_i$ bounded by an iso-surface $S_i$. The subspace may also cover several smaller subspaces with iso-surfaces

$$S^{sub} = \{S_0^{sub}, S_1^{sub}, \ldots, S_n^{sub}\}.$$

Each voxel in the subspace $V_i$ has 3 initial directions and each voxel on the iso-surfaces $S_i \bigcup S^{sub}$ has one direction as the constraint. The following smoothing step will modify the directions on each voxel while maintaining the constrained directions.

**Field smoothing.** The smoothness of a unit frame field can be measured as the integrated rotation differences between every two neighboring voxels. [110] have studied the energy of a 2D cross field and simplified it to a linear representation. In our 3D volume, the challenge lies at smoothing 3 vectors in separate directions while maintaining their orthogonality. Therefore, we take the local rotation matrix as the unknown variable. $F(v_i) = f_0, f_1, f_2$ is a frame with 3 orthogonal vector directions on each voxel $v_i$. We can also uniquely describe this frame by rotating from the origin fame to it. Each row of the rotation matrix $R(v_i)$ is a vector direction $R_{r:} = f_r, r = 0, 1, 2$. Now, the energy turns out to be the sum of all

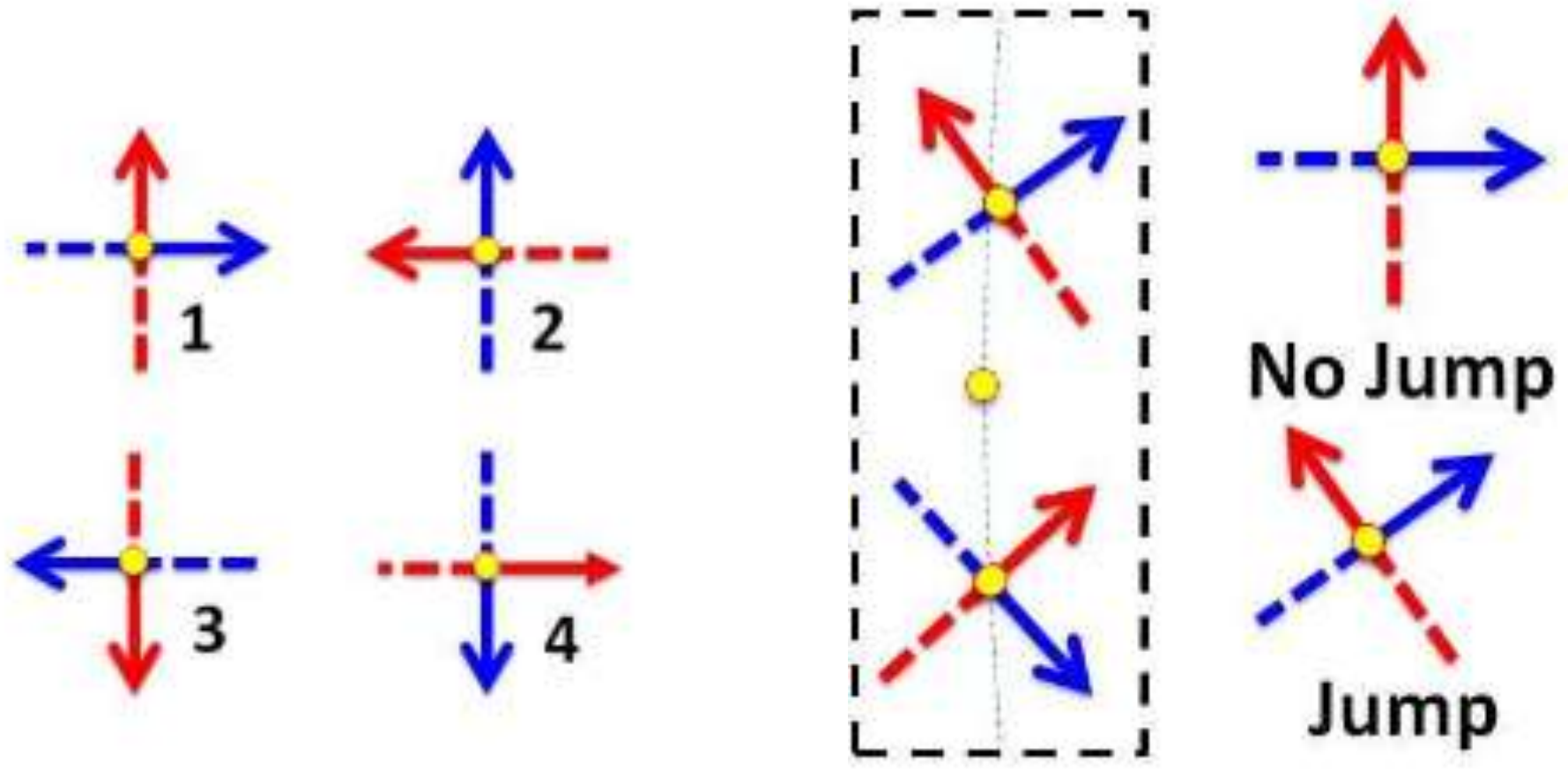

(a) (b)

Figure 5.3: (a) Jump matching: 4 frames has different principal directions along red and blue arrows. The smooth energy between them should be zero ideally. (b): The smoothing results with/without considering period jump.

corresponding vector differences between adjacent voxels

$$E_{smooth} = \sum_{e_{ij} \in E} \sum_{r=0}^{2} ||R_{r:}(v_i) - R_{r:}(v_j)||^2. \tag{5.6}$$

In order to solve unknown rotation matrix, we have to apply nonlinear solver (e.g.,Gaussian-Newton method) to minimize the energy function. Another difficulty is that Equation 5.6 predetermines the one-to-one mapping of 3 directions on two voxels, without considering "jump matching". "Jump matching" means all permutation cases of direction mapping. Fig. 5.3(a) shows all 4 "jump matching" cases for a 2-direction field. Similarly, we can have 24 "jump matching" cases for a 3-direction field. An ideal optimization algorithm should dynamically change direction mapping to get the best result. Fig. 5.3(b) shows a simple frame optimization on one voxel according to two adjacent voxels. Using jump matching we can get the perfect optimization result, while traditional method fails.

To overcome these problems, we design a novel optimization method. The key idea is that we compute the registration energy [4] between one voxel and its neighboring voxels. We extend 3 orthogonal principal directions into a length-normalized frame. Each frame gives 6 end positions $\{P(v_i)\} = \{p_0, \ldots, p_5\}$ at the end of 3 frame lines.

1. Get the union of all frame end positions on neighboring voxels: $\{S_2\} = \bigcup_{v_j \in N(v_i)} P(v_j)$.
2. The original point set $\{S_1\} = \{P(v_i)\}$ is the frame ending positions of $v_i$. Using the ICP-based registration [4], we compute a matrix T that approximately transforms voxels of $\{S_1\}$ to those of the approximated set $\{S_2\}$.
3. Decompose the transformation matrix T into a rotation matrix R and a shear matrix S using polar decomposition. Add the rotation R to the frame of $v_i$.

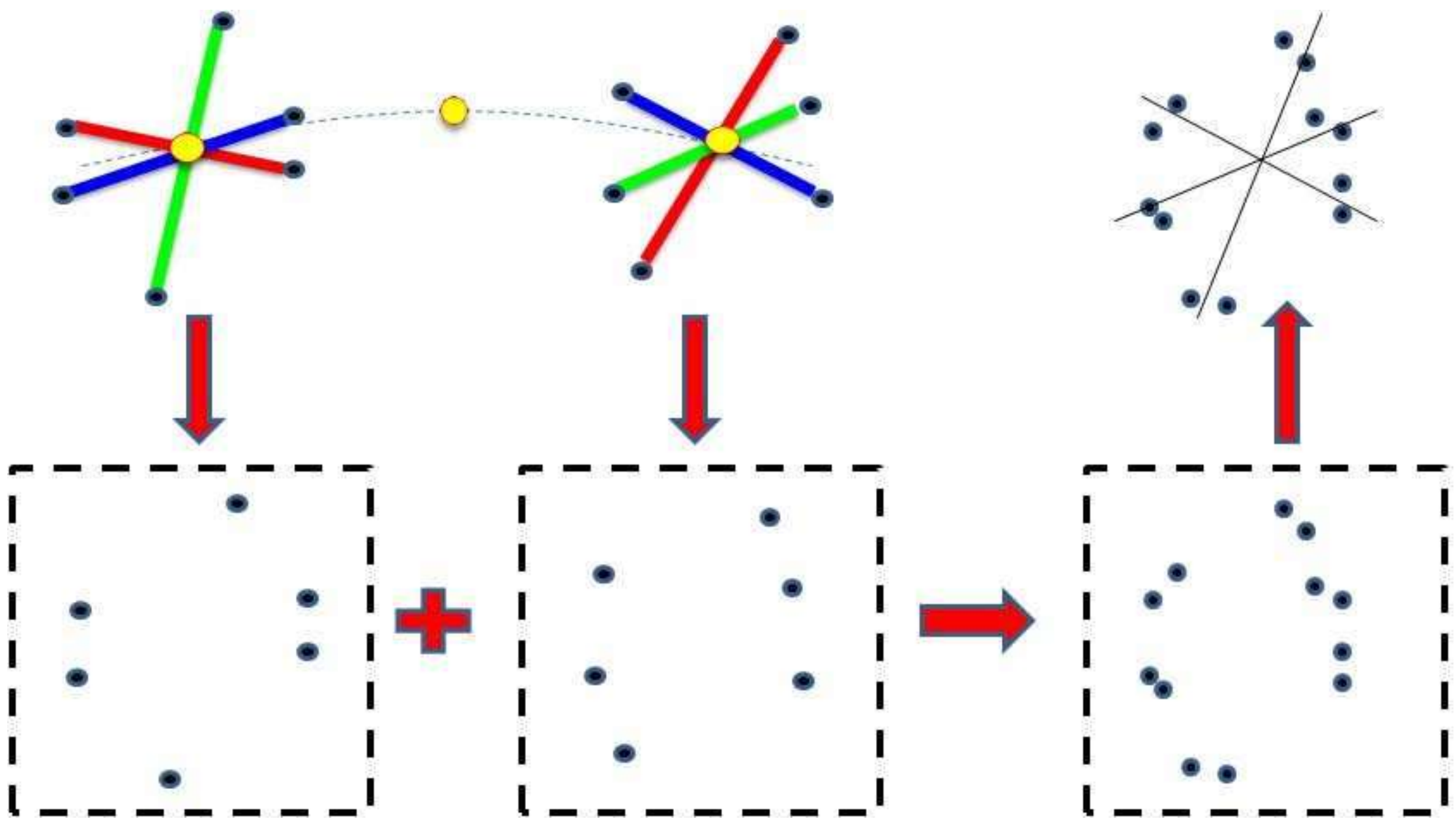

Figure 5.4: Major steps of optimization: (1) Union of ending points. (2) ICP-registration. (3) Compute rotation to get updated frame.

For an iso-surface voxel $v_i$ which has a constrained direction, we first apply the above algorithm without considering constraint. Then, in the updated frame, we search for the closest direction and project it to the constrained direction by a rotation. The whole frame is also rotated as the final updated result.

The above algorithm is computed on each voxel iteratively until we get a promisingly smooth field. Starting from initially smooth tensor field will make optimization converge quickly. Our optimization algorithm avoids solving non-linear equations; Moreover, we utilize jump matching to get a much better result.

## 5.3 Volumetric Parametrization

The parametrization should be locally oriented to the frame field from Section 5.2. Therefore, the parametrization is computed as a solution to the following energy minimization problem:

$$E_{param} = \sum_{v_i \in V} ||\nabla u_i - \mathbf{u}_i||^2 + ||\nabla v_i - \mathbf{v}_i||^2 + ||\nabla w_i - \mathbf{w}_i||^2, \tag{5.7}$$

where $u_i$, $v_i$, $w_i$ are the unknown parameters and $\mathbf{u}_i$, $\mathbf{v}_i$ and $\mathbf{w}_i$ are 3 frame field directions on each voxel. In practice, in order to respect the iso-surface and edge features, as well as preserving regularity in the final parametrization result, our parametrization algorithm has following steps:

1. Corner detection and selection: Determine all corner candidates from the frame field. Interactively select corner points from the candidates, serving as corners of the final parameter domain. These corner points directly determine the structure of the final parameter domain.

2. Energy minimization with constraints: Add parameter constraints on corner points and other points

if necessary. Add these parameter constraints into the energy minimization equation. Compute the minimization again to get the final parametrization result.

3. Remeshing: Guided by the generated parameter, trace and generate a small set of volumetric patches.

### 5.3.1 Corner Points

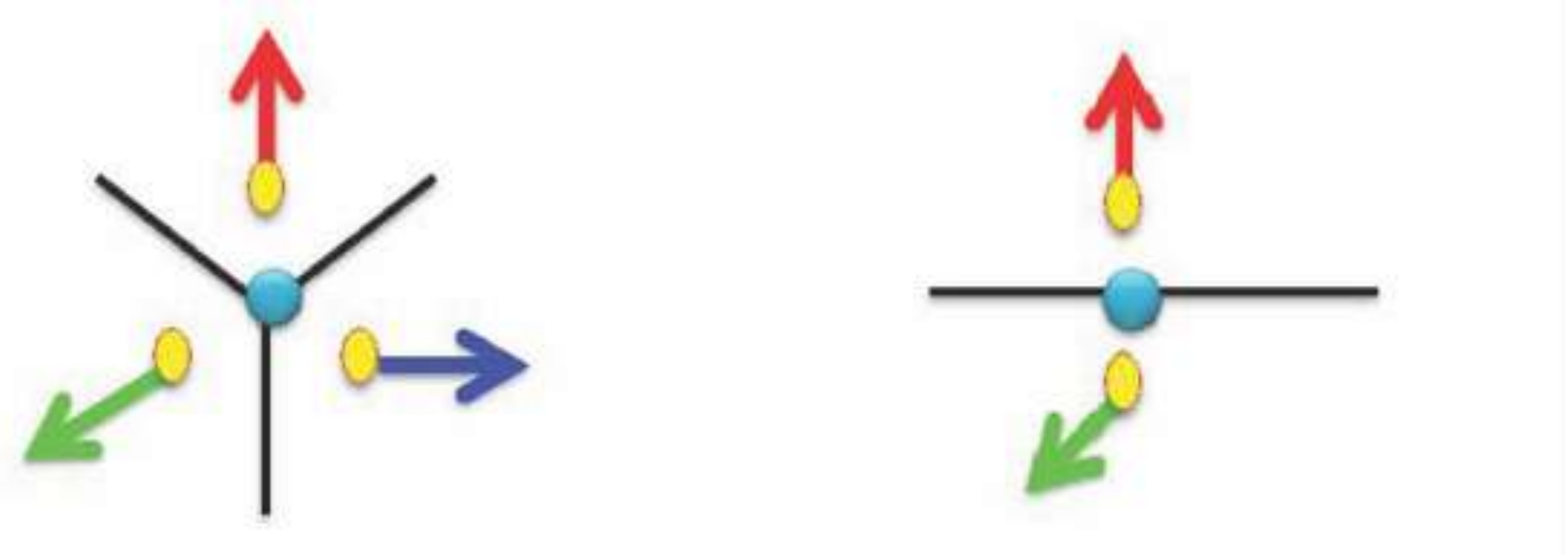

Figure 5.5: Corner point and edge point. Each vector is a constrained direction following the gradient of different scalar field u, v, w (red, green, blue) separately.

Intuitively, in a parameter domain as shown in Fig. 5.5, a corner candidate $v_i^c$ is the intersection point of 3 iso-parametric surfaces on u, v, w respectively. Consequently, some of its neighboring voxels should separately distribute on 3 iso-parametric surfaces and their normal vector follows 3 different parameter gradients. In practice, we define **Corner point** as the voxel that has neighboring voxels with constrained normal directions along 3 different gradients $\nabla u$, $\nabla v$ and $\nabla w$ separately. Similarly, we define **Edge point** in a similar way, but its neighboring voxels' constrained normal directions only follows 2 different gradients. For example in Fig. 5.6(a) 2D layout, 6 nodes are detected as the corners according to our definition.

From these corner candidates, we interactively choose several corners as the final corner points. These corners will be mapped to the corners of the parameter domain. Consequently, the edge points connecting a pair of corners will be mapped to the iso-parametric lines on the parameter domain. These edge points also partition the boundary surface into several patches. Intuitively, each patch should be mapped to an iso-parametric surface on the parameter domain.

**Frame field recomputing.** We notice that the original frame field tend to produce unnecessary singularities (Fig. 5.6(a)), making the parameter result and patch structure complicated [92]. To eliminate this problem, we can re-compute the frame field with new constraints: The normal direction of a voxel on an iso-parametric patch must be aligned to the parametric normal direction. As shown in Fig. 5.6(b), the normal directions on the left and right boundaries are forced to be aligned to the green direction, leading to a singularity-free frame field.

**Additional constraints.** Parameter constraints must be added into the energy minimization to make sure that any corner point we select will locate on a corner of the parameter domain. Thus, we associate each corner with a known parameter before solving the energy equation. However, this constraint may cause serious distortion in the solved parametrization. Therefore, we need to add more constraints to get a better parametrization. We observe that the distortion always happens on the geometric-complicated boundary surface patch which maps to an iso-parametric surface in the domain. Therefore, we can avoid this distortion by adding the additional constraints on the boundary surface patch if necessary.

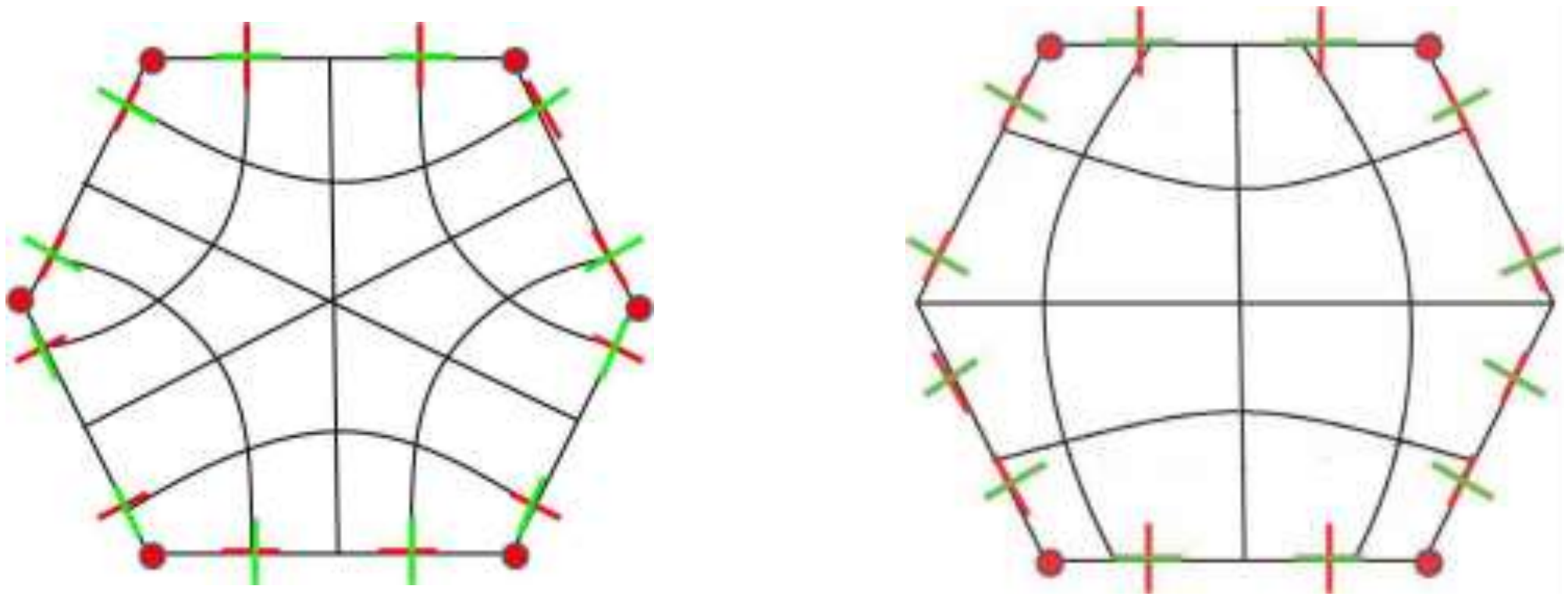

(a) (b)

Figure 5.6: (a) 2D layout of a frame field. It has 6 corners (red nodes) and one singularity. (b) Recomputed frame field. 4 nodes are selected as corners. The jump-match of the frame on the boundary is limited.

### 5.3.2 Energy Minimization

In order to minimize Equation 5.7, we have to design a linear formulation of the gradient operator $\nabla$ for any scalar field (i.e., U, V or W) on each voxel $v_i$. We notice that the gradient computing is invariant to the choice of parameter. Therefore, we again use the density function (Equation 5.1) and its derivatives to numerically describe the gradient operator $\nabla$. Equation 5.2 and 5.4 together describe the gradient operator on a voxel. For instance, we represent $\nabla u_i$ as :

$$\nabla u_i = (P_u C,\ P_v C,\ P_w C) = (P_u, P_v, P_w)(Q^T Q)^{-1} Q^T U^D, \tag{5.8}$$

where $U^D$ represents the vector of unknown scalar value u on $v_i$ and its neighboring voxels. Then, we substitute them into the energy equation, for example:

$$\sum_{v_i \in V} ||\nabla u_i - u||^2 = \sum_{v_i \in V} ||(P_u, P_v, P_w)(Q^T Q)^{-1} Q^T U^D - u_i||^2. \tag{5.9}$$

Equation 5.9 is a typical fitting problem, which can be converted into a linear system $AU^T = B$ through computing $\frac{\partial E}{\partial u} = 0$, where $U^T$ is the vector of unknown value u on all voxels. We can simply solve it by least square method.

**Modified norm.** It is obvious that feature orientation is more important than exact edge length. The orientation can be improved by less penalizing stretch which is in the direction of the desired iso-lines. In order to achieve this, [9] have introduced an anisotropic norm and we extend it to 3D vector computing:

$$||(u, v, w)||_{(\alpha,\beta,\gamma)} = \alpha u^2 + \beta v^2 + \gamma w^2.$$

This norm penalizes the deviation along the major directions with different weights. Then we modify the energy equation to the new form:

$$\sum_{v_i \in V} ||\nabla u_i - u_i||_{(\ ,1,1)} + ||\nabla v_i - v_i||_{(1,\ ,1)} + ||\nabla w_i - w_i||_{(1,1,\ )}, \tag{5.10}$$

with $\ \leq 1$.

## 5.4 Spline Approximation and Experimental Results

The previous steps generate a set of regular structured parametric patches thus it is very straight forward to define a regular high-order representation to approximate the shape and the density function of each patch. In our framework, we utilize T-splines for final approximation. A trivariate T-spline [151] can be formulated as:

$$F(u, v, w) = \frac{\sum w_i p_i B_i(u, v, w)}{\sum w_i B_i(u, v, w)}, \tag{5.11}$$

where (u, v, w) denotes parameter coordinates, $p_i = (X_i, Y_i, Z_i, I_i)$ denotes each control point, $w_i$ and $B_i$ are the weight and blending function sets. Each pair of $< w_i B_i >$ is associated with a control point $p_i$. Each $B_i(u, v, w)$ is a blending function given by $B_i(u, v, w) = N^3_{i0}(u) N^3_{i1}(v) N^3_{i2}(w)$, where $N^3_{i0}(u)$, $N^3_{i1}(v)$ and $N^3_{i2}(w)$ are cubic B-spline basis functions along u, v, w, respectively. We choose T-spline because it has two significant advantages: First, the refinement of control mesh is subdivided locally to reduce a large percentage of superfluous points and thus enhances the simplicity and accelerates the potential visualization applications; Second, T-spline scheme guarantees $\sum_i w_i B_i(u, v, w) \equiv 1$ across the entire space. Thus the computing of $F(u, v, w)$ and its derivatives can be much more efficient. We notice that, although our domain is globally consistent, each patch is treated as a single object and an independent T-spline in order to better approximate sharp feature.

### 5.4.1 Experimental Results

We introduce our experimental results in this section. A prototype system is implemented on a PC with 3.5GHz P4 CPU and 4GB RAM. We consider the Atom, Fuel, Ankle and Tooth as the test models, and use T-splines to approximate the density function based on our domain. Fig. 5.7 shows the continuous representation results. Compared with the original discrete data, reconstructed models perfectly preserve the shape and density information of the object. They also completely remove the background noise and simplify the procedure of transfer function design for the user. Fig. 5.8 shows more details about our parameterization: the corner points, parameter domain, surface parametrization and volumetric parametrization respectively. Table 5.1 summarizes the statistics of the performance of our processing on four models. These figures and tables showcase that our system effectively reconstruct the model with lower number of control points without sacrificing visual quality.

Table 5.1: Statistics of various test examples: $N_d$, # of voxels; RMS, root-mean-square fitting error (density only, $10^{-2}$); $N_c$, # of corners; $N_c^0$, # of control points.

| Model | $N_d$ | RMS | $N_c$ | $N_c^0$ |
|---|---|---|---|---|
| Atom | $256^3$ | 0.122 | 12 | $1.5 * 10^4$ |
| Fuel | $64^3$ | 0.877 | 16 | $7.2 * 10^4$ |
| Ankle | $128^3$ | 0.422 | 12 | $1.6 * 10^4$ |
| Tooth | $256^2 \times 161$ | 0.393 | 24 | $5.1 * 10^4$ |

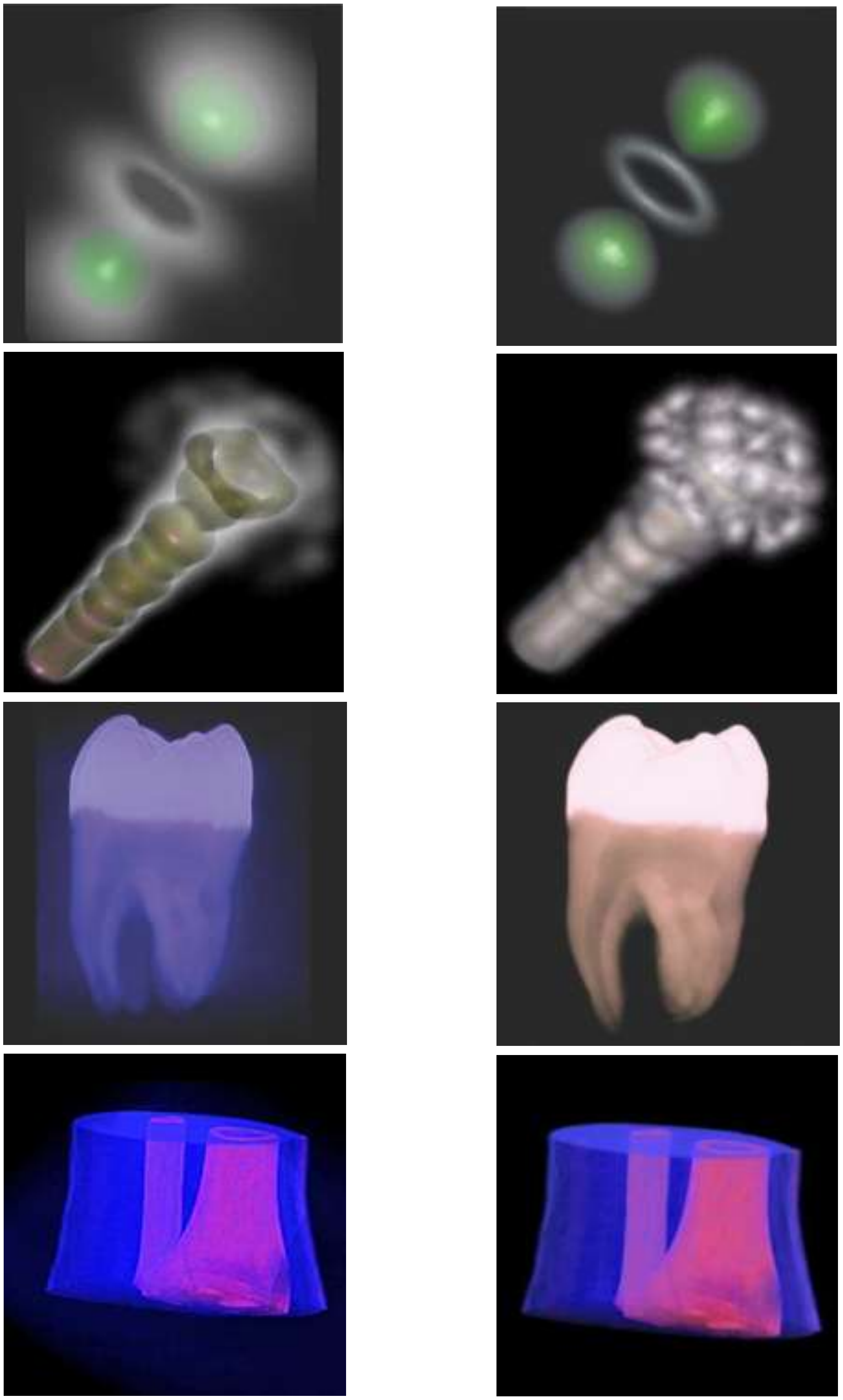

Figure 5.7: Left column: Volume visualization using input discrete models; Right column: Reconstructed models.

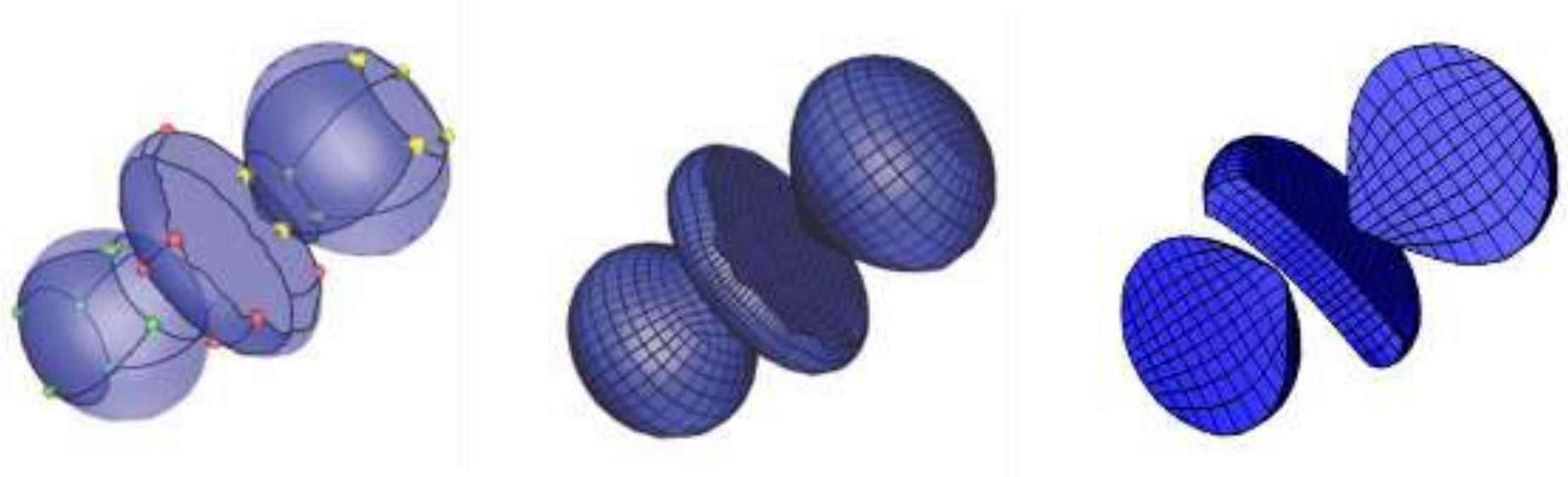

Figure 5.8: The atom model. Left column: Corner points and parameter domain. Middle column: Surface parametrization. Right column Interior parametrization.

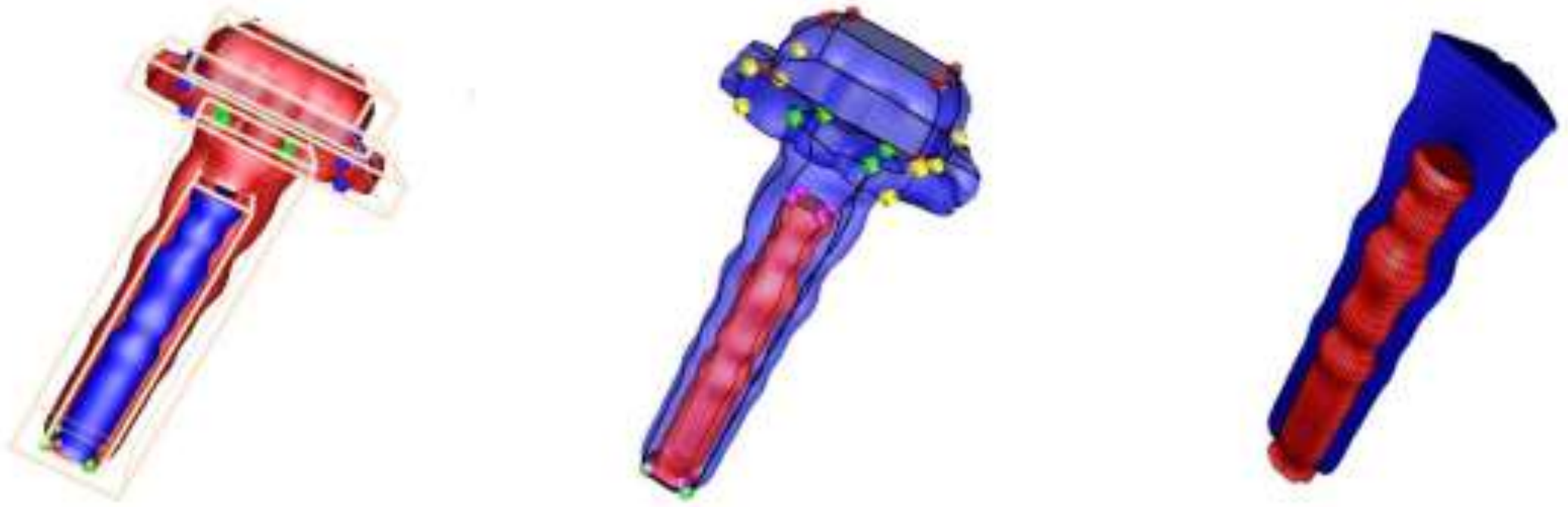

Figure 5.9: The fuel model. Left column: Corner points and parameter domain. Middle column: Surface parametrization. Right column Interior parametrization.

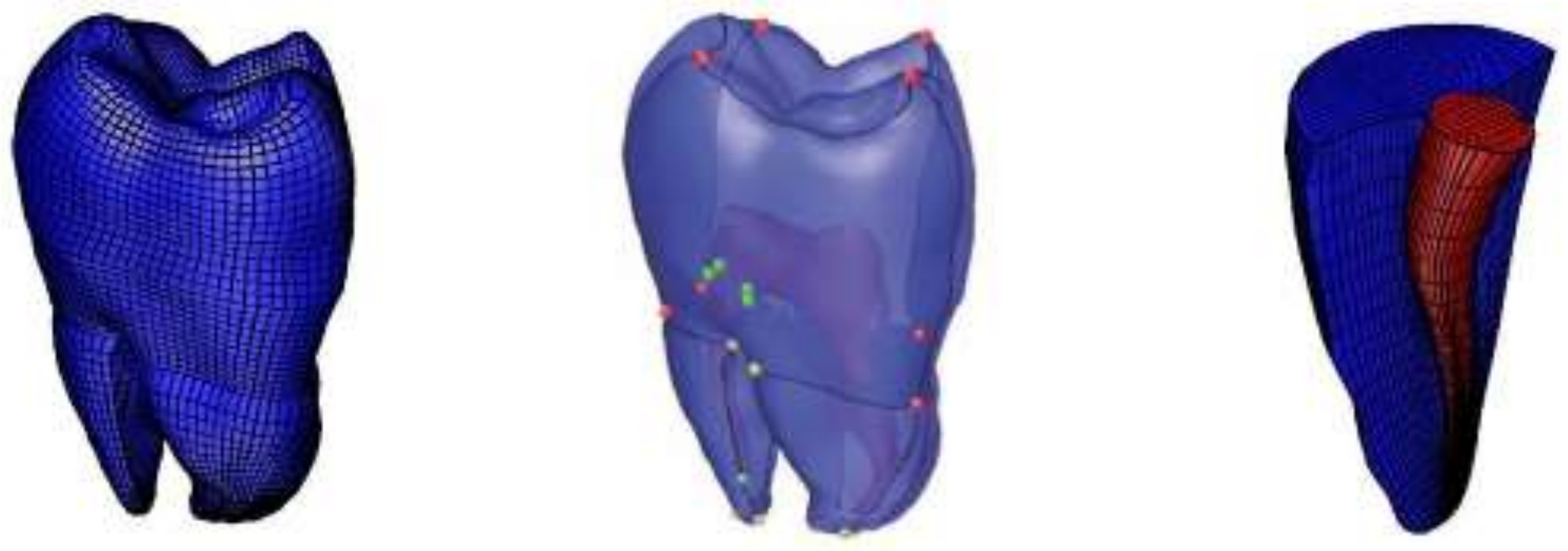

Figure 5.10: The tooth model. Left column: Corner points and parameter domain. Middle column: Surface parametrization. Right column Interior parametrization.

## 5.5 Chapter Summary

In this chapter, we have proposed a novel method that reconstructs the discrete volumetric data into the regular continuous representation. We start with the computing of principal curvatures on a hyper-volume and then find reliable feature-aligned constraints. Then we compute a smooth field respecting the most dominant shape features. Corner points are then computed and placed at geometrically meaningful locations. Based on the frame field, we can generate a regular parametrization which takes material feature-alignment constraints into account, producing a small number of regular patches. We construct trivariate T-splines on all patches to approximate geometry and density functions together. Our test results clearly verify our design.

Our framework perfectly promises a lot requirements in visualization such as feature-alignment, compactness, regular structure, high-order representation and as-homogenous-as-possible, etc. These modeling advantages naturally prompt us to explore its uncharted potential in the near future. We anticipate further novel GPU-accelerated isosurface direct visualization techniques based on our high-order regular representations. Meanwhile, the conjunctions between material-based physical analysis/simulation and our continuous hyper-volume shape functions are of great interest for potential physics-based applications.

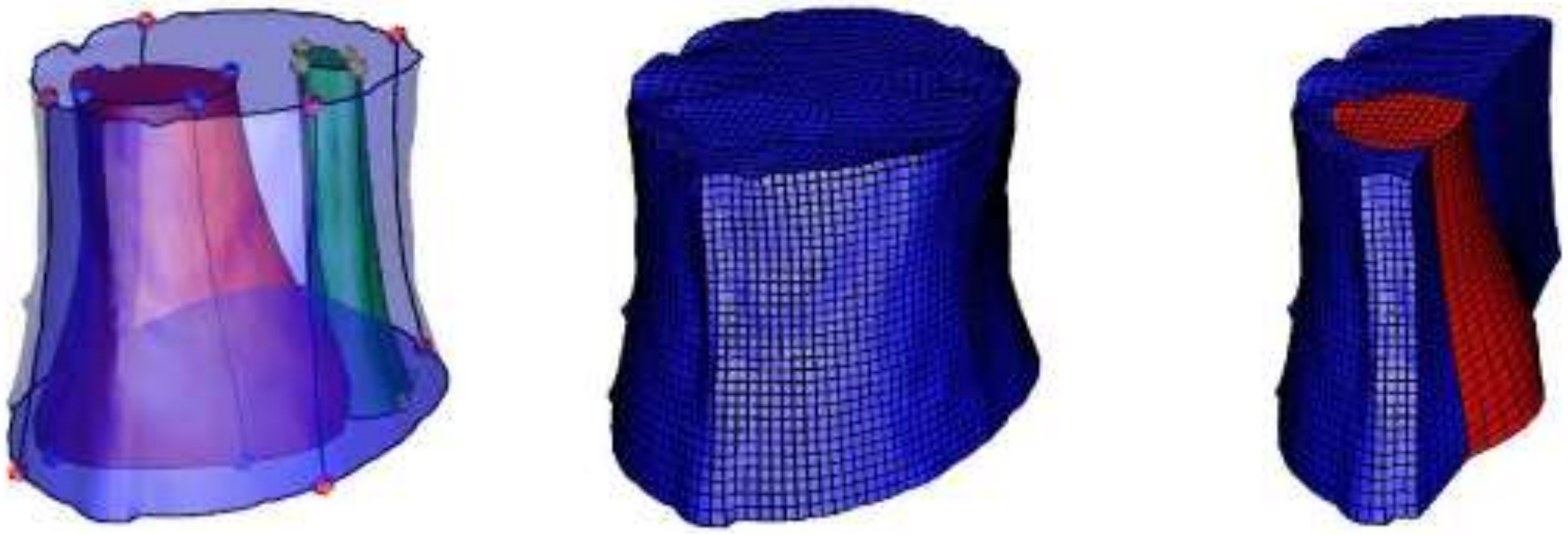

Figure 5.11: The ankle model. Left column: Corner points and parameter domain. Middle column: Surface parametrization. Right column Interior parametrization.

# Chapter 6

# Metrics-based Focus+Context Lens

In all previous chapters, we have discussed the techniques about "how to construct volumetric parameterization, spline construction and representation". Here, we argue that our volumetric parameterizations techniques also have various applications on computer graphics and visualization research. Therefore in this chapter, we attempt to study "how to apply the developed volumetric modeling techniques in other possible research areas".

As we introduced in Chapter 2, there is a stronger-than-ever need for visualizing large-scale datasets in various science/engineering applications. Meanwhile, with the explosive emergence of various types of portable devices (e.g., iPad), the industry frequently pursues as-large-as-possible data visualization on physically-small-sized screen of mobile device in recent years. Therefore, a careful tradeoff is required to deal with the potentially conflicting requirement of the inherent screen size limitation and ever-increasing data size. Focus+Context visualization offers a good strategy when tackling this problem.

Our ultimate goal is to design a flexible F+C methodology on 3D volume image. Therefore, we attempt to design a practical algorithm framework to support this idea. In this chapter, we first apply this framework onto 2D image data as the first step to 3D application. This choice is natural and necessary, because our idea is based on geometric modeling techniques and all relevant numerical computations on 2D manifolds are more mature, stable and robust than on 3D manifolds. Therefore, we decide to adapt it on image operations to test its efficiency.

In essence, we can view our core framework as a "reverse-parameterization" process. Instead of mapping a high-dimension object into a low-dimension space, we attempt to reversely map a low-dimension object into a high-dimension space, such that the visual information is enlarged. In the following sections we will discuss the algorithm in details.

## 6.1 Motivation

Focus+Context (F+C) visualization, as a natural solution, has gained much research momentum recently. In order to display regions of interest (ROIs) with high resolution, F+C allows the user to access and address the detail of interest ("Focus") while still keeping the overall content of the whole data to accommodate human cognitive custom ("Context"). Attractive F+C visualization should consider the following quality-centric aspects:

(1) **Shape-preserving.** Shape (such as angle, rigidity) plays a crucial role during magnification when improving the visual cognition. The improper magnification distortion may cause serious cognitive confu-sion.

(2) **Smooth transition.** Any visual gain from unifying the detail with the surrounding context may easily be lost if the transition between the focus and context regions is difficult to understand.

(3) **Flexibility.** For data with complex and multiple ROIs, the user may have preference for using different magnification methods or focusing on different shapes on the same input.

It is a tremendous challenge to optimize the output simultaneously with respect to all of these criteria. For example, many recent methods attempted to simulate optical lenses in depth (e.g., fish-eyes, bifocal lens) for magnification. The most challenging side effect is that, it rarely considers shape-preserving and smooth transition, thus lens distortions are intolerable when features become sufficiently intricate.

Inspired by recent image manipulation techniques such as resizing [27] [154], our new idea is to address the lens design and simulation problem using novel geometric modeling methods. The F+C visualization is then solved by a deformation metric design and optimization solution. This way, we examine this conventional 2D deformation task from a completely innovative perspective of 3D geometric processing. Rather than minimizing deformation energy on a 2D image/grid, we transform the 2D input to 3D mesh, and then conduct 3D deformations which minimize the shape distortions and magnify the ROIs. To achieve our goal, we design a novel deformation framework that functionally acts as a “lens”. We first build a special 3D mesh (*“Lens-Mesh”*) that magnifies any area of interest while keeping the rest of area with little distortion. Then, we automatically deform the lens-mesh back into 2D space for viewing. Both steps require us to find distortion minimization for each individual mesh element with an appropriate family of geometric metrics.

In this chapter we present a general theoretical and computational framework, in which 3D geometric modeling techniques can be systematically applied to the 2D lens simulation. The main contributions of our lens design and simulation include: (1) Our algorithm minimizes the geometric deformation metric distortion thus it is particulary suitable to satisfy the shape preserving property. Moreover, our deformation scheme lets the deformable mesh locally confine the resulting distortion with great flexibility rather than letting the distortion uniformly spread throughout the nearby spaces; The resulting transition between the focus and context regions is also smooth and seamless; (2) Instead of only using lenses with a regular circle or square shape, it is very easy to design an arbitrary shape of magnifiers using our lens-mesh to adapt various shapes; (3) The user can iteratively specify the geometric metrics, which allows easy production of visually pleasing effects. The whole algorithm is shown to be of high efficiency, because of the computation of a linear system with pre-processing.

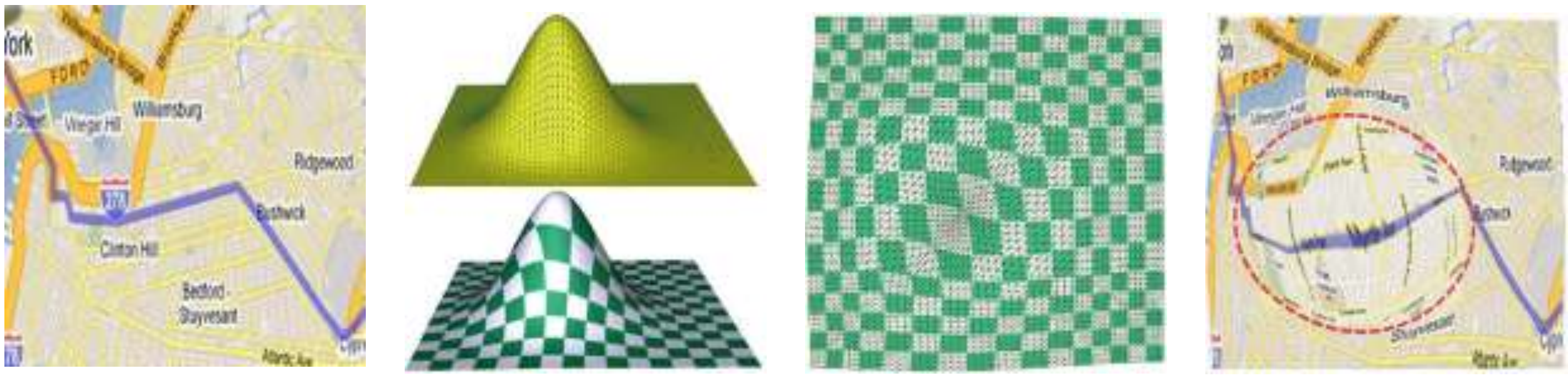

(a) (b) (c) (d)

Figure 6.1: An example of our entire framework: (a) The input image. (b) We generate a 3D lens-mesh to magnify the area of ROI. Then we transfer the texture from the input to the lens-mesh. (c) We deform the lens-mesh back into a 2D plane with minimized distortion. (d) Finally we get a new 2D image with area of ROI magnified.

## 6.2 Framework

This section gives a high level overview of our proposed framework. Our system takes as input a ready-to-display 2D image. For 3D dataset (e.g., volume datasets and 3D scanning models), we can convert them into the 2D format through volume rendering. In geometric deformation, we can consider our input as a 2D regular grid mesh G. Given that, the following steps generate as the output a 2D triangle mesh that contains colors (as texture) in each deformed triangle. Fig. 6.1 shows an example using our framework in a step-by-step fashion.

1. The user makes an initial choice about regions of interest (ROIs). The shape/boundary of a ROI can be determined by an automatic feature segmentation operation such as [127] or simple heuristic methods.
2. Generate a 3D lens-mesh $M_l$ based on the initial grid G in order to magnify the area of mesh on ROI.
   - (2.1) For each ROI, we deform the original 2D surface patch in ROI into a specified 3D surface, with the ROI boundary as constraints (and there is no changes outside the boundary). As a result, the area in the boundary is stretched/magnified.
   - (2.2) We transfer the texture from G to $M_l$ while satisfying the shape preserving property. To achieve this, we compute texturing coordinates $[u, v]$ on $M_l$ by solving the harmonic equation $\nabla^2 u = 0$ and $\nabla^2 v = 0$, with ROI's boundary as dirichlet boundary condition.
3. We deform $M_l$ back into a 2D plane without distortion. It includes two iteratively-executed phases. Starting from $M_l^0 = M_l$, each iteration generates a temporary mesh $M_l^t$ until the distortion of the final mesh $M_l^n$ is within the user-specified threshold.
   - (3.1) Metric registration. For each triangle $T_i$ in $M_l$, we compute its deformation metric $M_i$ (formulated as a $2 \times 2$ matrix) from its counterpart in $M_l^0$.
   - (3.2) Deformation. We determine the updated position of every vertex by solving the linear equation to approximate the above deformation metrics $M_i$ at each vertex.

The notations used throughout the article are: $v_i^t = (x_i, y_i, z_i)$ or $(x_i, y_i)$ represents a 3D/2D vertex in the lens-mesh $M_l^t$. The superscript $t$ denotes the index for the number of each iteration.

## 6.3 Lens-Mesh Generation

The input of our framework is the uniform 2D dataset. Aiming to effectively generate the 2D rendered image from the mesh model/volumetric dataset, we adapt the fragment program (initially proposed by Stemaier et al. [138]) for rendering, considering many parameters including depth, view angle, and camera position. The steps include: Cast the ray into the mesh model/volume dataset and composite the color based on the surface/volume data and transfer functions, and render the result into the frame buffer for display. The pixel color and the alpha of an image is controlled by a 1D transfer function.

### 6.3.1 Arbitrary Lens-Mesh Design

Now we generate the lens-mesh by magnifying the mesh shape inside each ROI. We have a mesh patch $\mathcal{M}_p$ for each ROI and $\partial\mathcal{M}_p$ is the patch boundary. The following algorithm seeks for stretching all vertices inside $\mathcal{M}_p$ to new positions while keeping other vertices unchanged. Our automatic algorithm consists of 3 steps.

**Step 1.** We conduct the medial axis transform for the patch $\mathcal{M}_p$, generating a central curved path $\mathcal{C}$ and each vertex $v_i$ in $\mathcal{M}_p$ a distance $d_i$ as the (normalized) shortest distance to the central path.

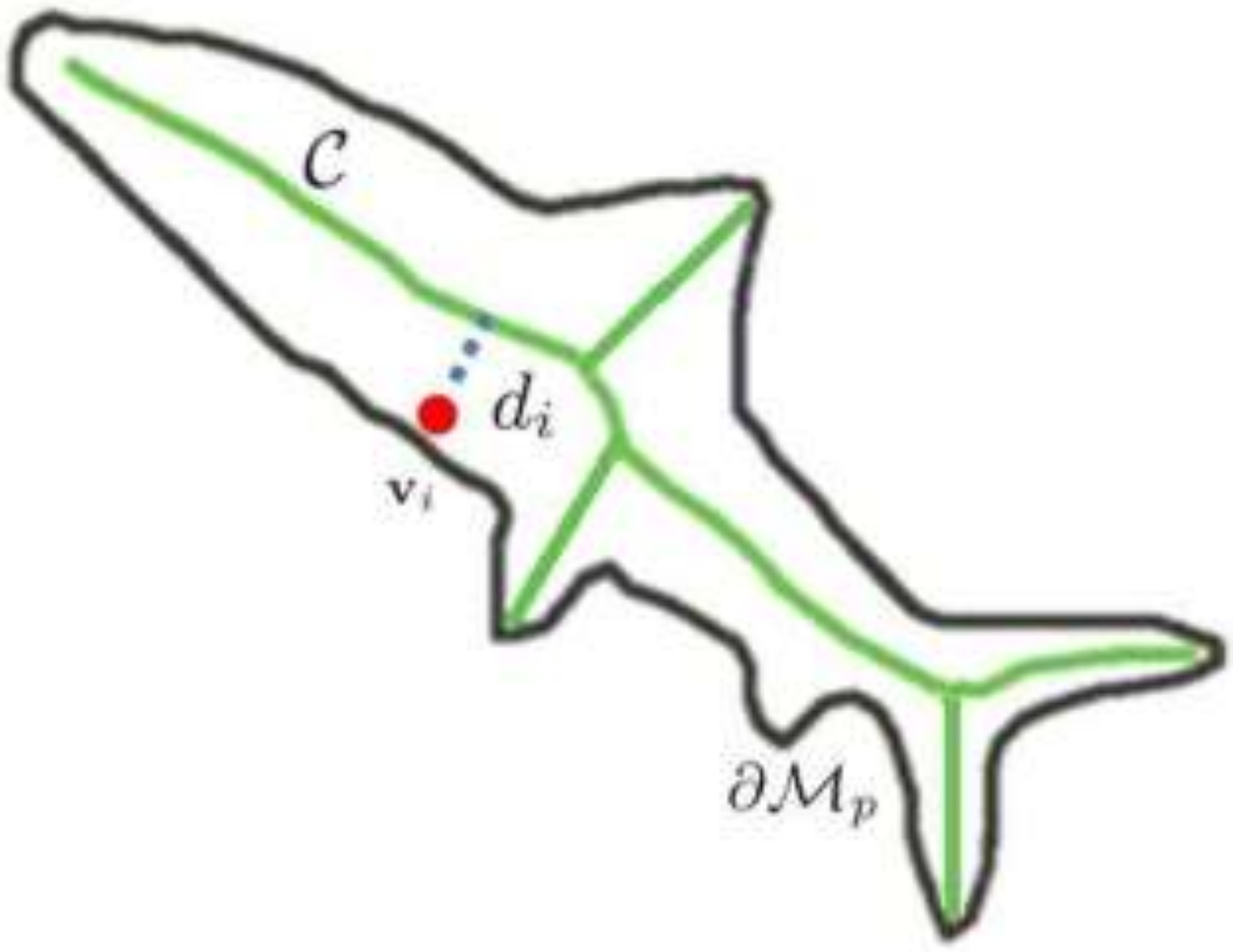

Figure 6.2: Illustration of lens-mesh design.

**Step 2.** The user decides the height $h_0$ of curved path $\mathcal{C}$. For each vertex $v_i$, we have its new position $(x_i, y_i, h_i)$, $h_i = g(1 - d_i)h_0$, where $g(x)$ is a standard gaussian function $e^{x^2}$.

**Step 3.** We subdivide the triangle in $\mathcal{M}_p$ if it is scaled or sheared too much after Step 2. Then we interpolate the locations, colors, distances and heights linearly for newly-inserted vertices.

Although the above automatic algorithm can handle versatile models very well, sometimes users still prefer to use several special shapes as the desirable lenses. For example, this choice of standard circle lens is specially natural and humans are more accustomed to it with better visual understanding. Fig. 6.3 shows different visual effects with different lens-meshes. We use the boundary of the teapot body to compute our lens in (d).

### 6.3.2 Lens-Mesh Texturing

The objective of this step is to texture the lens-mesh using the original dataset, while preserving the shape during texturing. Since both the input dataset and lens-mesh have squared boundary, we treat this problem as the energy minimization problem. We shall map the lens-mesh $\mathcal{M}_l$ to a uniform 2D domain by solving the harmonic functions $\nabla^2 u = 0$ and $\nabla^2 v = 0$, where $\nabla^2 = \frac{\partial^2}{\partial x^2} + \frac{\partial^2}{\partial y^2}$. In practice, solving equations for any but the simplest geometries must resort to an efficient approximation due to the lack of closed-form analytical solutions in the general setting, we shall use mean value coordinates to solve it numerically.

- We assign each vertex an initial coordinate, and we shall use its original 2D position $(u_i, v_i) = (x_i, y_i)$ for this purpose.

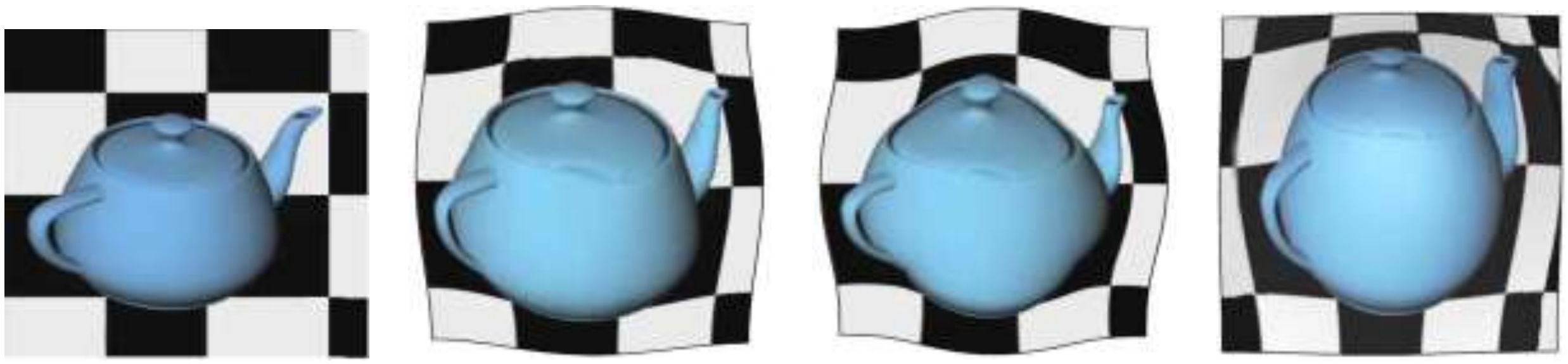

(a) (b) (c) (d)

Figure 6.3: Magnification results using different shapes of lenses for the 3D teapot mesh model. (a) Original teapot mesh model. (b-d) Magnification results using the circle-shaped, square-shaped, and automatically-generated lens-meshes, respectively.

- We iteratively update the coordinates for each vertex $(u_i, v_i) = \sum_{Ng(v_i)} w_j (u_j, v_j)$, and $Ng(v_i)$ is the one-ring neighbor of $v_i$, $(u_j, v_j)$ is a neighbor's coordinate, $w_j$ is the local mean value coordinate [32] computed on $M_l$. Two types of vertices serve as the Dirichlet boundary conditions (i.e., we avoid changing their coordinates): (1) The squared boundary only; (2) All regions outside any ROI.

## 6.4 Flattening

We search for a flattened mesh so that we can display the result on the popular flat screen (Note that, our algorithm also supports curved screen like "IMAX"). The key challenge in this problem is to preserve the important geometric deformation metric for each triangle in the $M_l$. The shape distortion can be measured as the total differences between the resulting triangles and the original triangles. We use the following algorithm to minimize the differences.

**Step 1 Standard 2D triangles.** For each triangle $T_i$ in 3D space, we reformulate it into a standard 2D triangle $T_i^d$ individually, while keeping its original shape. Suppose $v_1$, $v_2$, $v_3$ are 3 vertices of $T_i$ in 3D space, $e_1 = v_1 - v_2$, $e_2 = v_2 - v_3$, $e_3 = v_1 - v_3$ are 3 edge vectors of $T_i$. We recompute 2D positions of 3 vertices in $T_i^d$ as $v_1 = (0, 0)$, $v_2 = (||e_1||, 0)$ and $v_3 = (||e_3||\cos\theta, ||e_3||\cos\theta)$ (Fig. 6.4). $\theta$ is the angle between $e_1$ and $e_3$. Note that, a vertex in $M_l$ has different 2D positions in different $T_i^d$.

**Step 2 Flattening.** This step includes 2 iteratively computed phases. The output mesh $M_l^n$ has the same triangle mesh structure as $M_l$ while every vertex has only a 2D position. Initially, we set $M_l^0 = M_l$ and we reduce the dimension of vertices to 2D by projecting along axis-z: $v_i = (x_i, y_i)$.

(2.1) In this phase we compute the deformation metric for each triangle $T_i$. The metric represents the transformation from the localized standard $T_i^d$ to its t-th iteration counterpart $T_i^t$. We represent this transformation as a $2 \times 2$ matrix $M_i$. The computation of $M_i$ is detailed in Section 6.4.1.

(2.2) In this phase, we consider the resulting transformation matrix $M_i$ as the known parameters and update the position of each vertex.

$$E^t = \sum_{i=0}^{T} \sum_{j=1}^{3} w_{ij} ||e_i{}^t - M^t e_i{}^d||^2, \qquad (6.1)$$

j i j

where $e_{ij}^t$, $e_{ij}^d$ are edge vectors on the triangle $T_i^t$ and standard triangle $T_i^d$. We rewrite the function in terms of every edge vector:

$$E^t = \sum_{i,j} w_{ij} ||(v_i^t - v_j^t) - M_{T_k}(v_i^d - v_j^d)||^2, \tag{6.2}$$

where each pair of $(v_i, v_j)$ belongs to the triangle $T_k$ (Note that $(v_i, v_j)$ and $(v_j, v_i)$ are 2 different vectors that belong to different triangles). Setting the gradient to zero, we obtain the following linear equation:

$$WV^{t^T} = WMV^{d^T}, \tag{6.3}$$

where $W$ is the matrix depending only on $w_{ij}$, $M$ depends only on $M_{T_k}$, $V^t$ and $V^d$ are vectors of unknown/standard positions of vertices.

**Pre-factorization.** We observe that the above matrix $W$ depends only on the geometry of $M_l$. Thus this sparse matrix is fixed during iterations, allowing us to pre-factorize it with Cholesky decomposition and we can reuse the factorization many times throughout the algorithm in order to accelerate the process, which has a significant impact on algorithm efficiency. The total distortion error $E^t$ converges and we end the iteration when $||E^t - E^{t-1}||$ is smaller than the threshold $\alpha$ (we set $\alpha = 0.1\%$).

**Weights.** The choice of weight $w_{ij}$ in Eq.(6.2) depends on the importance of the triangle $T_k$. The triangles around the ROI center are more preferable to preserve the shape. The distortion on a large triangle is more visually confusing than that on the tiny ones. Therefore, we design the weight as $w_{ij} = (1 + h_k)A_k \cot(\theta)$, where $A_k$ is the area of the triangle $T_k$, $h_k$ is the averaged height (z-values) of the triangle, and $\theta$ is the opposite angle of the edge vector $(v_i, v_j)$ in $T_k$.

**Boundary control.** The user may prefer to get a resulting shape with a square boundary. We can add additional constraints into the linear equation to get a square boundary. For every vertex $v_i^0 = (x_i, y_i)$ on the boundary, we assign it with the final position directly $v^n = (\sqrt{A}x_i, \sqrt{A}x_y)$ and use it as one constraint during all iterations. Here, $A$ is the total area of the lens-mesh.

### 6.4.1 Computing Metrics

We determine the deformation metric $M_i$ for each triangle. Two most important shape features involve angle and rigidity. Angle-only metric [73] belongs to the family of conformal maps and may be computed easily by solving a simple linear system. However, it is not the most area-preserving one, thus it may have a negative effect on our magnification result. On the other hand, rigidity-only metric [133] preserves areas much better, but since it strives to be isometric, it might cause large angle distortion via this effort. Inspired by [82], we provide an iterative method that allows the user to specify a "mixed" metric that actually blends between two metrics. Note that both angle-only and rigidity-only metric involve $2\times2$ transformation matrix. Therefore, they can both be represented as

$$M = \begin{pmatrix} a & b \\ -b & a \end{pmatrix} = \begin{pmatrix} \sigma & 0 \\ 0 & \sigma \end{pmatrix} \begin{pmatrix} \cos\theta & \sin\theta \\ -\sin\theta & \cos\theta \end{pmatrix}, \tag{6.4}$$

where $\theta$ is the rotation angle, $\sigma$ represents the scaling degree along 2 orthogonal directions respectively. Rigidity-only metric requires that $\sigma = 1$. Our blended metric algorithm is as follows:

**Step 1.** We compute the rigidity-only metric $M_r$ by the algorithm [133]. Equivalent to [53], we first compute Jacobian matrix between $T_i^t$ and $T_i^d$.

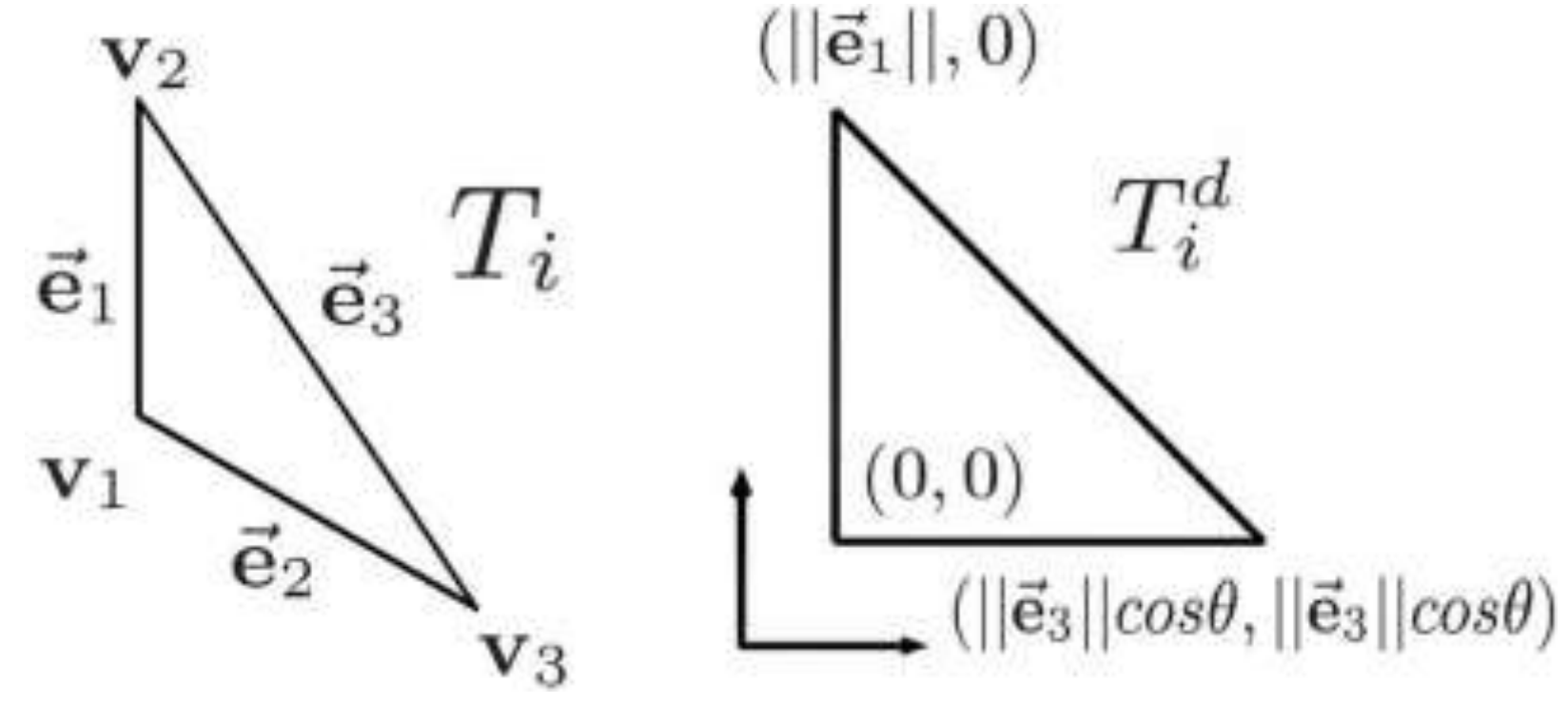

Figure 6.4: Generating a 2D standard triangle. Left: Original 3D triangle. Right: 2D standard triangle $T_i^d$.

$$\mathbf{J}(T_i^t) = \sum_{i=1}^{3} e_i{}^t (e_i{}^d)^T \tag{6.5}$$

Then the rigidity-only metric is the singular value decomposition (SVD) result of $\mathbf{J}$.

$$\mathbf{J}(T_i^t) = U\Sigma V^T, \mathbf{M}_r = UV^T. \tag{6.6}$$

Now we can reformulate it in the format like Eq.(6.4) and directly get $\mathbf{M_r}$'s rotation angle (denoted as $\theta_r$). Since $\mathbf{M_r}$ is rigidity-only, its two scaling values equal to one.

**Step 2.** We compute the angle-only metric $\mathbf{M_a}$ by solving the quadratic equation about a, b in Eq.(6.4):

$$E(a, b) = \sum_{i=1}^{3} ||e_i{}^t - \begin{pmatrix} a & b \\ -b & a \end{pmatrix} e_i{}^d||^2. \tag{6.7}$$

We can determine a, b by solving $\frac{\partial E}{\partial a} = 0$ and $\frac{\partial E}{\partial b} = 0$. Then we reformulate $\mathbf{M_a}$ like Eq.(6.4) and simply obtain the scaling and rotation parameters $\sigma_1$ and $\theta_a$ from $\mathbf{M_a}$ .

**Step 3.** The user selects a blending parameter $\alpha (0 \le \alpha \le 1)$. The resulting matrix is formulated as:

$$M = \begin{pmatrix} \sigma^b & 0 \\ 0 & \sigma^b \end{pmatrix} \begin{pmatrix} \cos\theta^b & \sin\theta^b \\ -\sin\theta^b & \cos\theta^b \end{pmatrix}, \tag{6.8}$$

where $\sigma^b = \alpha(\sigma_1 - 1) + 1$ and $\theta^b = (1 - \alpha)\theta_r + \alpha\theta_a$.

**Algorithm 5** The flattening algorithm.

```
Input: Initial lens-mesh M_l,
       Blending parameter α
       Fitting error threshold
Output: 2D mesh M^n
```

$W = BuildMatrix(M_l)$ `// See Eq.(3)`

$Cholesky - Decomposition(W)$

**for all** $T_i \in M_l$ **do**

  `//Compute the 2D standard triangle`

  $T_i^d = 2D - Standard(T_i)$

**end for**

$Initialize(M_l^0)$

**while** $||E^t - E^{(t-1)}|| > \epsilon_t$ **do**

  **for all** $T_i^t \in M_l$ **do**

    `//Compute metrics`

    $M_a = Matrix - Angle - Only(T_i^t, T_i^d)$

    $M_r = Matrix - Rigidity - Only(T_i^t, T_i^d)$

    $M_i^t = Blend(M_a, M_r, \alpha)$

  **end for**

  `// Build and solve Eq.(3)`

  $M_l^t = Fitting(M_l^t)$

  $E^t = FittingError(M_l^t, M_l^d)$

  $t = t + 1$

**end while**

## 6.5 Experimental Results

Our system can effectively provide F+C information to the user, allowing the user to get detailed focal region while maintaining the integral perception of the model. The results shown in the following figures demonstrate the power of our technique. Our experimental results are implemented on a 3GHz Pentium-IV PC with 4Giga RAM. In Fig. 6.6, we test our lens using several popular data structures such as graph, city, map, and text for information visualization. Graph is an abstract data structure representing relationships or connections. For access to relative nodes or to the particularly important nodes, our lens makes it easy to find and navigate toward these nodes, as shown in Fig. 6.6(e). Our framework also improves the magnification functions for routes and satellite maps. Fig. 6.6(f) and Fig. 6.6(g) show several results of multi-scale map/satellite magnification using our lens, which reveal and magnify the additional details (e.g., additional country names) that do not appear on Fig. 6.6(b-c). Fig. 6.6(d) showcases that our framework provides the efficient scanning function for the text reading. We can place the magnifier to zoom in the focus region while the remaining regions are evenly distributed to the context area.

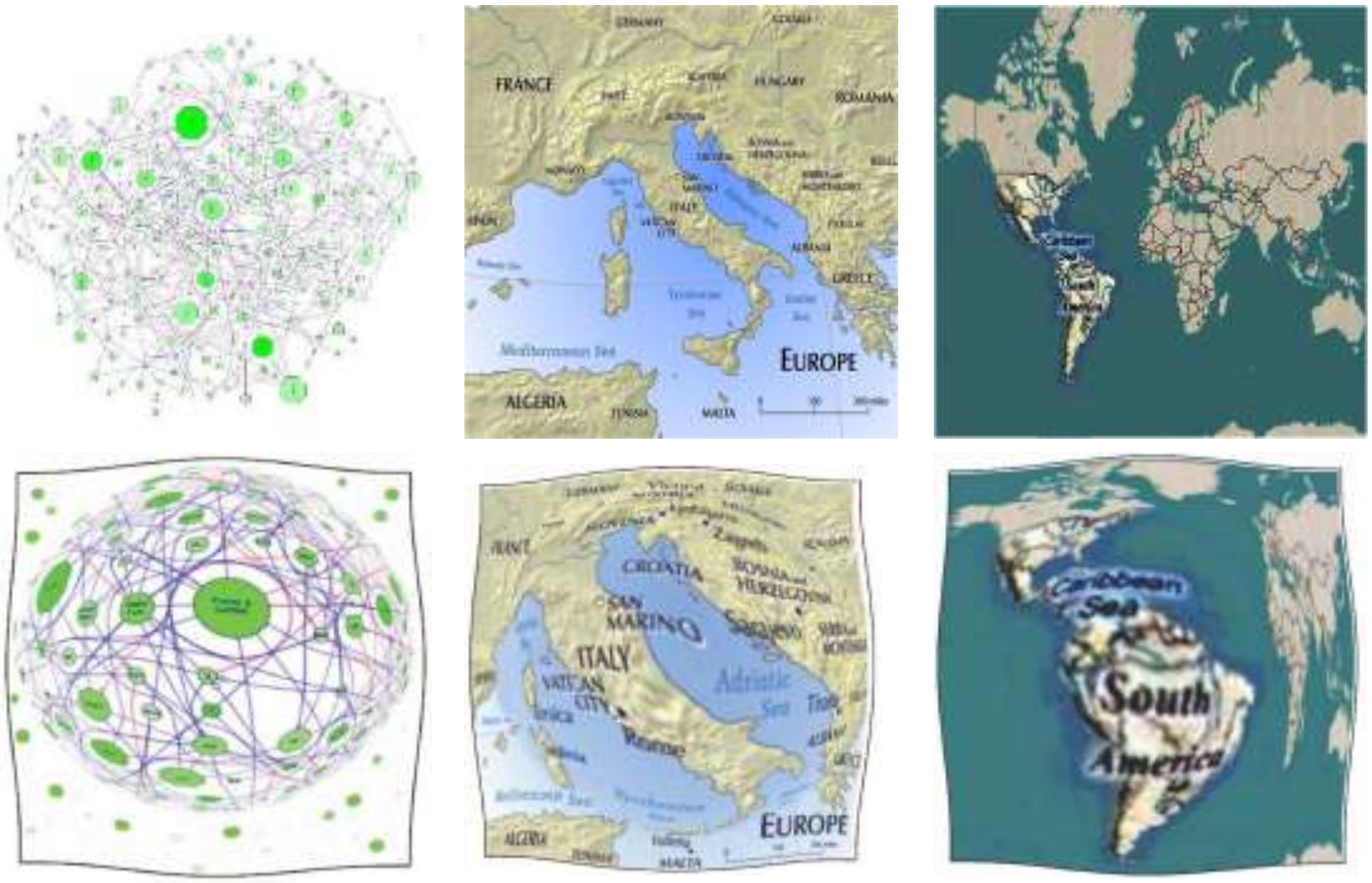

Figure 6.5: Applications of our lens simulation. Upper: Inputs. Lower: Graph of company relations, the connecting edges are revealed by the magnification; European map, major cities of Italy are revealed now; Satellite world map.

Fig. 6.1(d) is another excellent example to demonstrate that our technique offers a powerful lens for the route magnification. Normally, route display requires less disorientation to avoid getting lost, and the ability to see generals and specifics simultaneously. However, as an important widget for smartphones, the conventional interface (like “google map”) has two main shortcomings: (1) A long trip always requires multiple panning and zooming operations; (2) After magnification, the visual consistency is seriously damaged because of losing the context regions. Aiming to solve these problems, our result handles these issues very well. The global road distributions and orientations are preserved, and detailed streets are displayed around

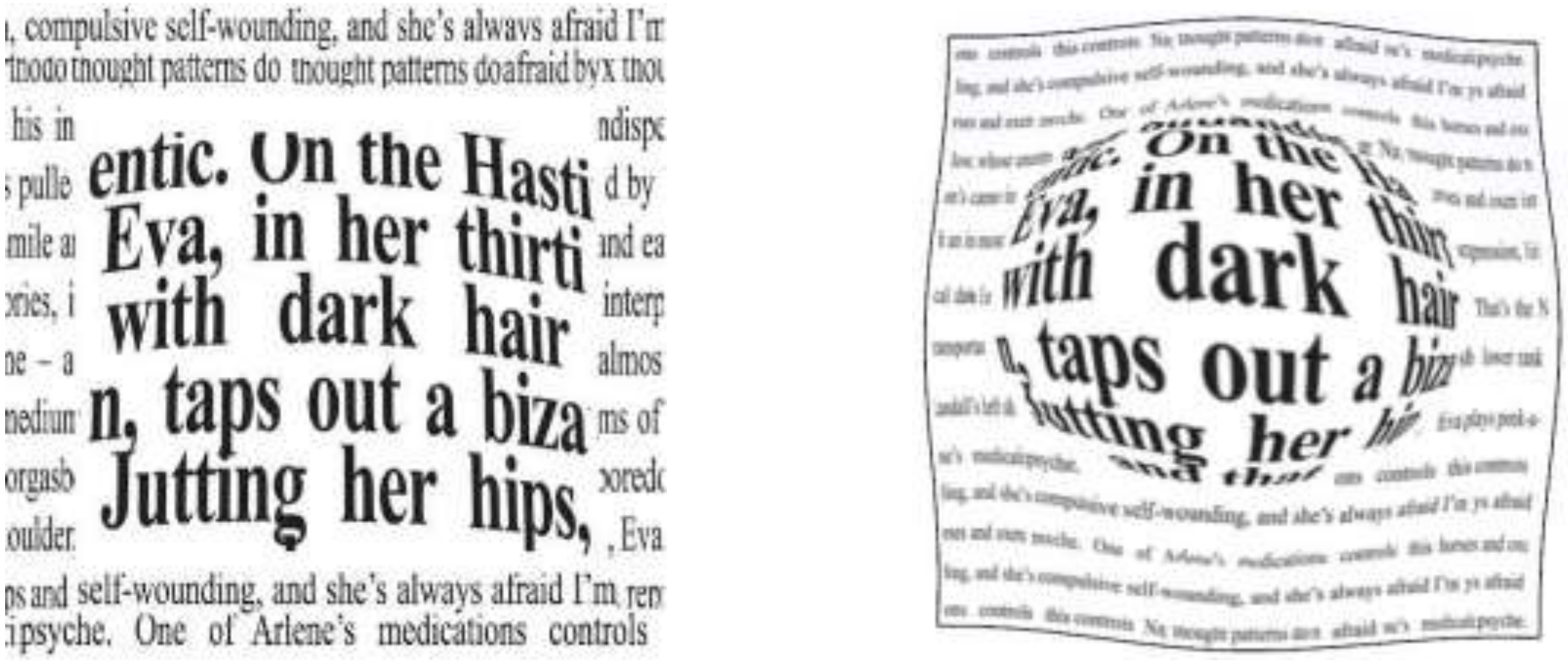

Figure 6.6: Left: Direct zoom-in. Right: F+C visualization.

ROI.

Fig. 6.7 shows a group of lenses with the same input lens-mesh but different geometric metrics, with the blending parameter $\alpha = 0, 0.001, 0.01, 1$. This blend metrics enrich the result and thus the user can modify the blending parameter to interactively change the visual effect until one result is satisfactory from the user's perspective.

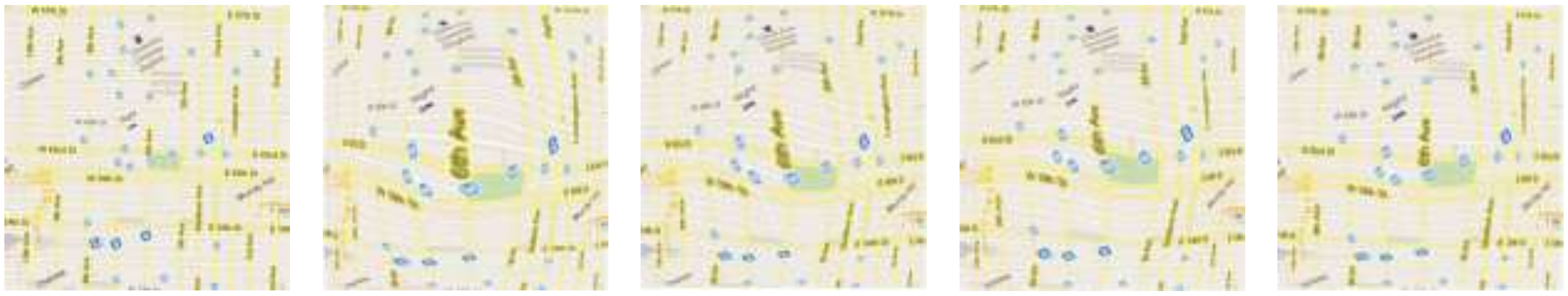

(a) (b) (c) (d) (e)

Figure 6.7: A group of different metrics with modified blending parameter $\alpha$.

**Performance.** Unlike other methods, the performance of our framework does not depend on the input image but the size of lens-mesh. So a conventional performance table ("model-by-model") is not necessary for the analysis purpose. The sample images we tested are all between $512 \times 512$ and $1024 \times 1024$. We provide two lens-meshes with sizes of $100 \times 100$ and $200 \times 200$ to handle small and large images separately. The smaller lens-mesh (10k vertices) uses only 0.3 second for one iteration and it always converges in $2 - 3$ iterations. We use the larger lens-mesh (40k vertices) to handle very high-detailed application and it uses 1.3 seconds for one iteration. The pre-processing (matrix assembling and pre-factorization) requires only about 1.0 second.

**Distortion.** Similar to Eq.(6.2), we apply the following term to measure the shape distortion on every triangle $T_i$:

$$E_i = \sum_{j=1}^{2} w_{ij} ||e_j - M_i e_j^d||^2 . \quad (6.9)$$

Fig. 6.8 compares the distortion between our lens and poly-focal lens [14]. The comparison is mean-ingful because both methods allow “free-boundary” to obtain better shape-preserving effects. To measure the distortion of poly-focal lens, we also consider their resulting image as a deformed mesh with each vertex/color moving to the new position. Thus we can also use the same criteria to measure the shape distortion. The color indicates that our method can reduce the shape distortion in a much better way. We use blue color to represent zero distortion and red the maximum (0.45 in our result).

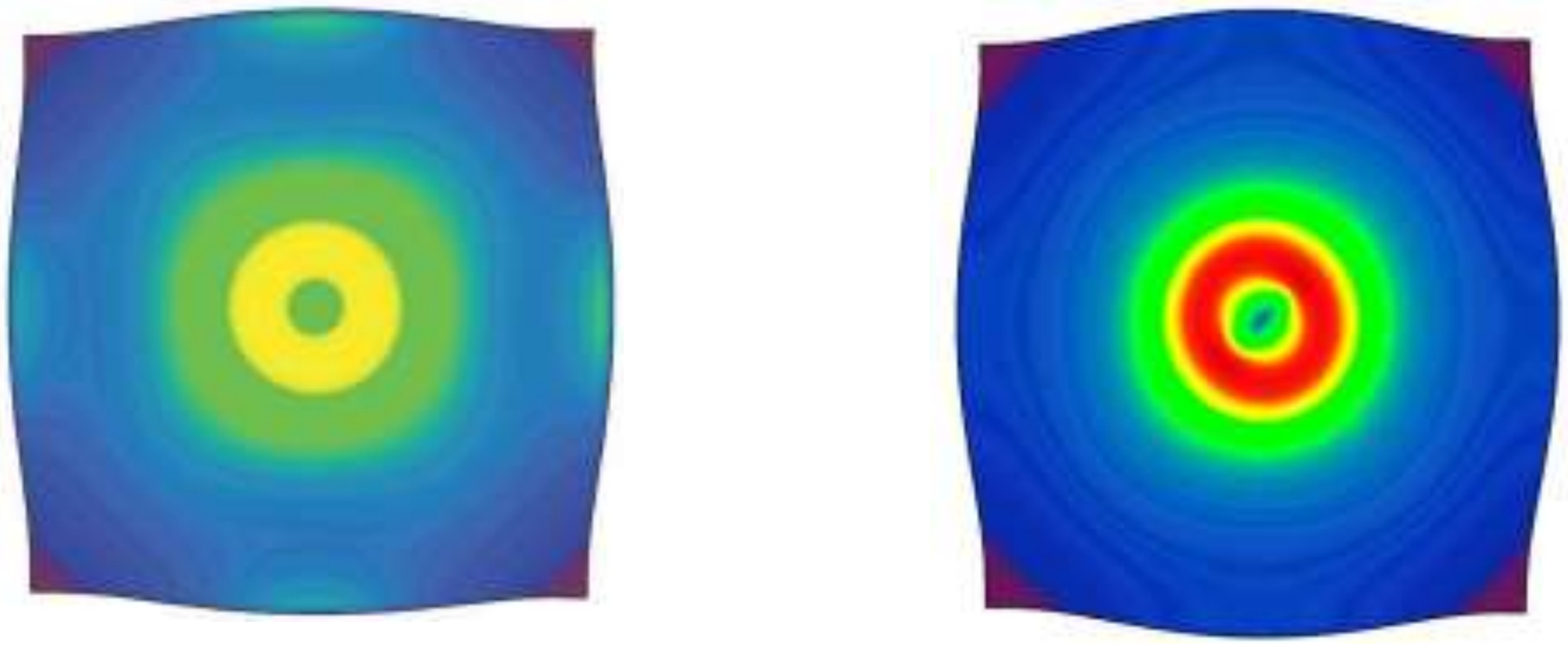

(a) (b)

Figure 6.8: (a-b) The distortion of our lens-mesh and poly-focal lens. The distortion is color-coded from blue (minimum) to red (maximum).

**Comparison for magnification results.** We apply our method to a volumetric colon dataset to verify the advantages of our lens and compare with others as shown in Fig. 6.10. Local shape preservation and smooth transition have important applications in the clinic education, diagnose, and even virtual surgery. In the normal clinic exam, the colonoscopy needle navigates along the colon axis and the lens is added along the same direction such that the clinicians are able to recognize polyps on the folds (the wrinkles on the colon wall, red circle). The folds in Fig. 6.10(b-c) are seriously distorted which may sabotage the clinicians' expertise on polyps detection. No matter how we modify their lenses in (b-c), the distorted folds always exist along the lens boundary. In sharp contrast, the fold details in (d) are better preserved and easy for recognition.

Fig. 6.9 compares our method with another technique [154]. After setting the user-selected focus region (red circle in (a)), the magnification result generated by Wang's method preserves structure/shape in the focus area, but seriously affects the context region (the upper body, red circle in (b)) and introduces visual artifacts for the proportion of body. Wang et al. [154] have also mentioned this problem as one of their major limitations. By comparison, our technique renders the result keeping upper/lower body proportion, without obvious shape confusion that may have negative influences on the accuracy of object cognition.

We compare our method with other approaches, like zoom-in, fish-eye, bi-focal, perspective wall, poly-focal [14] and cube deformation [153] in Table 6.1. Our method has advantages in the following aspects. First, our solution works well particularly with the complex shape, because it can flexibly design arbitrary shapes for lenses. Our method emphasizes angle and rigidity metrics for the shape-preserving purpose. Moreover, it allows the user to interactively design and blend various metrics.

**Limitations.** Our system flattens the lens-mesh to achieve F+C visualization, but potentially it may result in flip-over phenomenon (i.e., he resulting triangle covers another one or its orientation is reversed). Fortunately, this phenomenon always happens especially on a highly curved surface with complex topology. In contrast, our lens-mesh is relatively very simple compared with common models used in geometric mod-eling study and there are no flip-over triangles in all examples during our experiments. The texturing step (Section 6.3.2) also produces a fine mapping as a good initial guess. Meanwhile, we can always solve the flip-over problem using the existing algorithm [77].

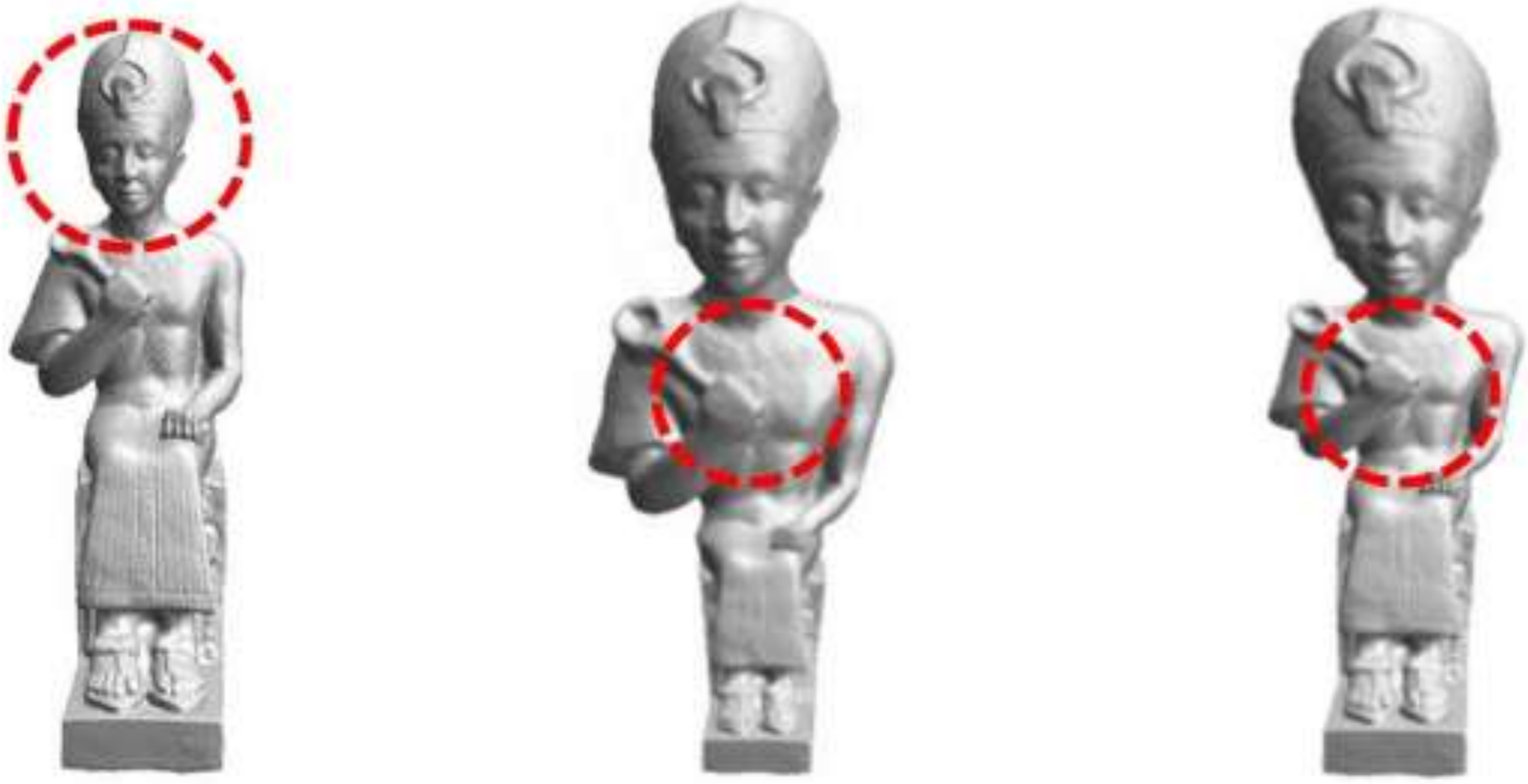

(a) (b) (c)

Figure 6.9: Comparison between Wang's method [154] and our lens.

Compared with the direct zoom-in and bi-focal methods, our method can not authentically keep exactly the same feature of a local region as the original input. Also, our metric lacks of the measurement to preserve the global structure, shape symmetry, or long straight lines. However, our human cognitive system for recognition is accustomed to automatically compensating these slight variations of a local region and thus it relieves possible disdisturbing experience for the user.

Table 6.1: Comparison with the existing approaches.

| Method | zoom-in | fish-eye | bi-focal | perspective wall | poly-focal | deformation | our method |
|---|---|---|---|---|---|---|---|
| Shape preserving | yes (focus region) | no | yes (focus) no(transition) | no | no (focus) yes (transition) | yes | angle+ rigidity |
| Smooth transition | no | yes | no | no | yes | yes | yes |
| Arbitrary lens shape | no | no | no | no | no | no | yes |
| Interactive metric design | no | no | no | no | no | no | yes |

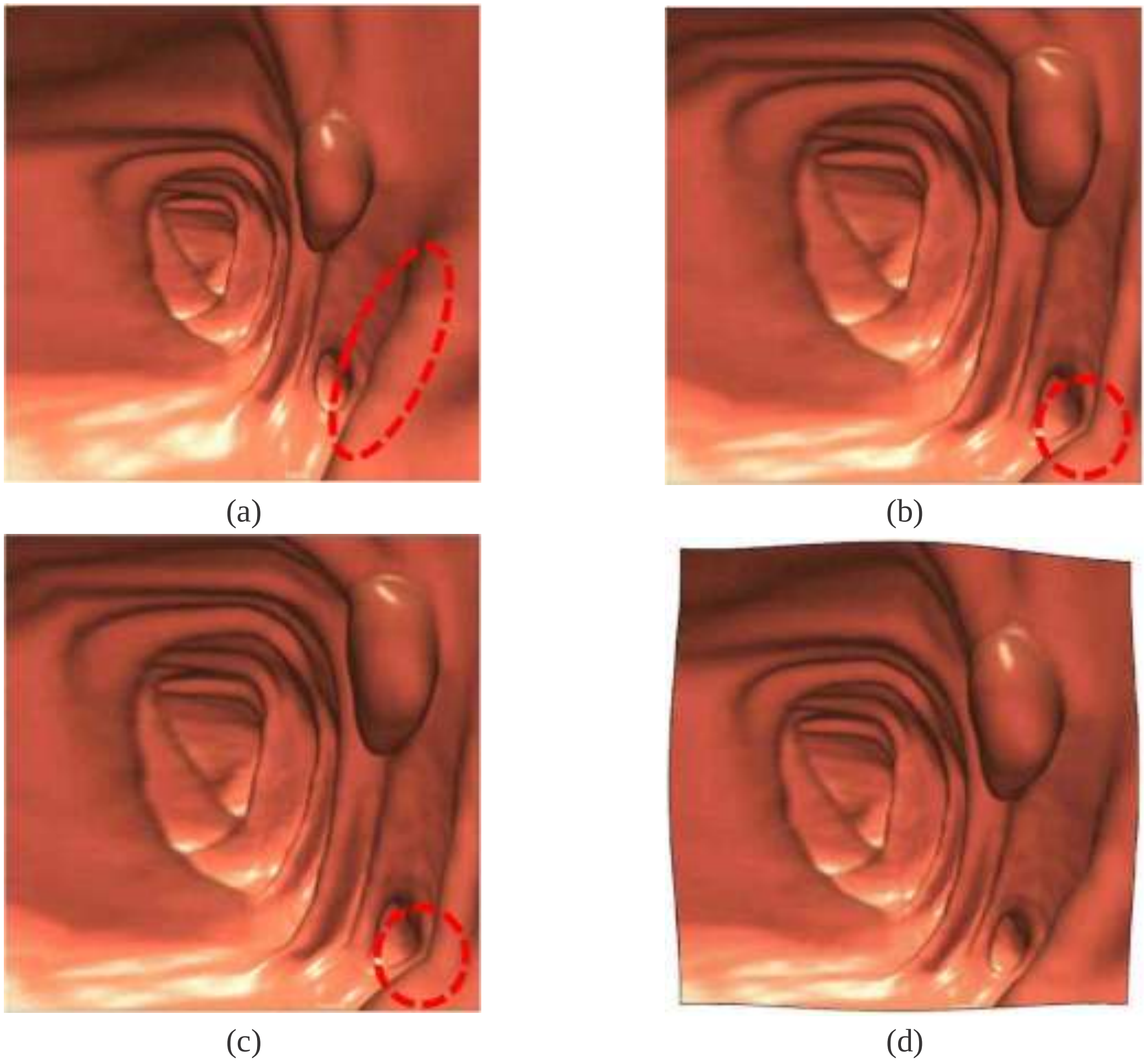

Figure 6.10: Magnification results using different lenses for volumetric colon dataset. (a) Original colon dataset. (b-d) Magnification results using bifocal, polyfocal, and our lenses. By comparison, the folds on the interior colon surface are seriously distorted by all the other lenses because of the sharp transition between the focus and context regions, while our lens shows the accurate shapes/features of the interior colon surface without any obvious distortion.

## 6.6 Chapter Summary

We have developed a novel and interactive technique to achieve Focus+Context visualization based on geometric deformations. Specifically, we develop from the input a 3D lens-mesh and magnify the ROIs through deformation on the lens-mesh. Our lens design methodology and the prototype system manifest that the geometric deformation metrics greatly enhance the F+C visualization, and our approach is expected to transcend the traditional boundary of geometric modeling and will benefit data visualization and visual analytics.

The important features of our framework can be summarized as: (1) **Shape-preserving.** The geometric deformation metrics are minimized so that the resulting details appear similar to their original counterparts. Geometric deformation also generates a continuous transition region where the user can get a smooth viewing transition from the highly-magnified interior region to the non-magnified exterior region; (2) **Robustness.** It enables the user to design arbitrary number/shape of magnifiers to effectively display the entire

ROIs for visualization of multiple and complex features. It also allows the user to interactively specify geometric metrics for various visual effects; (3) **Efficiency.** The computation is very efficient because of our pre-factorization processing. Our experimental results have demonstrated that our lens, as a novel F+C technique, has great potentials in many visualization applications.

# Chapter 7

# Conclusion and Ongoing Work

## 7.1 Thesis Proposal Summary

In this thesis proposal, we have investigated and presented a spline-based volumetric modeling framework to solve 3D objects modeling problems. Particularly, we emphasize our research interest on regular domain ("cuboid") tensor-product splines, because of their favorite advantages. Combining volumetric decomposition, parameterization with trivariate splines, we successfully and effectively solve a variety problems in the areas of geometric shape design and modeling.

Our specific contributions include:

1. We propose a new concept of *"Generalized poly-cube"* (GPC). A GPC comprises a set of regular cube domains topologically glued together. Compared with conventional poly-cubes (CPCs), GPC is much more powerful and flexible and has improved numerical accuracy and computational efficiency. We propose an automatic method to construct a GPC domain and we develop a novel volumetric parameterization and spline construction framework based on the resulting domain, which is an effective modeling tool for converting surface meshes to volumetric splines (Chapter 3).

2. We design a novel component-aware shape modeling methodology based on tensor-product trivariate splines for solids with arbitrary topology. Instead of using conventional top-down method, our framework advocates a divide-and-conquer strategy: The model is first decomposed into a set of components and then each component is naturally modeled as tensor-product trivariate splines. The key novelty lies at our powerful merging strategy that can glue tensor-product spline solids together subject to high-order global continuities, meanwhile preserving boundary restriction and semi-standardness (Chapter 4).

3. We propose a systematic framework that transforms discrete volumetric raw data from scanning devices directly into continuous spline representation with regular tensor-product structure. To achieve this goal, we propose an novel volumetric parameterization technique that constructs an as-smooth-as-possible frame field, satisfying a sparse set of directional constraints and compute a globally smooth parameterization with iso-parameter curves following the frame field directions. The proposed method can efficiently reconstruct model with multi-layers and heterogenous materials, which are usually extremely difficult to be handled by the traditional techniques 5.

4. Aiming to promote new applications of our powerful modeling techniques in visual computing, we present a novel methodology based on geometric deformation metrics to simulate magnification lens that can be utilized for Focus+Context (F+C) visualization. Compared with conventional optical lens design (such as fish-eyes, bi-focal lens), our geometric modeling based method are much more capable of preserving shape features (such as angles, rigidities) and minimizing distortion 6.

Practically, we demonstrate their power in many valuable applications, and show their great potential as enabling tools serving for research in broad areas of computer graphics, geometric modeling and processing. Our spline-based framework is endowed with many advantageous properties for modeling continuous quantities defined over multiple domains. Through our extensive experiments, we demonstrate that our framework is more efficient and effective in solving a variety of problems in computer graphics, image processing and other engineering applications.

## 7.2 Ongoing Work

At present, we have proposed a systematic framework for volumetric modeling. Theoretically, it only brings a fundamental starting progress in understanding, analysis and application on the research of volumetric modeling. It is desirable to keep investigating the great potential of developing better volumetric modeling techniques and utilizing results into more valuable applications. Particulary, we can conclude and explore the possible ongoing research topics by carefully studying the limitations of our existing research.

First, our existing volumetric parameterization techniques in Chapter 3, 4 focus on mapping to the component-aware domains. However, an ideal spline-friendly parameterization technique also requires the property of feature-aware. We have attempted to extract the "layer" feature in our volumetric modeling algorithm like Chapter 5. However, as we have argued in Chapter 2 we also seek to design new approach to offer a flexible mechanism to allow extra input, through the definition of alignment features (especially sharp edges) that are respected during the mesh generation process. We briefly explore our idea in Section 7.2.1.

Another ongoing work comes out immediately from Chapter 6: We have already studied the metrics-based F+C lens simulation on 2D images. As we mentioned at the beginning of the chapter, our ultimate goal is to achieve a practical F+C framework on 3D volume datas. Therefore, the existing methods in Chapter 6 are only partial work to our goal and we may naturally generalize the same framework onto 3D volume data as discussed in Section 7.2.2.

### 7.2.1 Template-based Feature-Friendly Volume Data Reconstruction via Trivariate Splines

Our primary goal to construct a volumetric parameterization/remeshing that mapping to a conventional poly-cube domain, while mapping important features (sharp lines/feature points) to the cuboid domain edges/corners. Currently some existing papers only focus on feature-aware quadrangulation on surface mesh. We will strengthen current framework to facilitate volumetric poly-cube generation, and more importantly, enforce points-to-corners mapping and lines-to-edges mappings.

(1) We argue that theoretically, constructing a poly-cube mapping equals to generating a quad-mesh with/only with certain points mapping to a valence-3/5 singularities. This is because a poly-cube mapping result has only convex corners (valence-3) and concave corners (valence-5) and no other singularities anymore. Computationally, current quad-mesh generation techniques (e.g., [9]) only generate singularities without nay control. We desire to attempt new constraints during parameterization optimization such that we can set selected points with required valence number. Consequently, we can select feature points as cuboid corners by deciding their valences.

(2) We attempt to propose new template-based technique for feature curves' preserving. First, we build a base domain tetrahedral mesh constrained to a few feature curves. We seek to use one tetrahedra to cover each feature curve. As the result, all feature curves become the edges of the resulting dual mesh. Using tetrahedra as the domain of tensor-product splines will definitely lead to serious distortions. Thus in the second step, we merge two neighboring tetrahedral elements into a hexahedra, which is now very suitable for generalized poly-cube parameterization, with feature curves mapping to the edges.

### 7.2.2 4D Space Volume Data Focus+Context Lens

Similar as Chapter 6, we can view our core framework as a "reverse-parameterization" process. This can be used for F+C lens design [175-177]. Instead of mapping a high-dimension object into a low-dimension space, we attempt to reversely map a low-dimension object into a high-dimension space, such that the visual information is enlarged. Since we have demonstrated efficiency and robustness of our metric-based framework, now we generalize the same idea onto 3D volume images. Specifically, we notice that the lens-mesh now becomes a 3-manifold subspace embedded into a 4D space.

(1) The user makes an initial choice about regions of interest (ROIs). The shape/boundary of a ROI can be determined by an automatic contour abstraction algorithm like "March Cube" or simple heuristic methods.

(2) Generate a 4D lens-mesh based on the initial grid in order to magnify the area of mesh on ROI. We first deform the original 3D grid patch in ROI into a specified 4D-mesh, with the ROI boundary as constraints. Then, we construct the mapping between the original 3D grid and the new generated 4D-mesh. We expect to design a mapping technique similar to [63].

(3) Unlike 2D image visualization, the 4D-mesh cannot be directly visualized. Therefore, we must "flatten" it back into 3D space. We desire to design new metric modeling representation on each 4D-mesh element and then generalize the deformation energy minimization method by adapting this new metric.